\documentclass[twocolumn]{aastex63}
\usepackage{graphicx}
\usepackage{ascii}
\usepackage{amsmath}
\usepackage{multirow}
\usepackage{enumitem}

\newcommand{\dclusterfigsize}{3.5cm}
\newcommand{\dclustergap}{0.3cm}
\newcommand{\tabledash}{\textemdash}

\newcommand{\hst}{\emph{HST}}
\newcommand{\jwst}{\emph{JWST}}
\newcommand{\spitz}{\emph{Spitzer}}
\newcommand{\desk}{{\asciifamily DESK}}

\newcommand{\agb}{346,623}
\newcommand{\dustyagb}{4,802}
\newcommand{\clusteragbs}{1,356}
\newcommand{\dustyclusteragb}{17}
\newcommand{\oagb}{20,441}
\newcommand{\odustyagb}{284}
\newcommand{\oclusteragbs}{101}
\newcommand{\cagb}{346}
\newcommand{\cdustyagb}{45}
\newcommand{\cclusteragbs}{1}
\newcommand{\dustyappended}{78}
\newcommand{\clusterchem}{102}

\newcommand{\clustersourcesyoungerthaneight}{304}
\newcommand{\fullcat}{2,325,605}
\newcommand{\chemnohst}{6,282}
\newcommand{\chembelowtrgb}{4,648}
\newcommand{\RHeB}{2,453}

\newcommand{\splashc}{76}
\newcommand{\splashcn}{34}
\newcommand{\additionalinclusion}{2,597}
\newcommand{\onlyadditionalinclusion}{1,976}

\newcommand{\renrsg}{1933}

\newcommand{\rgbs}{1,058,694}

\newcommand{\agbscaled}{1 million}
\newcommand{\boyerpercentagerecovered}{75}
\newcommand{\leopercentagerecovered}{73}
\newcommand{\secondclosestspitzmatch}{3,453}
\newcommand{\thirdclosestspitzmatch}{183}
\newcommand{\secondclosestspitzmatchpercent}{3.6\%}
\newcommand{\thirdclosestspitzmatchpercent}{0.2\%}
\newcommand{\commonfiltercut}{1,010}
\newcommand{\commonfiltercutpercent}{1.1\%}

\newcommand{\ofullcat}{20,441}
\newcommand{\cfullcat}{346}
\newcommand{\boyermedbandcat}{20,787}

\newcommand{\khaninphat}{201,381}
\newcommand{\gscinphat}{137,717}

\graphicspath{{./}{figures/}}

\received{\today}
\revised{}
\accepted{}

\submitjournal{ApJS}
\shorttitle{A Census of Metal-rich TP-AGB Stars in M31}
\shortauthors{Goldman et al.}

\begin{document}
\title{{\bf \large A Census of Thermally-Pulsing AGB stars in the Andromeda Galaxy and a First Estimate of their Contribution to the Global Dust Budget}}

\correspondingauthor{Steven Goldman}
\email{steven.r.goldman@nasa.gov}

\author[0000-0002-8937-3844]{Steven R. Goldman}
\affil{Space Telescope Science Institute, 3700 San Martin Drive, Baltimore, MD 21218, USA}

\author[0000-0003-4850-9589]{Martha L. Boyer}
\affil{Space Telescope Science Institute, 3700 San Martin Drive, Baltimore, MD 21218, USA}

\author[0000-0002-1264-2006]{Julianne Dalcanton}
\affil{Department of Astronomy, Box 351580, University of Washington, Seattle, WA 98195, USA}

\author[0000-0003-0356-0655]{Iain McDonald}
\affil{School of Physics \& Astronomy, University of Manchester, Manchester, M13 9PL, UK}
\affil{Department of Physics \& Astronomy, Open University, Walton Hall, Milton Keynes, MK7 6AA, UK}

\author[0000-0002-6301-3269]{L\'eo Girardi}
\affil{Padova Astronomical Observatory, Vicolo dell’Osservatorio 5, Padova, Italy}

\author[0000-0002-7502-0597]{Benjamin F. Williams}
\affil{Department of Astronomy, Box 351580, University of Washington, Seattle, WA 98195, USA}

\author[0000-0002-2996-305X]{Sundar Srinivasan}
\affil{Instituto de Radioastronom\'ia y Astrof\'isica, UNAM.\ Apdo.\ Postal 72-3 (Xangari), Morelia, Michoac\'an 58089, Michoac\'{a}n, M\'{e}xico}

\author[0000-0001-5340-6774]{Karl Gordon}
\affil{Space Telescope Science Institute, 3700 San Martin Drive, Baltimore, MD 21218, USA}

\begin{abstract}
We present a near-complete catalog of the metal-rich population of Thermally-Pulsing Asymptotic Giant Branch stars in the northwest quadrant of M31. This metal-rich sample complements the equally complete metal-poor Magellanic Cloud AGB catalogs produced by the SAGE program. Our catalog includes \hst\ wide-band photometry from the Panchromatic Hubble Andromeda Treasury survey, \hst\ medium-band photometry used to chemically classify a subset of the sample, and \spitz\ mid- and far-IR photometry that we have used to isolate dust-producing AGB stars. We have detected \agb\ AGB stars; these include \dustyagb\ AGB candidates producing considerable dust, and \clusteragbs\ AGB candidates that lie within clusters with measured ages, and in some cases metallicities. Using the \spitz\ data and chemical classifications made with the medium-band data, we have identified both carbon- and oxygen-rich AGB candidates producing significant dust. We have applied color--mass-loss relations based on dusty AGB stars from the LMC to estimate the dust injection by AGB stars in the PHAT footprint. Applying our color relations to a subset of the chemically-classified stars producing the bulk of the dust, we find that $\sim$\,97.8\% of the dust is oxygen-rich. Using several scenarios for the dust lifetime, we have estimated the contribution of AGB stars to the global dust budget of M31 to be 0.9--35.5\%, which is in line with previous estimates in the Magellanic Clouds. Follow-up observations of the M31 AGB candidates with the \jwst\ will allow us to further constrain stellar and chemical evolutionary models, and the feedback and dust production of metal-rich evolved stars. \\
\end{abstract}


\section{Introduction}
Intermediate-mass stars (0.8--8\,M$_{\odot}$) go through a short Thermally Pulsing Asymptotic Giant Branch (TP-AGB) phase in the final stages of their evolution. During this period, they undergo large thermal pulses and contribute a considerable amount of their material (up to 80\%) back to the interstellar medium (ISM) via dense stellar winds \citep[e.g.][]{Groenewegen2018}. As these stars lose mass, much of the circumstellar material condenses into dust grains. Collectively, AGB stars\footnote{For simplicity, we use the term ‘AGB’ to refer to both Thermally-pulsing and early-AGB stars.  Ultimately, our catalog is dominated by TP-AGB stars since most early-AGB stars are fainter than our selection criteria.  The term ‘RGB’ star is used for objects fainter than AGB stars in our catalog, though we note that these include early-AGB stars and core He-burning stars as well. See  \S3 for a description of the selection criteria for these objects.} are rivaled only by supernovae in terms of their total dust production. They contribute significantly to the dust and metals in nearby galaxies \citep{Matsuura2009,Riebel2012,Schneider2014,Srinivasan2016,Boyer2017}, and may have a similar impact in high-redshift galaxies.

\begin{figure*}
 \centering
 \includegraphics[width=\linewidth, trim=4 4 4 4,clip]{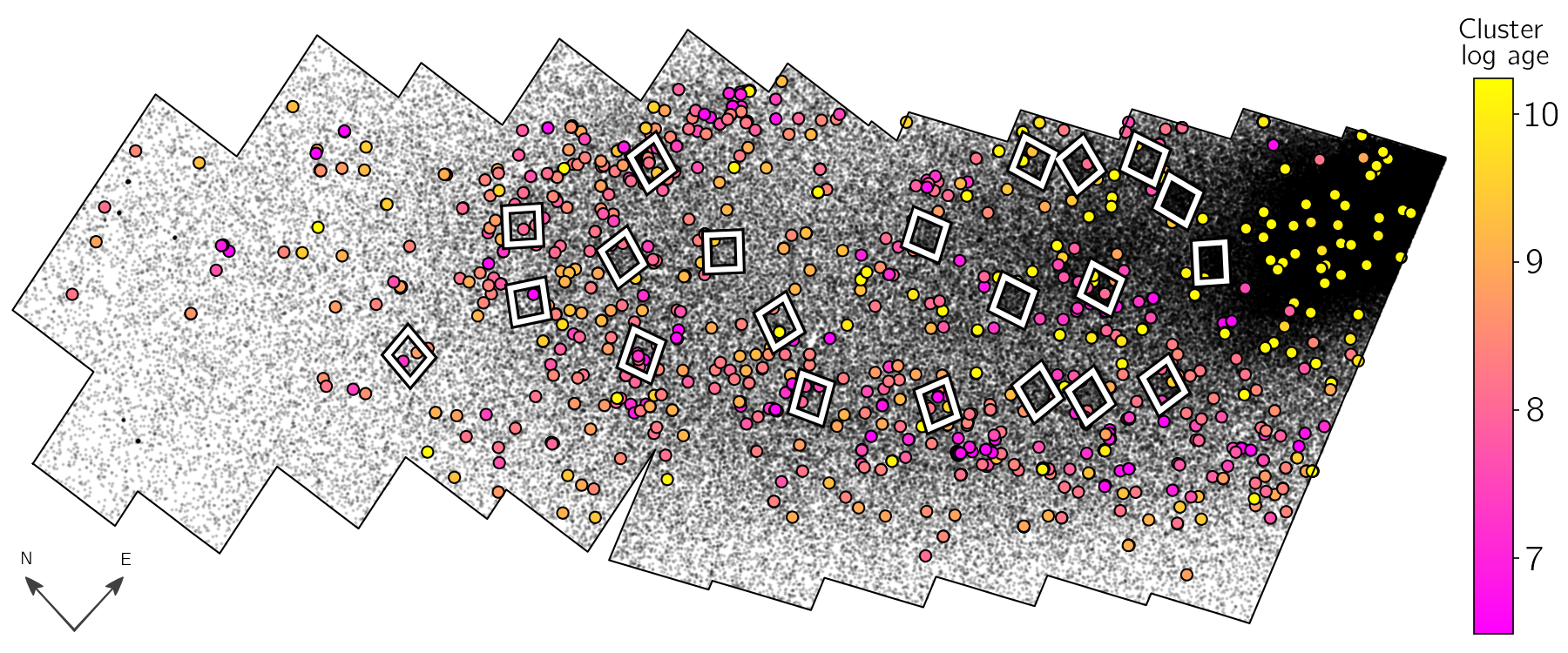}
 \caption{The spatial distribution of the AGB candidates identified here (black), outlined by the PHAT footprint. Clusters with known ages and detected AGB candidates are marked with circles (see \S4), and the medium-band \hst\ footprints where TP-AGB chemical types (carbon- or oxygen-rich) have been determined by \citet{Boyer2019} are marked with white squares. While our \hst\ data are limited to the PHAT footprint, \spitz\ imaging data exists covering the entirety of M31's disk. \\}
 \label{fig:spatial_distribution}
\end{figure*}

While it is clear that these stars affect galaxies in a number of ways, their complex circumstellar environments make quantifying their impact a challenge \citep{Karakas2014a,Hoefner2018}. Convection, mass loss, and other internal processes introduce degeneracy into radiative transfer models, stellar evolutionary models, and cosmological simulations. Uncertainties in AGB dust properties limit our ability to understand the dust budgets of galaxies. All of these uncertainties are further compounded by the unknown effects of metallicity. In this paper, we will explore the characteristics of metal-rich AGB stars in M31 and compare them with metal-poor samples in the Magellanic Clouds (MCs) to look for clues to how metallicity affects the observational characteristics of AGB stars, as well as their mass loss, dust production, and evolution.

\subsection{M31} \label{sec:m31}
M31 is one of the few nearby, metal-rich galaxies that we can resolve with current instruments and is the best nearby example of a star-forming massive galaxy like our own. It is an \emph{L}$^*$ galaxy\footnote{These are galaxies above the flattening of the central galaxy luminosity-halo mass relation, and generally representative of the mean luminosity of central galaxies within larger halos \citep{Cooray2005}.} hosting a diversity of stellar populations, a spiral structure, a traditional spheroidal component, and disk and bulge components. M31's stellar population is primarily metal-rich ([M/H]$\sim$-0.2--0.1), but has stars that span an order of magnitude in iron abundance \citep{Gregersen2015}.

M31's size, distance, and unobscured view make it an excellent target for studying metal-rich stellar populations. Its stellar population has a known common distance, measured using multiple methods and with little disagreement \citep[776\,kpc;][]{Freedman1990,Riess2012,Dalcanton2012,li2021}. It is similar in size to the Milky Way \citep{Chemin2009,Kafle2018}, and is largely unobstructed by foreground dust, with little extinction limiting our sensitivity to sources in the optical. Its internal dust has also been well-mapped \citep{Dalcanton2012,Draine2014}. This is in stark contrast to the Galactic samples, which are incomplete, due to foreground extinction, and have highly uncertain distances. \emph{Gaia} is also only able to measure distances to nearby Galactic AGB stars given the migration of the photocenters, as a result of convective cells \citep{Chiavassa2011,Chiavassa2018}. M31 therefore allows us to study a statistically large sample of AGB stars in a single galaxy with minimal issues caused by extinction, crowding, and uncertain distances.

The Panchromatic Hubble Andromeda Treasury \citep[PHAT;][]{Dalcanton2012} program covered a third of M31's disk (Figure \ref{fig:spatial_distribution}), with UV, optical, and NIR imaging. This data set has been used to study stellar and galaxy evolution in great detail \citep[e.g.][]{Rosenfield2012,Lewis2015,Williams2017}. With M31's large stellar mass, and abundance of resolved stars \citep[120 million;][]{Williams2014}, this data set offers the opportunity to study different stellar processes in a metal-rich environment without being limited by Poisson statistics. Clusters have been detected in M31 \citep{Krienke2007,Krienke2008,Barmby2009,Hodge2009,Perina2010,Johnson2015}. Using the superior sensitivity and angular resolution of the PHAT data, we can identify AGB candidates in clusters with associated ages, masses and metallicities.

Previous ground-based observations targeting AGB stars in M31 have been limited to shallow observations, low angular resolution, and/or small fields \citep{Brewer1995,Brewer1996,Kodaira1998,Davidge2001,Battinelli2005,Davidge2005,Boyer2013,Hamren2015,Boyer2019,Ren2021,Massey2021,Wang2021}. By characterizing M31's AGB population, we will have the confidence of uniformity and statistical power need to compare it to other nearby galaxies. We will then be able to probe how metallicity affects AGB dust properties and the global dust budget.

\subsection{Metallicity \& Dust}\label{sec:metallicity_and_dust}
AGB dust and dust production have been studied in the MCs and other nearby metal-poor dwarf galaxies \citep[e.g.;][]{Sloan2009}. While carbon stars produce carbonaceous dust species like graphite, oxygen-rich AGB stars produce silicates made up of different species of olivines and pyroxenes \citep{Gail2009,Jones2012}. The exact properties of the dust (e.g. composition, shape, grain size, grain size distribution, porosity, and refractory metal content), however, are still unknown, and likely change depending on the elemental abundances and the environment, leaving the impact of metallicity unclear.

It is expected that the dust production of oxygen-rich AGB stars should be affected by metallicity. A decrease in metals should limit the number of available sites for dust-seed nucleation \citep{Lagadec2008,Nanni2018}. This conclusion is supported by evidence of decreased dust production from the Large Magellanic Cloud (LMC) to the Small Magellanic Cloud (SMC) and nearby globular clusters \citep{Sloan2008,McDonald2009,McDonald2011b,Sloan2010}. However, studies have found evidence for \citep{Loon2000,vanLoon2005} and against \citep{McDonald2011,Sloan2012,Sloan2016} the claim that carbon-rich AGB dust production depends on metallicity. Theoretical models have attempted to predict the effect of metallicity on the dust production of both oxygen-rich and carbon-rich AGB stars \citep[c.f.][]{Nanni2013,Nanni2014,Ferrarotti2006}. While uncertain, these models favor a small impact of metallicity on the total AGB dust output.

There have also been empirical studies suggesting that metallicity can have a dramatic impact on the wind speeds of AGB mass outflows, with differing effects for carbon- \citep{Groenewegen2012} and oxygen-rich \citep{Goldman2017,McDonald2019a,McDonald2020} AGB stars. While based on a limited range in metallicity, and small samples, metallicity seems to have a much larger impact on the wind speeds of oxygen-rich AGB stars.

While sufficient numbers of AGB stars have been discovered and studied in metal-poor environments \citep{Boyer2011,McDonald2010,Riebel2012,Srinivasan2016,Boyer2017,Goldman2019a,Goldman2019b,Karambelkar2019}, to begin assessing the role of metallicity on AGB evolution, we need data in the metal-rich regime to fully leverage previous studies. In this paper, we take advantage of the exquisite sensitivity and resolution of space-based archival data from \hst\ and \spitz\ to identify the AGB population in M31. We produce the most complete catalog of metal-rich AGB stars to date, with particular focus on dust-production, complementing the metal-poor samples already identified in the MCs. Section 2 outlines the data, Section 3 describes our catalog matching method and classification criteria, Section 4 discusses the catalog results, and Section 5 the impact of AGB stars on the dust budget of M31.

\section{Data}\label{sec:data}
In this work, we use archival imaging of M31 from \hst\ \citep{Dalcanton2012,Boyer2013,Boyer2019} and \spitz\ \citep{Barmby2006} as well as classifications from archival stellar spectra from Keck \citep{Guhathakurta2006}. These data cover the UV through the IR with coverage in both wide and narrow bands that probe molecular features. Here we combine these data to identify the evolved-star population in the PHAT footprint of M31 (Figure \ref{fig:spatial_distribution}).

\subsection{Broad-band photometry}\label{sec:broad-band_photometry}

\subsubsection{HST/PHAT}\label{sec:HST/PHAT}
The PHAT survey resolved $\sim$\,120 million stars in M31. The observations were split into 23 sub-regions referred to as ``Bricks'' that collectively cover $\sim$\,0.5 square degrees of M31's disk. These regions were imaged with the \hst\ in ultraviolet (F275W \& F336W), optical (F475W \& F814W), and near-infrared (F110W \& F160W) filters using the WFC3/UVIS, ACS/WFC3, and WFC3/IR instruments, respectively. The \hst\ data are sensitive down to the red clump (F160W\,$\sim$\,24 mag) in each of these filters in the outskirts of the galaxy, where crowding is minimal. In the most crowded regions near the Bulge (Bricks 1 \& 3), the depth is closer to F160W\,$\sim$\,21.5 mag. Point-spread function (PSF) photometry was performed using {\asciifamily DOLPHOT} \citep{Dolphin2002} after which we applied photometric quality cuts. These cuts require the F110W and F160W photometry to satisfy the good-star ``GST'' sharpness and crowding criteria  \citep[outlined in][]{Williams2014} used to limit contamination. As the PHAT data only covers around a third of M31's disk, and we have restricted our catalog to stars in this region.

\subsubsection{Spitzer}\label{sec:spitzer}
\citet{Barmby2006} and \citet{Gordon2006} observed M31 with the \emph{Spitzer Space Telescope} using both the Infrared Array Camera \citep[IRAC;][]{Fazio2004} and the Multiband Imaging Photometer for \emph{Spitzer} \citep[MIPS;][]{Rieke2004}, respectively, in bands centered at 3.6, 4.5, 5.8, 8.0 and 24\,$\mu$m. These observations covered the entirety of the disk including the PHAT footprint. The \spitz/IRAC observations had limited spatial resolution ($\lesssim$\,2\arcsec) compared to \hst\ ($0\overset{\prime\prime}{.}07$ -- $0\overset{\prime\prime}{.}15$). The IRAC data are also less sensitive, with a limit near the tip of the red-giant branch (TRGB; [3.6]\,$\sim$\,21.2 mag). Most TP-AGB stars are brighter than the TRGB, allowing us to identify a near-complete TP-AGB population in M31 from the optical to mid-IR (discussed further in \ref{sec:catalog_completeness}). We measured PSF photometry on the individual dithered IRAC exposures using the DUST in Nearby Galaxies with Spitzer (DUSTiNGS) pipeline \citep{Boyer2015a}, which uses {\asciifamily DAOphot II} and {\asciifamily ALLSTAR} \citep{Stetson1987}. We also require our \spitz\ IRAC photometry to meet photometric quality criteria (GST) that remove extended objects and image artifacts. The photometry are required to be detected over a level of $5\sigma$, meet {\asciifamily DAOphot} sharpness ($-0.4<S_{\lambda}<0.4$) and $\chi^2>4$ thresholds, and must be detected in both the [3.6] and [4.5] filters. We will refer to this data as the \spitz\ good-star catalog (GSC).


We have matched our IRAC GSC sources with longer-wavelength \spitz\ data ([5.8], [8.0], and [24]) from \citet{Khan2017}. While the \citet{Khan2017} catalog includes the IRAC [3.6] and [4.5] data, the DUSTiNGS pipeline provides higher fidelity data for brighter sources in the 3.6 and 4.5\,$\mu$m filters \citep[see ][for details]{Boyer2015a}. This provides more precise AGB classifications using the IR data. The DUSTiNGS pipeline measures the co-added frames for the fainter sources ([3.6] $>$ 14.7 mag) to obtain the deepest possible photometry. Using these mosaicked and subsampled images, however, can distort the PSF, for example, if it includes a rotation between frames. Since brighter sources are more sensitive to changes in the PSF, the DUSTiNGS pipeline performs the photometry for the brighter sources on the individual images.

The photometric catalog from \citet{Khan2017} also appears to have limited calibration. Given the spatial distribution of the sources, the catalog does not include many of the necessary corrections including the array location correction, pixel-phase correction, pixel solid angle variation correction, and color correction. Here, we include the lower-resolution longer wavelength (5.8, 8, and 24\,$\mu$m) photometry from \citet{Khan2017}, but these data are not used for stellar classification or any other measurements or derived values in this work.

\subsection{HST medium-band photometry}\label{sec:hst_medium-band photometry}
\citet{Boyer2019} observed M31 in 21 fields (white squares in Figure \ref{fig:spatial_distribution}) in  three near-IR medium-band \hst\ filters, F127M, F139M, and F153M. These filters probe features from water in oxygen-rich AGB stars and CN+C$_2$ in carbon stars \citep{Boyer2013,Boyer2017,Boyer2019}, and allow for the classification of the chemical types (discussed further in \S\ref{sec:chemical_types}). These data were also run through the PHAT photometric pipeline \citep[described in ][]{Williams2014}, and used the same sharpness and crowding cuts to exclude blended and extended sources\footnote{The DUSTiNGS pipeline was modified to include an additional {\asciifamily chi} $<$ 4 constraint for the
F127M band to limit spurious detections \citep{Boyer2019}.}. The photometry are complete in each of the three bands down to around 22--23 mag, several magnitudes below the TRGB, ensuring completeness in this subset of the sample.

\subsection{SPLASH}\label{sec:splash}
Optical spectra for 1,867 AGB stars are available from the Spectroscopic and Photometric Landscape of Andromeda’s Stellar Halo (SPLASH) survey \citep{Guhathakurta2006}. These data were taken using the DEIMOS instrument on the Keck II 10\,m telescope, with a spectral range of 4800--9500\,{\AA} that spans different molecular features in both oxygen- and carbon-rich AGB stars. The survey targeted sources primarily in the outer disk where crowding is low. The spectra were used to classify AGB stars using color-magnitude diagrams (CMDs) and a suite of spectral templates \citep{Hamren2015,Prichard2017}. Carbon-rich AGB stars were identified using the C$_2$ Swan bands and oxygen-rich AGB stars using the TiO bands and near-IR Ca {\sc ii} triplet. We have matched our AGB candidates with the SPLASH classifications and have included them in our catalog.

\begin{figure}[]
    \centering
\includegraphics[width=\linewidth]{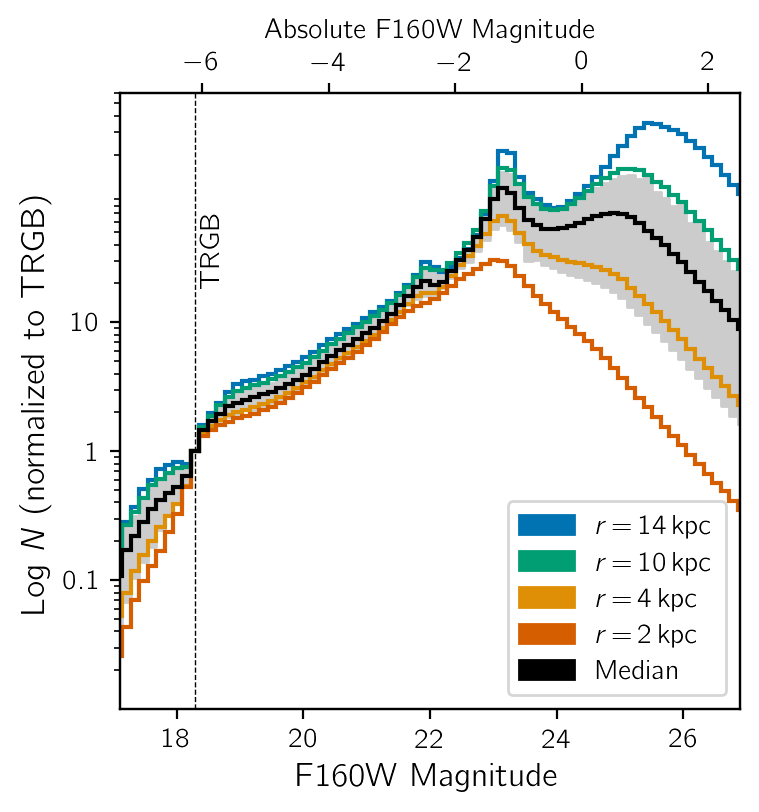}
\includegraphics[width=\linewidth]{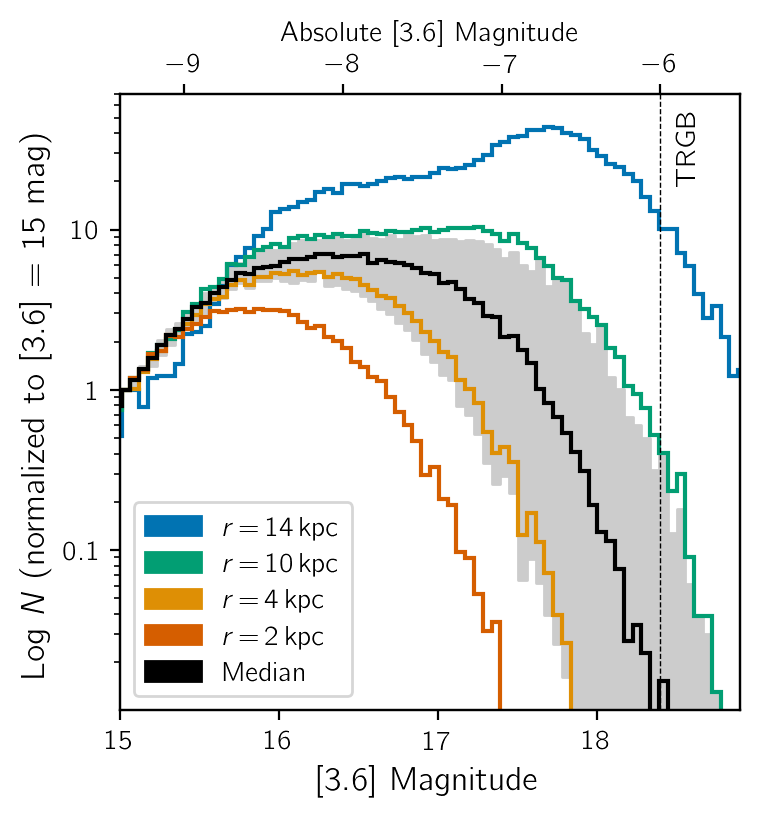}
    \caption{Luminosity functions illustrating the completeness of our M31 data for \hst\ (Top) and \spitz\ (Bottom) in radial arcs across the PHAT footprint. Each radial bin includes all photometry at a specified deprojected radius ($r$) $\pm$ 1\,kpc, so long as they meet the data quality sharpness and crowding cuts. The TRGB is shown for both filters. Also shown is the median functions of the completeness calculated at 1\,kpc intervals, as well as the 68\% confidence level around the median calculated by bootstrapping (shown in gray). \\}
    \label{fig:ir_brick_completeness}
\end{figure}

\subsection{Catalog Completeness} \label{sec:catalog_completeness}

The completeness of our data is limited by crowding and thus depends on the radial distance to the galaxy center. Figure \ref{fig:ir_brick_completeness} shows the completeness of the \hst\ and \spitz\ data in concentric radial bins, with each bin consisting of all catalog sources at the specified deprojected radius ($r$) $\pm 1$\,kpc. We show four example regions that demonstrate the range of completeness from the most ($r=2$\,kpc) to least crowded regions ($r=14$\,kpc).

Our \hst\ data are complete in even our most-crowded regions down to F160W$ \sim 22$ mag, four magnitudes below the TRGB. This ensures that we can detect even the faintest and lowest-mass evolved stars with \hst. We are limited, however, in detecting those stars that are so dusty that they are obscured in the optical and near-IR; we target these sources with \spitz. Our \spitz\ photometry are complete down to $[3.6] \sim 15.9$ mag (M$_{[3.6]} = -8.5$) for most of our sample and down to $[3.6] \sim 15.4$ mag in our most crowded regions. While not as deep as the \hst\ data, this level of completeness is sufficient for detecting most of the dustiest evolved stars in the Magellanic Clouds, that we estimate at 92\% and 98\% of the x-AGB samples in the SMC, and LMC, respectively. This ensures that we have completeness for the bulk of the AGB population using our combined \hst\ and \spitz\ catalog.

\subsection{Contamination}\label{sec:contamination}
Before identifying evolved stars in our catalog, we will discuss potential sources of contamination. We expect some contamination from a variety of IR-bright sources. This includes foreground stars and background galaxies, image artifacts and blended sources, and ISM features confused as stars.

\paragraph{Foreground Contamination}
We expect some contamination from foreground stars that may be masquerading as our most luminous AGB candidates, distributed around the disk. We expect most of these foreground stars to be M dwarfs with a small fraction of red giants \citep{Massey2016,Ren2021}. We have modeled the expected foreground contamination using the \textsc{trilegal} code \citep{Girardi2012}, centered on, and with the same size, as the PHAT footprint. We estimate that 367 foreground sources, ($<$\,0.1\% of the AGB sample) would fall within our AGB selection criteria. We also estimated contamination based on the data from \citet{Boyer2019}. The medium-band filters are able to isolate AGB stars from foreground stars, and suggests the contamination may be as high as 2.6\%. However, the \citet{Boyer2019} data also includes other contaminants, especially massive stars, so some of that 2.6\% may be due to these other stellar types belonging to M31.

\paragraph{Foreground diffraction spikes}
Spurious detections within the wings of diffraction spikes from foreground stars can be confused with AGB stars in the PHAT data, while still passing the GST criteria. To identify these, we identify concentrations of stars not associated with clusters. From this sample subset we identify sources based on pixel statistics of F160W image cut-outs. We use the standard deviation ($\sigma$) and max counts per pixel (V\textsubscript{max}) within $3\arcsec \times 3\arcsec$ F160W image cutouts and select all sources within our clustered subsets with V\textsubscript{max} $>$ 100 and $\sigma$ $>$ 10. We then remove foreground diffraction spikes by inspecting the image cutouts that meet these criteria.

\paragraph{Background Contamination}
Nearby samples of evolved stars like those in the MCs are far more luminous than typical background galaxies. In more distant samples like M31, however, the apparent brightness of the AGB stars is closer to the brightness of background galaxies. Regardless, the high resolution of the \hst\ observations ensures that these would be removed by our photometric quality cuts.

\paragraph{Blending Contamination}
The \hst\ data is not strongly affected by blending for bright stars, except near M31's bulge and cluster centers. We have mitigated these effects using crowding cuts in the \hst\ data \cite[outlined in][]{Williams2014}. In contrast, the lower-resolution \spitz\ imaging is likely to be affected by unresolved blended stars, which would over-inflate the measured brightness. We have also used crowding cuts in our \spitz\ photometry (see \S\ref{sec:spitzer}), but remain crowding-limited throughout the disk. For the dustiest sources detected with \hst\ and \spitz\ (described later in \S \ref{sec:x-AGB_stars}), the fraction of the sample that have a dusty (F110W$-$F160 $>$ 2 mag and F110W $<$ 20 mag) PHAT source within \spitz 's full width at half maximum ($\sim\,2\arcsec$) is eight sources or roughly $0.17\%$. This ensures that most of the dusty sources are not significantly affected by blending.

\paragraph{ISM confusion}
IR-bright ISM features also have the potential to create spurious detections that may appear as our dustiest TP-AGB stars. The MIPS 24\,$\mu$m photometry has a much lower angular resolution than our \hst\ or IRAC photometry, making it difficult to distinguish ISM from stellar sources morphologically. While no crowding or sharpness cuts have been made to the MIPS data \citep{Khan2017}, it is not used in the AGB selection criteria, and thus should not affect the source selection of the AGB catalog. We expected minimal contamination from ISM features in the [3.6] and [4.5] filters.

\subsection{Chemical Types}\label{sec:chemical_types}
AGB stars have circumstellar environments of either carbon- or oxygen-rich chemistry, or, in some rare cases, both \citep{Merrill1922}. The circumstellar chemistry is determined by the relative abundances of carbon and oxygen after the majority of these elements have combined to form CO. These abundances depend on internal processes referred to as third dredge-up events (TDUs) and hot-bottom burning (HBB). The C/O boundaries in stellar properties where these processes dominate, however, are still unclear. Models predict that the C/O boundaries vary dramatically with metallicity and initial mass \citep{Marigo2013,Karakas2014,Marigo2017} and observations have shown that carbon-star formation drops dramatically (and may even be prevented) in metal-rich environments \citep{Brewer1995,Boyer2013,Boyer2019}.

\begin{figure}[b]
\centering
    \includegraphics[width=\linewidth]{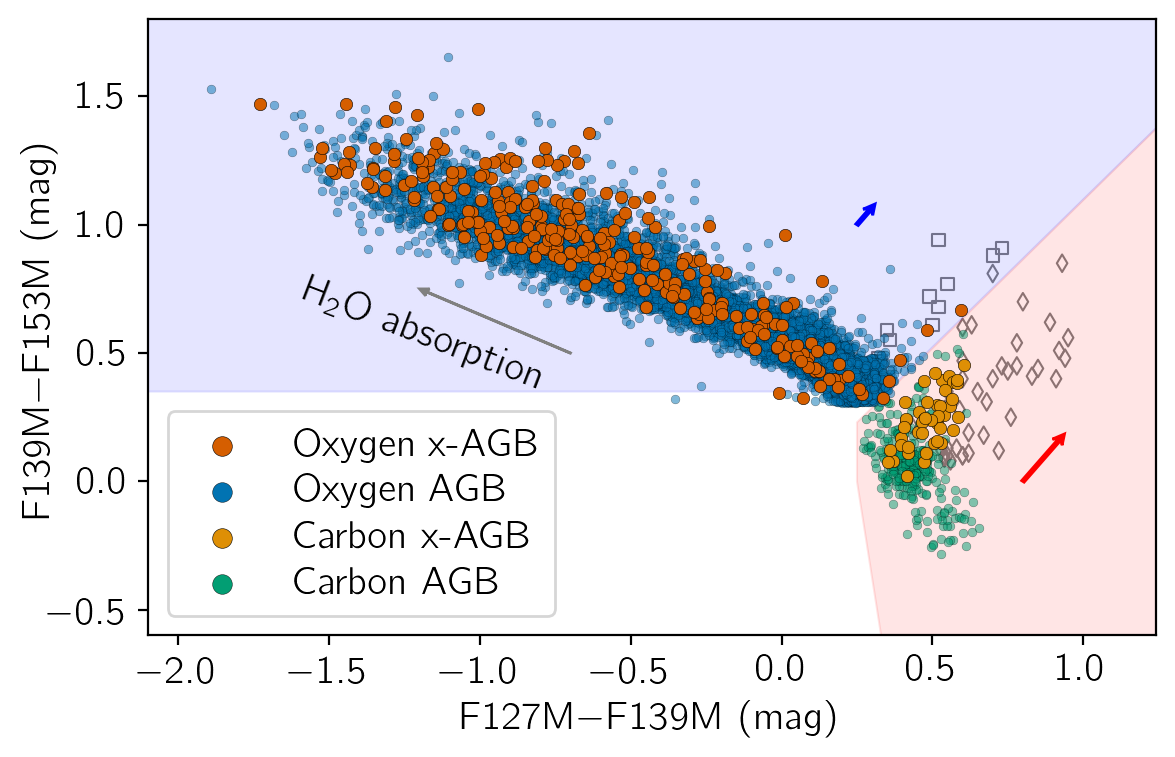}
    \caption{Our M31 AGB candidates that were previously classified as oxygen- or carbon-rich \citep{Boyer2019}, with the subsets we have subsequently classified as the dustiest, extreme (x-)AGB candidates (discussed further in \S \ref{sec:x-AGB_stars}). Also shown are the dustiest carbon- (diamonds) and oxygen-rich (squares) sources in the LMC \citep{Groenewegen2018}. Extinction vectors with an $E$($J-K_{s}$) = 1 mag are shown for oxygen-rich (blue arrow) and carbon-rich (red arrow) dust species, with 60\% silicates and 40\% AlO$_{\rm x}$ for the oxygen-rich dust and 70\% amorphous carbon and 30\% SiC dust for the carbon-rich dust.\\}
    \label{fig:knee_plot}
\end{figure}

We can use chemically-classified AGB stars in M31 to probe the conditions that favor production of either carbon- or oxygen-rich AGB atmospheres. The chemical subtypes of carbon (C) or oxygen (M) have already been determined for a subset of the AGB stars in small regions of M31. \citet{Hamren2015} spectroscopically identified 103 carbon-rich and 736 oxygen-rich AGB stars. \citet{Boyer2019} identified 346 carbon stars and 20,441 oxygen-rich AGB stars photometrically using medium-band \hst\ filters (Figure \ref{fig:knee_plot}). While both of these datasets only cover small areas across the PHAT footprint, we use these classifications to calibrate our AGB classification criteria for the full M31 sample; we will discuss this further in \S \ref{sec:agb_criteria}.

\subsection{Clusters}\label{sec:clusters}

Thanks to their well-defined ages and metallicities, clusters have been critical for calibrating complex phases of stellar evolutionary models like the TP-AGB. For decades these calibrations have relied primarily on small samples of AGB stars in 31 clusters in the MCs \citep{Frogel1990}. These clusters, however, span a limited range of ages and metallicities, and suffer from stochastic sampling. While stochastics can be overcome by binning \citep[e.g.;][]{Girardi2007}, LMC clusters with ages $\sim 1.6$ Gyr have also shown to have derived quantities that are not proportional to the AGB lifetimes. This effect, referred to as ``TP-AGB boosting'', compromises a large fraction of the MC clusters' utility as AGB calibrators \citep{Girardi2013}. As clusters in M31 span a larger range in age (and metallicity) than the LMC, we do not expect TP-AGB boosting to affect so much the M31 cluster sample. The larger range of cluster ages and metallicities also provides data near the unexplored boundaries in parameter space where we expect transitions in circumstellar chemistry. A handful of AGB candidates have also been identified near clusters in other nearby galaxies \citep[e.g.][]{Karambelkar2019}.

Thousands of clusters have been identified in the M31 PHAT data \citep{Johnson2015}, providing pockets of stars with known ages and metallicities. The compilation of cluster ages and masses combines results from CMD fitting for younger clusters ($<$300 Myr) as presented in \citet{Johnson2016} and integrated-light estimates of the ages and masses using the method by \citet{Fouesneau2014}, compiled by \citet{Beerman2015}. \citet{Girardi2020} identified 937 AGB candidates, and we have identified a similar slightly-higher number of AGB candidates ($N$=\clusteragbs). The discrepancy in these numbers is related to the differences in the selection criteria (discussed later in \S \ref{sec:cluster_statistics}). Cluster membership was determined in \citet{Girardi2020} and here by having on-sky positions within the apparent radius \citep[measured by][]{Johnson2015} of each cluster.

\section{Compiling the AGB catalog}\label{sec:compiling_the_AGB_catalog}
We have culled and then matched the PHAT photometry to \hst\ medium-band and \spitz\ photometry based on their positions and brightness. The catalog matching process includes the following four steps:

\begin{enumerate}
    \item \emph{PHAT}: We include an initial cut (described in the following section) to remove faint and bluer (non-AGB) PHAT sources ($\sim$\,120 million $\rightarrow$ $\sim$\,23 million stars).
    \item \emph{Other HST}: We match the selected subset of the PHAT catalog ($\sim$\,23 million) to the medium-band \hst\ data using the stellar positions (within 1\arcsec). The F814W band data from \citet{Williams2014} was re-reduced by \citet{Boyer2019} simultaneously with the medium-band data (same data, but performed independently) and was used to assess the quality of matches; we will call this {\tt cat1}.
    \item \spitz : We match our \spitz\ IRAC [3.6] and [4.5] ``GSC'' catalog to the \spitz\ catalog with longer-wavelength ([5.8], [8.0], and [24]) data from \citet{Khan2017} using the source positions and brightness. Within the PHAT footprint, the Khan catalog also included 3.6 and 4.5\,$\mu$m data, reduced independently. We use that photometry for matching sources (see below) but not for subsequent catalogs; we will call this {\tt cat2}.
    \item \emph{Final catalog}: We match the resulting two catalogs ({\tt cat1}, {\tt cat2}) based on the stellar positions (within 1\arcsec) and refer to this as {\tt cat3}.
\end{enumerate}

Each of these steps is described further in the following sections.

\begin{figure}
    \centering
    \includegraphics[width=\linewidth]{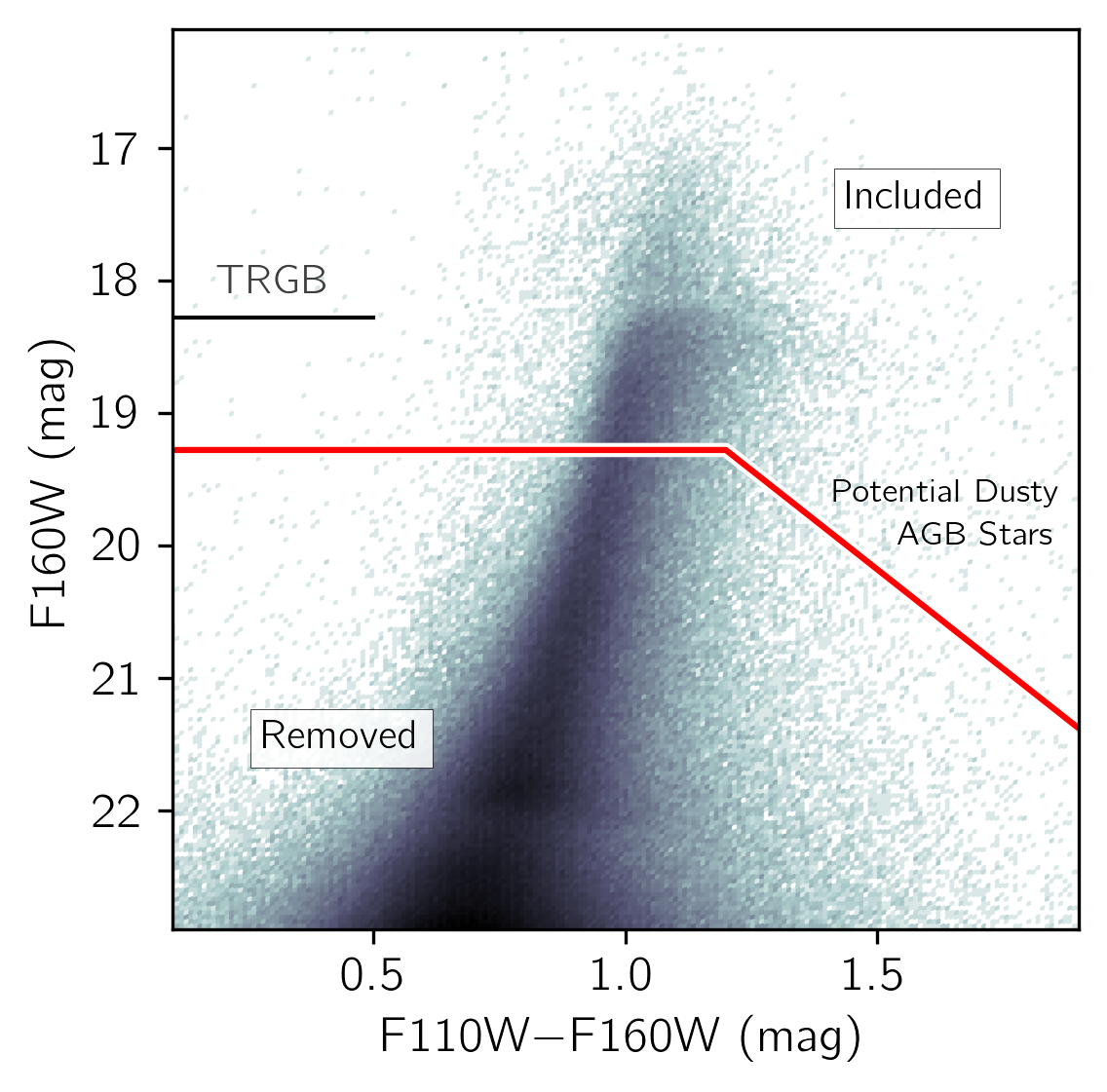}
    \caption{An \hst\ CMD showing the initial cuts to the PHAT catalog. These cuts remove fainter non-AGB sources while still including fainter reddened sources. This cut minimizes spurious matches between the \hst\ and \spitz\ data. \\}
    \label{fig:initial_selection_hst}
\end{figure}

\begin{figure*}
    \centering
    \includegraphics[width=\linewidth]{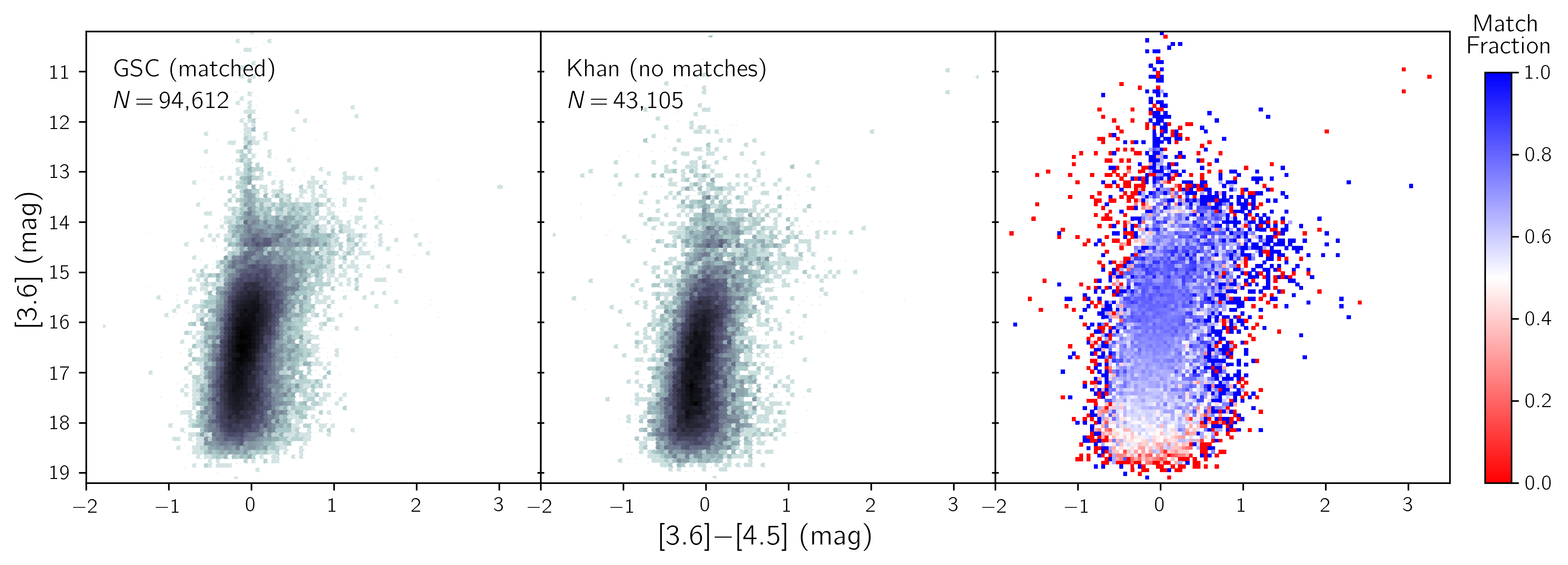}
    \caption{CMDs showing the \spitz\ GSC data with matches to the \citet{Khan2017} catalog (Left), the \citet{Khan2017} catalog with no matches to GSC (Center), and binned regions with the fraction of the GSC sources with matches to \citet{Khan2017} shown in color (Right). We see no clear biases in the matched-fraction CMD indicating that we are not missing any parts of the IR CMD in our matching process. \\}
    \label{fig:match_spitz}
\end{figure*}

\subsection{Culling the PHAT Data}\label{sec:culling_the_data}
Given the differences in the depth, angular resolution, and wavelength coverage of our \hst\ and \spitz\ data, we must first rid our \hst\ data of sources unlikely to be AGB stars to achieve better matching of the catalogs. The \hst\ data is far more sensitive than the \spitz\ photometry, which is only sensitive down to the TRGB, and has a higher spatial density of sources due to the higher resolution. With the differences in sensitivity we run the risk of matching the highly reddened \spitz\ photometry with other spurious sources in the PHAT data. To mitigate this, we have used several \hst\ cuts to reject the fainter and bluer sources from the \hst\ catalog of plausible AGB candidates.

The \hst\ selection of candidate AGB stars is (any of the following):
\begin{enumerate}[label={\Alph*)}]
    \setlength\itemsep{-0.25em}
    \item F110W\,$<$\,20.28 mag, to restrict to luminous stars.
    \item F160W\,$<$\,19.28 mag, to restrict to luminous stars.
    \item F110W--F160W\,$>$\,1.2 and above the line $(F160W) = 2 \times (F110W-F160W) + 16.88$ mag, to recover dustier AGB candidates.
  \end{enumerate}

As shown in Figure \ref{fig:initial_selection_hst}, we limit the \hst\ data to all of the PHAT sources that are no more than 1 mag below the TRGB \citep[measured in][]{Boyer2019} in either the F110W or F160W filters (Requirements A \& B). We also include fainter stars if they are significantly reddened (Requirement C), to allow for dusty AGB candidates whose circumstellar extinction can make them quite faint even in the near-IR filters \cite[e.g.,][]{Boyer2017}. We will now compile the IR component of our catalog.

\begin{figure*}
\centering
\includegraphics[height=6.5cm]{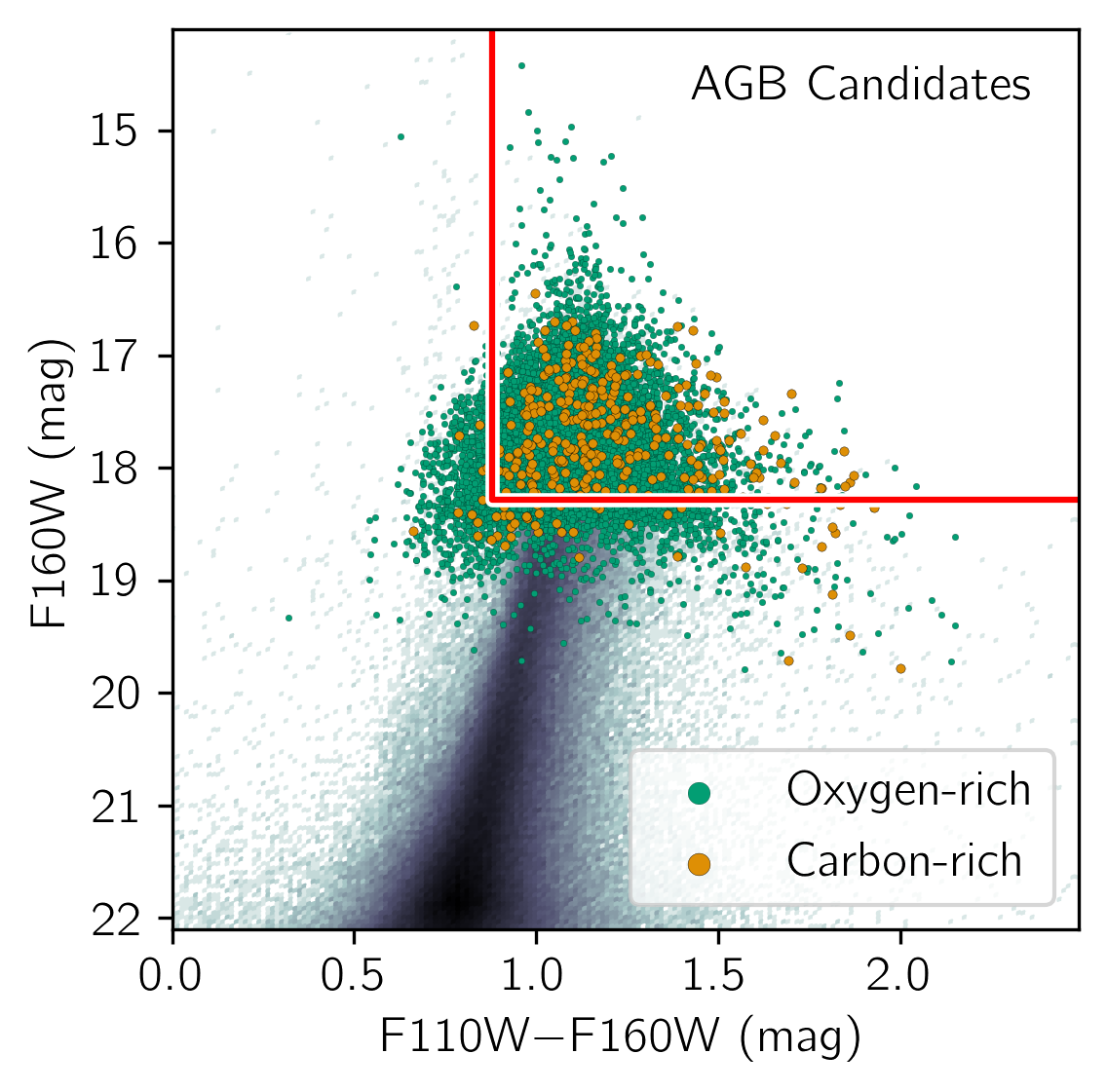}
\includegraphics[height=6.5cm]{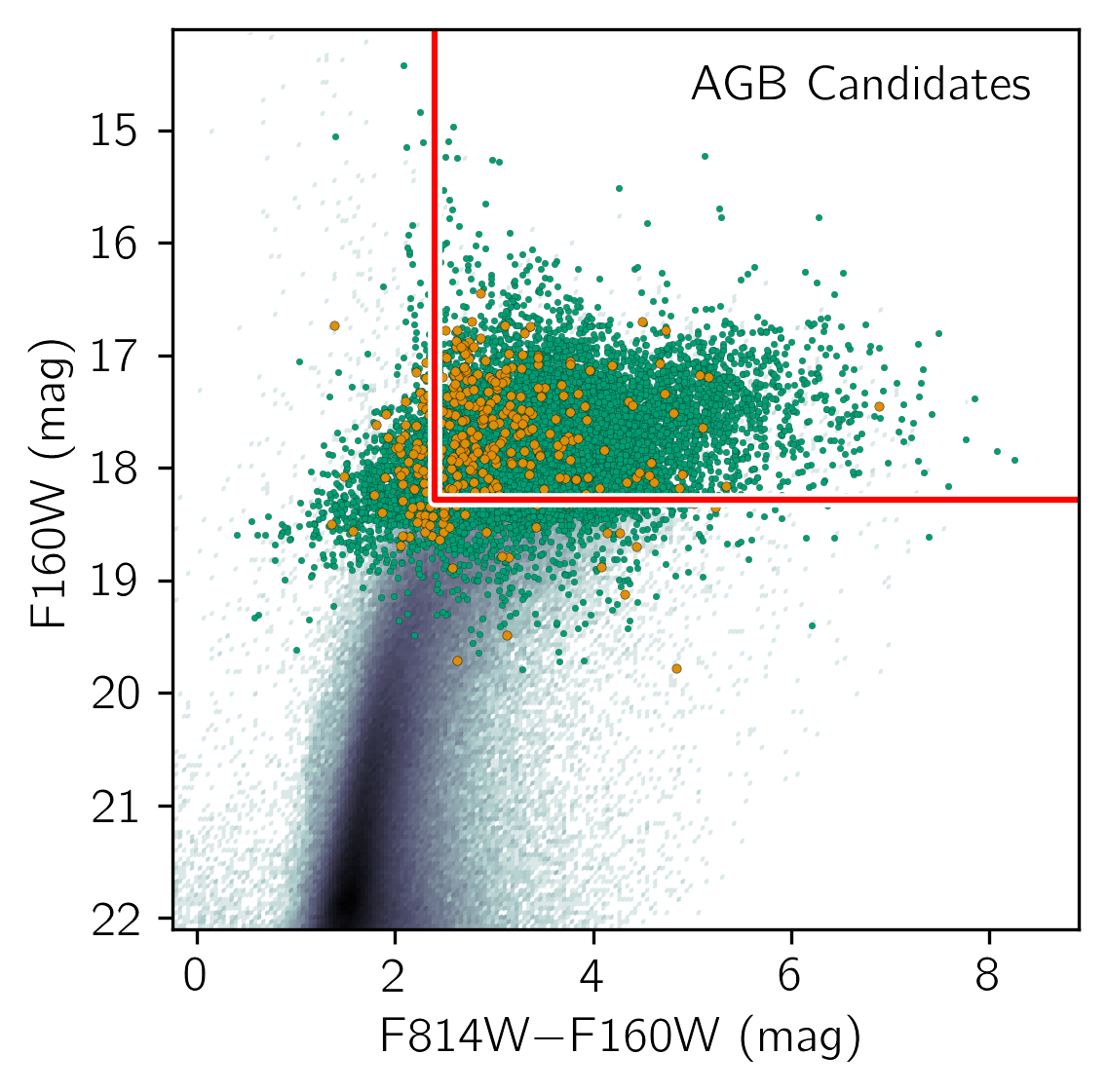}
 \caption{CMDs showing the chemically-classified AGB sample from \citet{Boyer2019} that was used to calibrate the AGB classification criteria (red lines). The oxygen- and carbon-rich stars from \citet{Boyer2019} are shown as green and orange points, respectively. The PHAT photometry for Brick 9 is shown in black, for reference. \\ }
 \label{fig:both_cuts}
\end{figure*}

\subsection{Matching with Spitzer}\label{sec:matching_with_spitzer}

In Step 3, above, we match our \spitz\ catalog, with IRAC [3.6] and [4.5] data, to the \spitz\ catalog by \citet{Khan2017}, which includes data in the [3.6], [4.5], [5.8], [8.0], and [24] filters. The \citet{Khan2017} catalog is larger that our \spitz\ catalog due to our additional crowding and sharpness cuts, and thus we match only 52\% of the \citet{Khan2017} catalog within 1\arcsec. The \citet{Khan2017} sources without matches are not restricted to any region of the IR CMD (Figure \ref{fig:match_spitz}). This gives us more confidence that our matching routine isn't biased against specific stellar types.

We have used both position and brightness in matching our \spitz\ catalogs. We match catalogs by taking each of our \spitz\ sources within the PHAT footprint  ($N= \gscinphat$) and examine the three nearest sources from the larger \citet{Khan2017} catalog within 1\arcsec\ ($N=\khaninphat$ within the PHAT footprint). We then compare each of the three 3.6\,$\mu$m magnitudes to our source's 3.6\,$\mu$m magnitude, and select the best-fitting [3.6] match based on brightness, so long as the data is available. This resulted in sources where the second-closest ($N=\secondclosestspitzmatch$; \secondclosestspitzmatchpercent) and third-closest ($N=\thirdclosestspitzmatch$; \thirdclosestspitzmatchpercent) spatial match was chosen on the basis of being a better magnitude match. If none of the three sources within 1\arcsec\ have a magnitude match within 3\%, we do not include any of the [5.8], [8.0], or [24] data from \citet{Khan2017}; this excluded \commonfiltercut\ (\commonfiltercutpercent) of the sources. We use a relatively conservative 3\% brightness matching limit as the data are the same, but the photometry is performed independently. Our brightness matching ensures that for matches that do differ significantly at 3.6\,$\mu$m (some by as much as 10\%), further differences are not included in the less-sensitive, longer-wavelength filters.

In Step 4 we match the \hst\ ({\tt cat1}) and \spitz\ ({\tt cat2}) catalogs using their positions. We have 7,023 \spitz\ sources from {\tt cat2} (5\%) without any \hst\ matches. We omit the \spitz\ data in these cases. In the case of multiple \spitz\ matches to the same \hst\ source, we use the nearest positional match.

\subsection{AGB criteria}\label{sec:agb_criteria}

Using the matched and culled catalog described in the previous sections, we applied color and magnitude cuts to identify AGB candidates spanning the PHAT footprint. Our classification criteria were guided using the AGB sample already identified by \citet{Boyer2019}, and are illustrated in Figure \ref{fig:both_cuts}. We start with the \hst\ data, including sources above the TRGB. We also exclude blue supergiants and main sequence stars by limiting our selection to red colors. \\ \smallskip

1. \hst\ AGB candidate (all of the following):
\begin{itemize}
    \setlength\itemsep{-0.25em}
    \item F110W\,$<$\,19.28 mag or F160W\,$<$\,18.28 mag
    \item F110W--F160W\,$>$\,0.88 mag$^{*}$
    \item F814W--F160W\,$>$\,2.4 mag$^{*}$ \\
    $^*$applied only if data available.
\end{itemize}

The slight magnitude difference between the initial PHAT selection and AGB criteria will become more clear when we discuss our additional AGB criteria in the IR (\S\ref{sec:dusty_agb_stars}).

\begin{figure*}
    \centering
    \includegraphics[width=\linewidth]{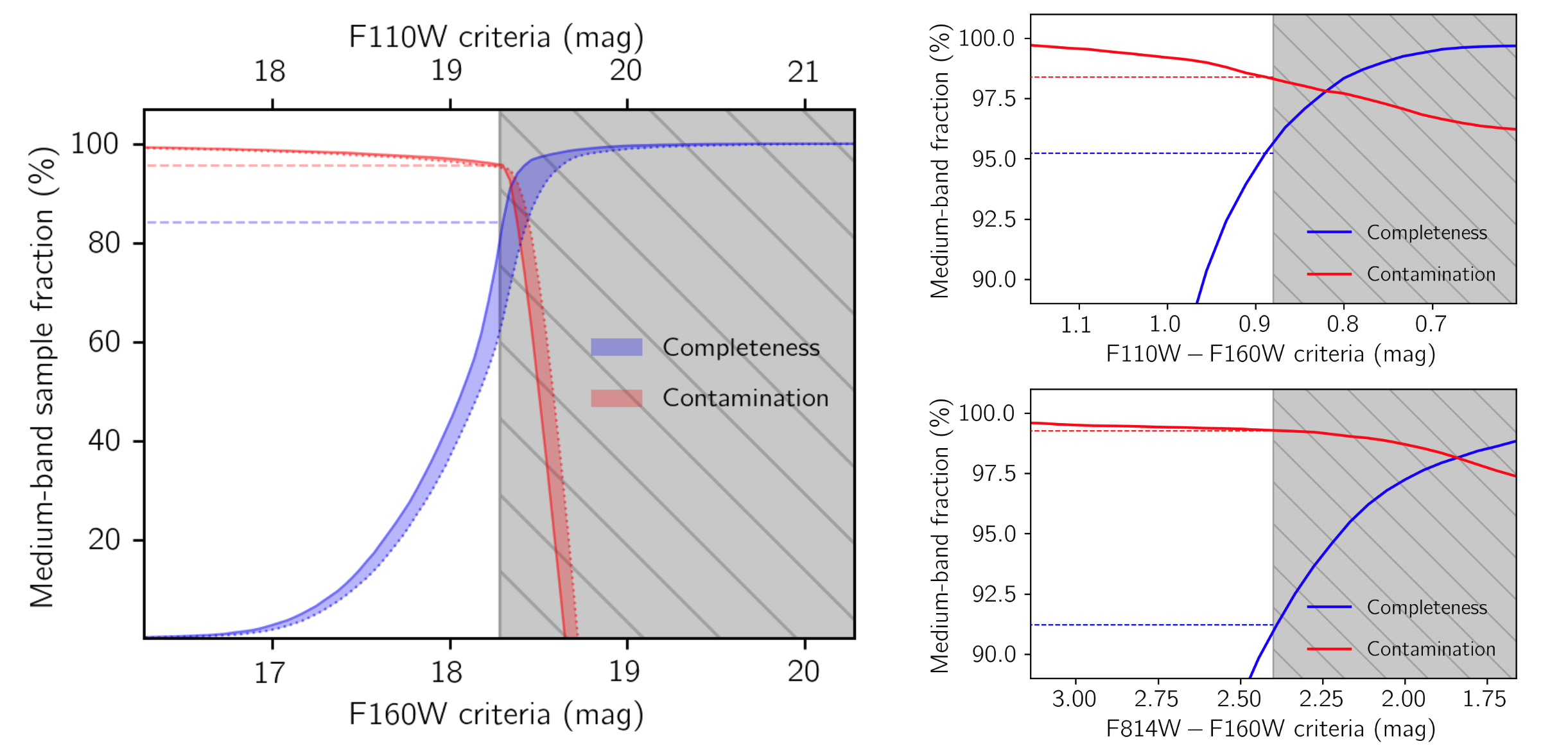}
    \caption{Our completeness vs. contamination (or rather, sample purity) recovering AGB stars classified with medium-band data \citep{Boyer2019}, using our AGB criteria. Left: The AGB magnitude criteria completeness vs. contamination showing an increasing fraction of the sample having sources with medium-band data but not classified as AGB stars towards fainter magnitudes and below our criteria (shown in gray); our criteria require photometry above the TRGB in the F110W (top axis) or F160W (bottom axis) filters. Shown are the F110W (dotted blue/red lines) and F160W (solid blue/red lines) filter criteria and the fraction of the sample (dashed lines) at our more inclusive magnitude requirement (F110W or F160W) for the completeness and contamination. Right: Similar figures showing the AGB color criteria completeness vs. contamination for our color criteria. These figures are limited to sources that also meet our AGB magnitude criteria.\\}
    \label{fig:completeness_contamination}
\end{figure*}

\subsubsection{AGB criteria effectiveness} \label{sec:agb_criteria_effectiveness}

We have used our previously classified AGB stars from \citet{Boyer2019} to determine the effectiveness of our AGB criteria. Of the sources classified as AGB stars using medium-band \hst\ photometry (\boyermedbandcat), 25\% (\chemnohst) did not meet our \hst\ AGB criteria; this is split up between 14\% (\chembelowtrgb) that fall below the TRGB in F110W and F160W, 11\% (2,281) that we remove using the F814W--F160W color cut, 5\% (1,074) that we remove with the F110W--F160W color cut, and 0.3\% (74) that did not have any match to the PHAT data within 1\arcsec. The previous medium-band color cuts have better discretionary power for identifying AGB stars. The TRGB requirement in \citet{Boyer2019} applied to any of five filters (as opposed to our two) providing more opportunities to classify sources as AGB stars. We expect these removed sources in the medium-band regions to be AGB candidates and they are re-introduced into the catalog based on the previous classification.

Our color cuts were selected to balance recovering chemically-classified sources from \citet{Boyer2019} and removing warmer red supergiants. The completeness and contamination levels for different magnitude and color cuts, as well as the criteria we chose, are shown in Figure \ref{fig:completeness_contamination}. Our magnitude requirements were chosen to be in line with those of \citet{Boyer2019} and we use their estimates for the TRGB for out criteria. For our color cuts, we chose our F110W--F160W color threshold to be in line with that of \citet{Girardi2020}, which aims to remove RSGs and foreground stars. Figure \ref{fig:completeness_contamination} illustrates our balance in recovering the chemically-classified sample while limiting contamination. The figure does not show our potential contamination from warmer supergiants, which are not isolated in broad-band filters, and which can be more challenging to identify. We have added additional AGB criteria in the IR to recover the dustiest AGB stars too obscured to be detected with \hst, as discussed in the following sections.

\subsection{Dusty AGB Stars}\label{sec:dusty_agb_stars}
An intermediate-mass star in the core He-burning phase produces a modest amount of dust, but still loses mass in a form of stellar winds driven by acoustic and chromospheric processes \citep{Dupree1984}. As it evolves from the core He-burning to AGB phase, the star begins to pulsate, levitate more material out to larger radii, and begin to produce significant dust. As mass in the envelope is lost and surface gravity decreases, the strength of the pulsations and mass loss increases.

\begin{figure}[]
    \centering
    \includegraphics[width=\columnwidth]{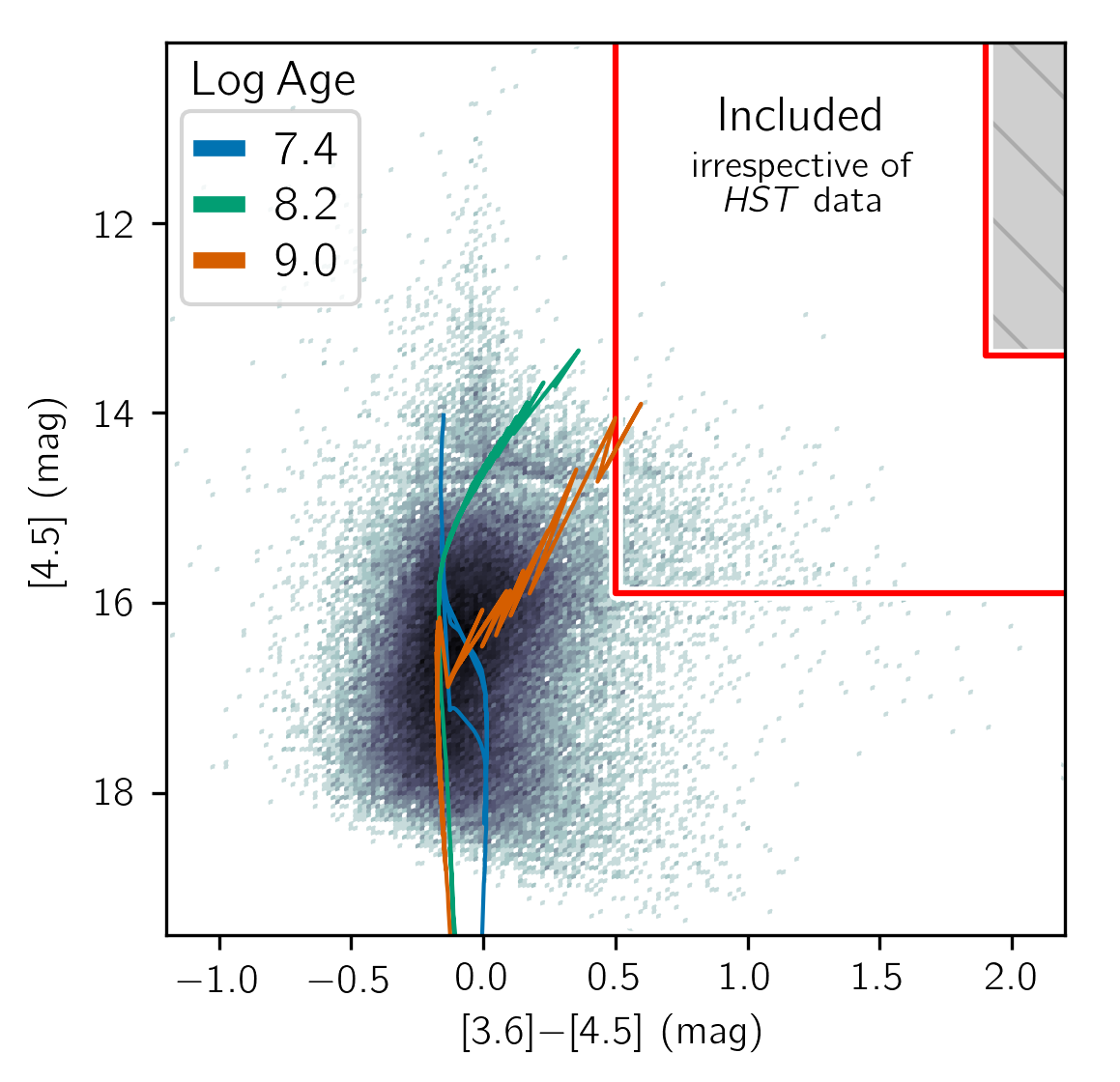}
    \caption{A CMD isolating \spitz\ sources that are bright and reddened in the IR. These have been included in the AGB catalog irrespective of their \hst\ (or lack of \hst) data. PARSEC isochrones \citep{Marigo2017} are shown to indicate the end stage of evolution before the onset of AGB mass loss and considerable dust production. \\}
    \label{fig:ir_criteria_cmd}
\end{figure}

Even with the inclusion of the fainter red sources in the PHAT data, it is possible we are missing some of these dusty AGB stars. We therefore add stars back into the catalog if their \spitz\ [3.6]--[4.5] color puts them in the region of the CMD where dusty AGB stars are likely to be located \citep{Boyer2011,Boyer2017}; with this ``additional inclusion'' criteria we recover \dustyappended\ dusty AGB candidates (Figure \ref{fig:ir_criteria_cmd}), that would have been missed using our \hst\ criteria alone. While a small number, these are among the dustiest sources (see \S \ref{sec:x-AGB_stars}), so it is important to include them in the final sample.\\

2. Additional inclusion (either of the following):
\begin{itemize}
    \setlength\itemsep{-0.25em}
    \item {$[3.6] - [4.5] > 0.5$} \hspace{0.1cm} and \hspace{0.1cm} $[4.5] < 16.4$ mag; \\ to recovered the dusty AGB stars too faint or obscured to be detected in the near-IR.
    \item We also add stars back in that were identified as AGB stars by \citet{Boyer2019} using medium-band filters.
\end{itemize}

Based on the AGB photometry in the LMC \citep{Riebel2012}, we expect that we are not missing many dusty AGB stars that are too obscured and faint for our \hst\ criteria, yet not sufficiently reddened in the IR to meet our \spitz\ criteria. We can use the much deeper 2MASS \emph{J} and \emph{H} photometry for the LMC AGB sample as a proxy for our \hst\ F110W and F160W filters as they share a similar wavelength coverage. Applying our near-IR and \spitz\ AGB criteria to the LMC AGB sample, we would only have missed three AGB stars (0.017\% of the full LMC AGB sample).

This additional inclusion step is used to recover dusty AGB stars that were either not detected in \hst\ or did not meet the criteria due to extreme circumstellar dust extinction. The \hst\ criteria, which requires \hst\ photometry above the TRGB, differs from the initial \hst\ selection (one mag below TRGB). This is done to include \hst\ photometry for sources classified as AGB candidates based on their \spitz\ photometry, but that have fainter \hst\ photometry. We also include an ``additional removal criteria'' that removed four background galaxies.\\

3. Additional removal (either of the following):
\begin{itemize}
    \setlength\itemsep{-0.25em}
    \item {$[3.6] - [4.5] > 1.9$} and $[4.5] < 13.4$ mag;\\ to remove background galaxies.
    \item manually identified as foreground or imaging artifact.
\end{itemize}

\subsection{AGB catalog results}\label{sec:agb_catalog_results}
We have identified AGB candidates within the PHAT footprint using data from \hst\ and \spitz. The results of our classification process are shown in Table \ref{table:stats}. In total, we find \agb\ AGB candidates. With the \spitz\ data, we add an additional \onlyadditionalinclusion\ AGB candidates based only on their \spitz\ photometry. We have also cross-matched our AGB candidates with the clusters identified by \citet{Johnson2015}. We have identified \clusteragbs\ AGB candidates within the measured radii of M31 PHAT clusters. This number is slightly higher than those identified with similar color and magnitude cuts in \citet{Girardi2020}. Given that the PHAT footprint covers around a third of M31's disk, we expect around \agbscaled\ AGB stars in M31's full disk. We find AGB candidates associated with \splashc\ carbon-rich and \splashcn\ oxygen-rich AGB stars from SPLASH, and \renrsg\ matches with red supergiant candidates identified in \citet{Ren2021}; we expect these may be RSGs or massive AGB stars.

\begin{deluxetable}{lrrr}[b]
\tablewidth{\columnwidth}
\tabletypesize{\small}
\tablecolumns{4}
\tablecaption{M31 source statistics. \label{table:stats}}

\tablehead{
&
\colhead{$N_{\rm All}$} &
\colhead{$N_{\rm O-rich}$} &
\colhead{$N_{\rm C-rich}$}}

\startdata
Initial selection & \fullcat & \ofullcat  & \cfullcat \\
AGB & \agb & \oagb & \cagb \\
x-AGB & \dustyagb & \odustyagb & \cdustyagb \\
Cluster AGB & \clusteragbs & \oclusteragbs & \cclusteragbs \\
Cluster x-AGB & \dustyclusteragb & 0 & 0 \\
\enddata
\tablenotetext{}{{\bf Note.} The cuts used for the initial selection of the PHAT catalog (Initial selection) are discussed in \S\ref{sec:culling_the_data}. The oxygen (O-) and carbon (C-) rich classifications are from \hst\ \citep{Boyer2019} and are only for a small subset of the sample in small regions across the PHAT footprint. Also listed are results for extreme (x-)AGB stars (described later in \S\ref{sec:x-AGB_stars}). We expect additional cluster x-AGB stars to exist outside of the regions that were covered by PHAT.}
\end{deluxetable}

We have included the AGB candidates and their photometry in a catalog (Table \ref{table:photometry}) which also includes the previous chemical classifications, and cluster properties. The estimated foreground extinction toward M31 is \emph{E(B$-$V)} = 0.062 \citep{Schlegel1998ApJ}, far less than the levels of differential extinction within M31's disk \citep{Dalcanton2015}. We have used foreground extinction-corrected photometry for matching source brightness in the medium-band \hst\ and PHAT catalogs and for the AGB classification, but present the uncorrected photometry in Table \ref{table:photometry}. We find 539 AGB candidates (0.16\%) within $1\arcsec$ of stars with positive measured parallaxes from {\it Gaia} EDR3 \citep{Gaia2021}. We expect some of these may be foreground M dwarfs as parallaxes in AGB stars are highly affected by their convective cells. We leave these stars in the catalog, but flag them as potential foreground contamination.

\begin{deluxetable*}{lll}
\tablewidth{\linewidth}
\tabletypesize{\normalsize}
\tablecolumns{3}
\tablewidth{\linewidth}
\tablecaption{The photometry of the M31 AGB sample \label{table:photometry}}
\tablehead{
\colhead{Col \#} &
\colhead{Col name} &
\colhead{Description}}
\startdata
{\footnotesize 1} & ID & ID for this catalog\\
{\footnotesize 2} & RA & Position in degrees (J2000) \\
{\footnotesize 3} & Dec. & Position in degrees (J2000)\\
{\footnotesize 4} & Brick    & PHAT Brick \citep[1\,$-$\,23;][]{Williams2014}\\
{\footnotesize 5} & R   & Deprojected radius from galaxy center (kpc)\\
{\footnotesize 6} & Radius $Z$ & Estimated [M/H] based on R \citep[relation from][]{Gregersen2015}  \\
\multicolumn{3}{l}{\tabledash\ Stars within clusters \citep{Johnson2015} \tabledash}\\
{\footnotesize 7} & Cluster ID & ID from cluster catalog \\
{\footnotesize 8} & Cluster Z & Cluster [Fe/H] measured spectroscopically by  \citet{Caldwell2011} \\
{\footnotesize 9} & Cluster Log Age & log age of cluster \\
{\footnotesize 10} & Cluster Mass & Mass of cluster \\
\multicolumn{3}{l}{\tabledash\ PHAT photometry \citep{Dalcanton2012} \tabledash} \\
{\footnotesize 11} & F275W    & WFC3/UVIS magnitude \\
{\footnotesize 12} & F336W    & WFC3/UVIS magnitude\\
{\footnotesize 13} & F475W    & ACS/WFC magnitude\\
{\footnotesize 14} & F814W    & ACS/WFC magnitude\\
{\footnotesize 15} & F110W    & WFC3/IR magnitude\\
{\footnotesize 16} & F160W    & WFC3/IR magnitude\\
 \multicolumn{3}{l}{\tabledash\ Medium-band \hst\ photometry for small regions \citep[][Fig. \ref{fig:spatial_distribution}]{Boyer2019} \tabledash}\\
{\footnotesize 17} & F127M    & WFC3/IR magnitude\\
{\footnotesize 18} & F139M    & WFC3/IR magnitude\\
{\footnotesize 19} & F153M    & WFC3/IR magnitude\\
\multicolumn{3}{l}{\tabledash\ IRAC photometry \citep[performed here following][]{Boyer2015a}) \tabledash}\\
{\footnotesize 20} & IRAC1    & IRAC [3.6] magnitude\\
{\footnotesize 21} & IRAC2    & IRAC [4.5] magnitude\\
\multicolumn{3}{l}{\tabledash\ Additional \spitz\ photometry \citep{Khan2017} \tabledash}\\
{\footnotesize 22} & IRAC3    & IRAC [5.8] magnitude\\
{\footnotesize 23} & IRAC4    & IRAC [8.0] magnitude\\
{\footnotesize 24} & MIPS24   & MIPS [24] magnitude\\
\multicolumn{3}{l}{\tabledash\ Classifications for Full PHAT Footprint \tabledash}\\
{\footnotesize 25} & AGB  & Classified here as AGB candidate based on \hst\ criteria \\
{\footnotesize 26} & x-AGB  & Classified here as dusty-AGB candidate based on \spitz\ criteria\\
{\footnotesize 27} & RHeB Candidate & Classified here as RHeB candidates \\
{\footnotesize 28} & RSG Candidate & Classified as Red supergiant candidate \citet{Ren2021} \\
\multicolumn{3}{l}{\tabledash\ Classifications for Small Subset \tabledash}\\
{\footnotesize 29} & Chem Type & AGB photometric chemical type for small regions from \hst\ \citep[][Fig. \ref{fig:spatial_distribution}]{Boyer2019}\\
{\footnotesize 30} & SPLASH Type & AGB spectroscopic chemical type for small subset from SPLASH \citep{Hamren2015} \\
\enddata

\tablenotetext{}{\small{{\bf Note.} Positions are from the PHAT catalog \citep{Williams2014}, otherwise from the IRAC catalog, which is aligned to 2MASS astrometry \citep{Cutri2003}. Metallicites are inferred from cluster data (Cluster Z, compiled from the literature for globular clusters) or estimated (Radius Z) using the gradient measured by \citet{Gregersen2015}. The method for determining cluster values depends on the optimal method for the age of the cluster \citep[``Best'' in][]{Beerman2015}. Photometry and errors (included but not shown) are in Vega magnitudes. Chemical types were determined by \citet{Boyer2019} in 21 fields using the medium-band \hst\ photometry (C, M) included in this catalog, or from the additional DEIMOS/Keck II optical spectra (C, M) from the SPLASH survey \citep{Hamren2015}.}}
\end{deluxetable*}

\begin{figure}[]
    \centering
    \includegraphics[height=0.75\textheight]{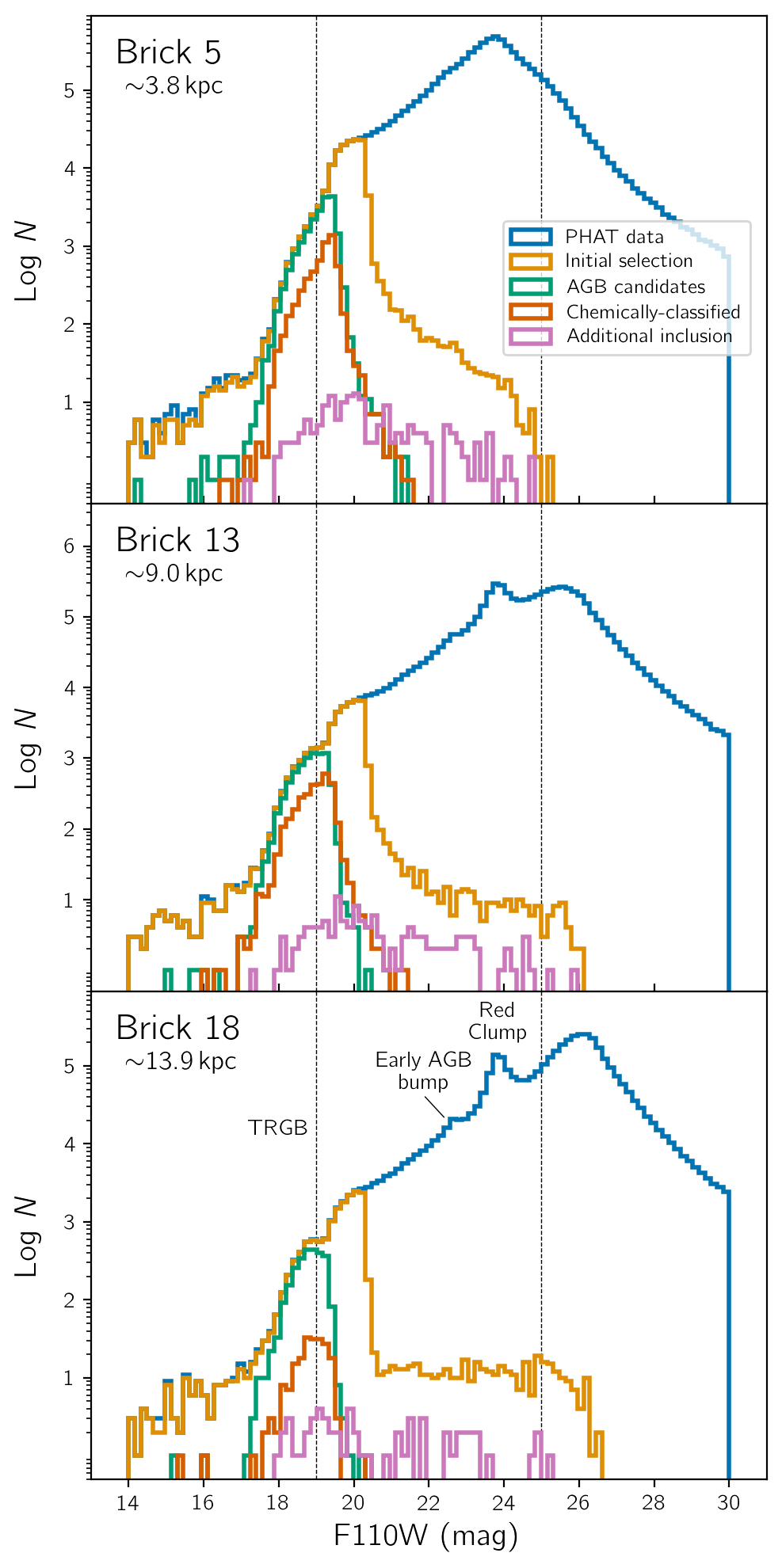}
    \caption{Luminosity functions showing the PHAT catalog, the initial selection of the PHAT catalog (discussed in \S\ref{sec:culling_the_data}), the AGB candidates, the chemically-classified AGB candidates, and the AGB candidates classified by \spitz\ (additional inclusion criteria) for three Brick regions of increasing distance from the galaxy center. For the chemically-classified sources in smaller regions within the PHAT footprint we use all sources $\pm$1\,kpc of the deprojected radius value listed below the brick number in each panel. Also shown are the evolutionary features like the red clump and early-AGB bump, and the TRGB (left dashed line). An additional dashed lines at 25 mag is plotted for relative comparison.\\}
    \label{fig:f110w_lum_func}
\end{figure}

\citet{Gregersen2015} found a metallicity gradient in M31 by fitting the red giant branch (RGB) population in the PHAT catalog with sets of isochrones. The gradient is smooth excluding an asymmetric metallicity enhancement between 3--6\,kpc related to M31's bar. In addition to the photometry, cluster properties, and classifications, Table \ref{table:photometry} includes the deprojected radius ($R_{\rm deproj}$) for each source, as well as the estimated metallicity ([M/H]) based on this metallicity gradient. We calculate the deprojected radius as the on-sky distance from the galaxy center (00$^{\rm h}$42$^{\rm m}$44$^{\rm s}$.330, +41$^{\circ}$16\arcmin07\arcsec.50), assuming a position angle of 38$^{\circ}$, and inclination of 74$^{\circ}$ \citep{Barmby2006}, and using the relation [M/H]\,=\,$-0.02 \times R_{\rm deproj}$ (kpc)\,+\,$0.11$.

\subsubsection{AGB Completeness} \label{sec:agb_completeness}

Within our \hst\ data, we are not significantly affected by crowding and completeness, as most of our sample is brighter than the TRGB. Figure \ref{fig:f110w_lum_func} shows the F110W luminosity distribution of the photometric catalog and our various cuts at different deprojected radii from the center of M31. In the initial PHAT data, we can identify features like the TRGB around 19 mag, the early-AGB bump at 22.5 mag, and the red clump around 24 mag; completeness begins to drop around 25 mag. Our final AGB catalog has isolated sources primarily above the TRGB in the F110W filter, or those reddened in the \hst\ or \spitz\ filters. The similarity of the chemically-classified and AGB candidates samples gives us confidence that we are not missing fainter AGB stars in our \hst\ data due to completeness issues. For our AGB stars classified using \spitz\, data, we may be more affected by crowding, especially near the galaxy center.

\begin{figure}
    \centering
    \includegraphics[width=\linewidth]{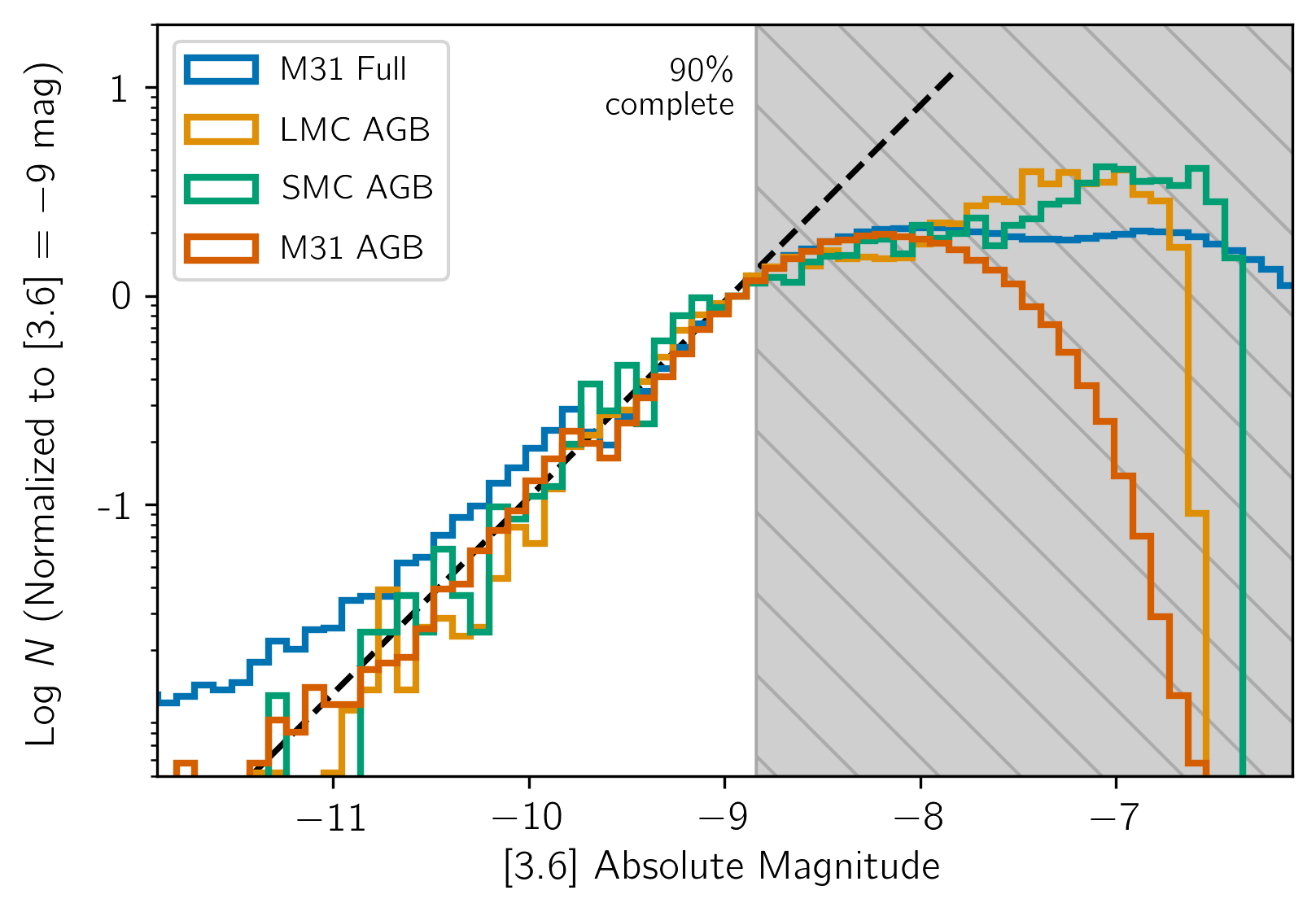}
    \includegraphics[width=\linewidth]{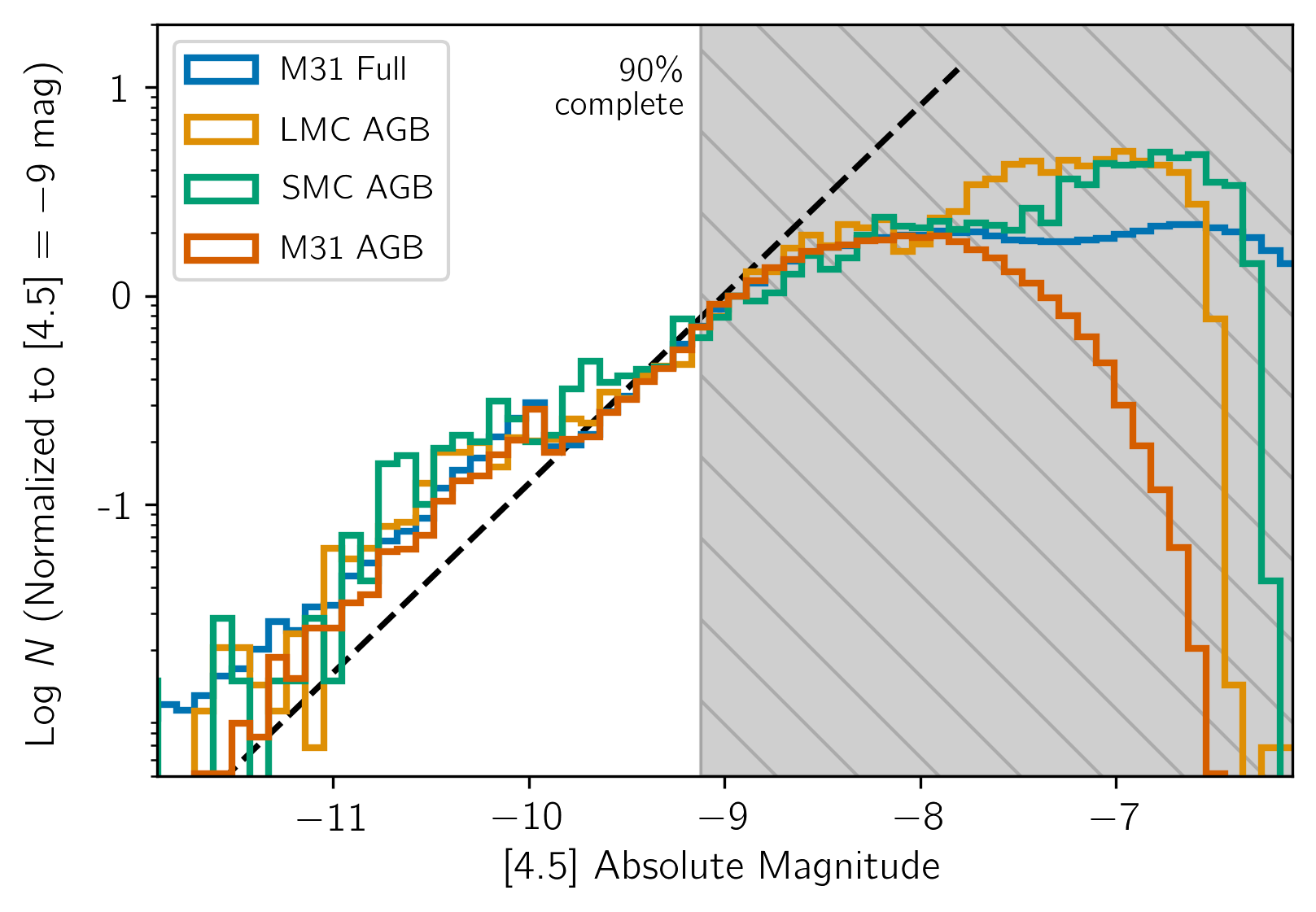}
    \caption{Luminosity functions showing the completeness of our AGB candidates in the IR. The full \spitz\ IRAC [3.6] and [4.5] data for M31 \citep{Barmby2006}, the LMC \citep{Meixner2006}, and SMC \citep{Gordon2011} are included for comparison. While the IR data in the MCs are nearly complete, the M31 IRAC data begin to lose sensitivity around $-9$ mag, shown in the flattening of the luminosity function. The slope of the fainter AGB sources (dashed line) has been used to estimate the completeness of the data (see \S\ref{sec:agb_completeness}). \\}
    \label{fig:ir_lum_func}
\end{figure}

Comparing our \spitz\ IR data with that of the MCs, we can see the sensitivity limits of M31 AGB candidates more clearly. Figure \ref{fig:ir_lum_func} shows \spitz\ absolute magnitudes of the M31 sample\footnote{We assume distance moduli of $m - M = 24.4$, 18.477, and 18.96 mag for M31 \citep{Dalcanton2012}, the LMC \citep{Pietrzynski2019}, and the SMC \citep{Scowcroft2016}, respectively}. The full M31 \spitz\ catalog as well as our AGB candidates can be seen to deviate from a power law around $[3.6] = 15.4$ mag (M$_{[3.6]} \sim -9$ mag). The \spitz\ photometry for the MCs are very near complete \citep{Meixner2006,Gordon2006} and show a smooth bump in the luminosity function where the dustiest carbon-rich stars lie, and where we begin to lack completeness in our \spitz\ photometry. The change in the shape of the luminosity function (at M$_{[3.6]} \sim -10$) is the result of our two-part photometric pipeline (see \S2). We have fit the slope of the luminosity functions between $-$9 and $-$10 mag and extrapolated the function (dashed line) towards fainter magnitudes to estimate our completeness. We estimate a $\sim$90\% completeness down to $-$8.84 and $-$9.12 mag for the [3.6] and [4.5] bands, respectively, which includes the majority of the dustiest AGB stars within the Magellanic Clouds \citep{Riebel2010,Boyer2011}. There is significant incompleteness down to M$_{[3.6]} = -6$ mag; as a result, moderately dusty AGB candidates classified with \hst\ data may not have an associated \spitz\ magnitude.

\subsubsection{CMDs} \label{sec:cmds}

\begin{figure*}
    \centering
    \includegraphics[height=20cm]{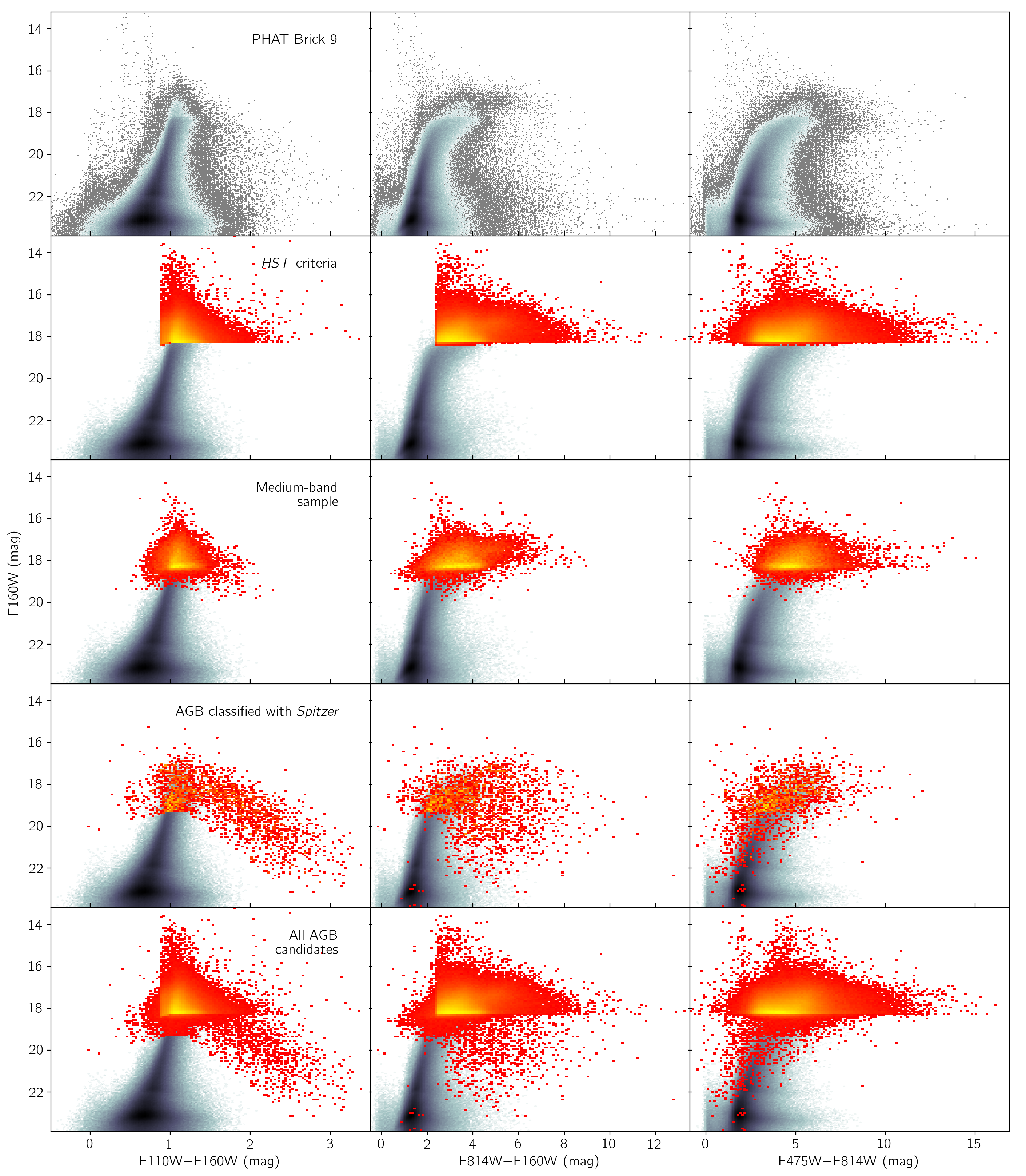}
    \caption{CMDs of all of the AGB candidates within the PHAT footprint meeting our various criteria. Shown are Row 1: an example of the PHAT photometry in Brick 9, Row 2: the AGB candidates classified with \hst, Row 3: the chemically-classified subset from \citet{Boyer2019}, Row 4: the AGB candidates classified with \spitz, Row 5: all criteria combined. The most-yellow regions for each row starting from the \hst\ criteria  correspond to densities of around 12, 5, 2, and 12, respectively. \\}
    \label{fig:torch_plot}
\end{figure*}

CMDs that include the AGB candidates, the chemically-classified AGB candidates, and the x-AGB candidates are shown in Figure \ref{fig:torch_plot}.

\begin{figure*}[]
\centering
 \includegraphics[width=0.48\linewidth]{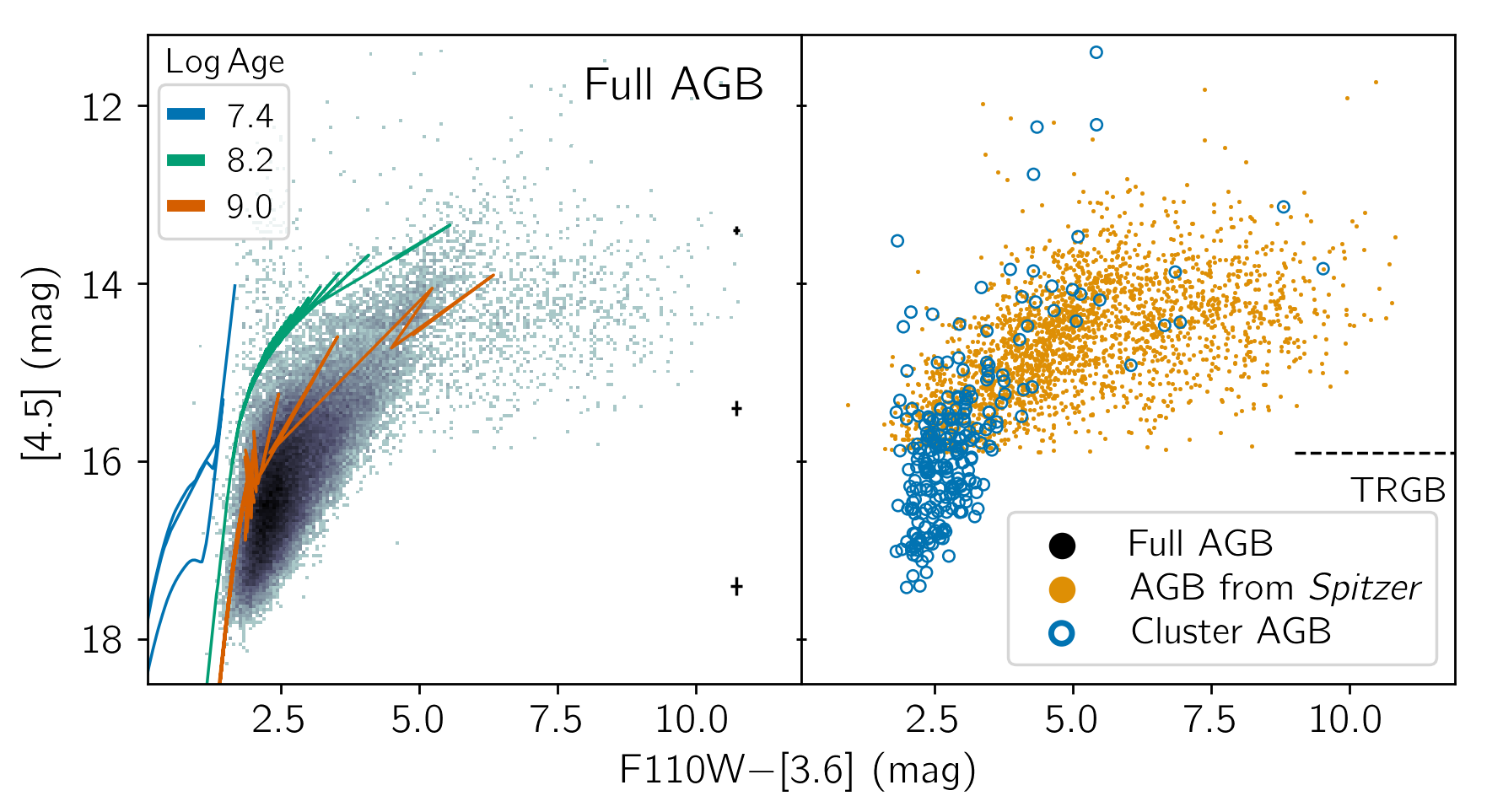}
 \includegraphics[width=0.48\linewidth]{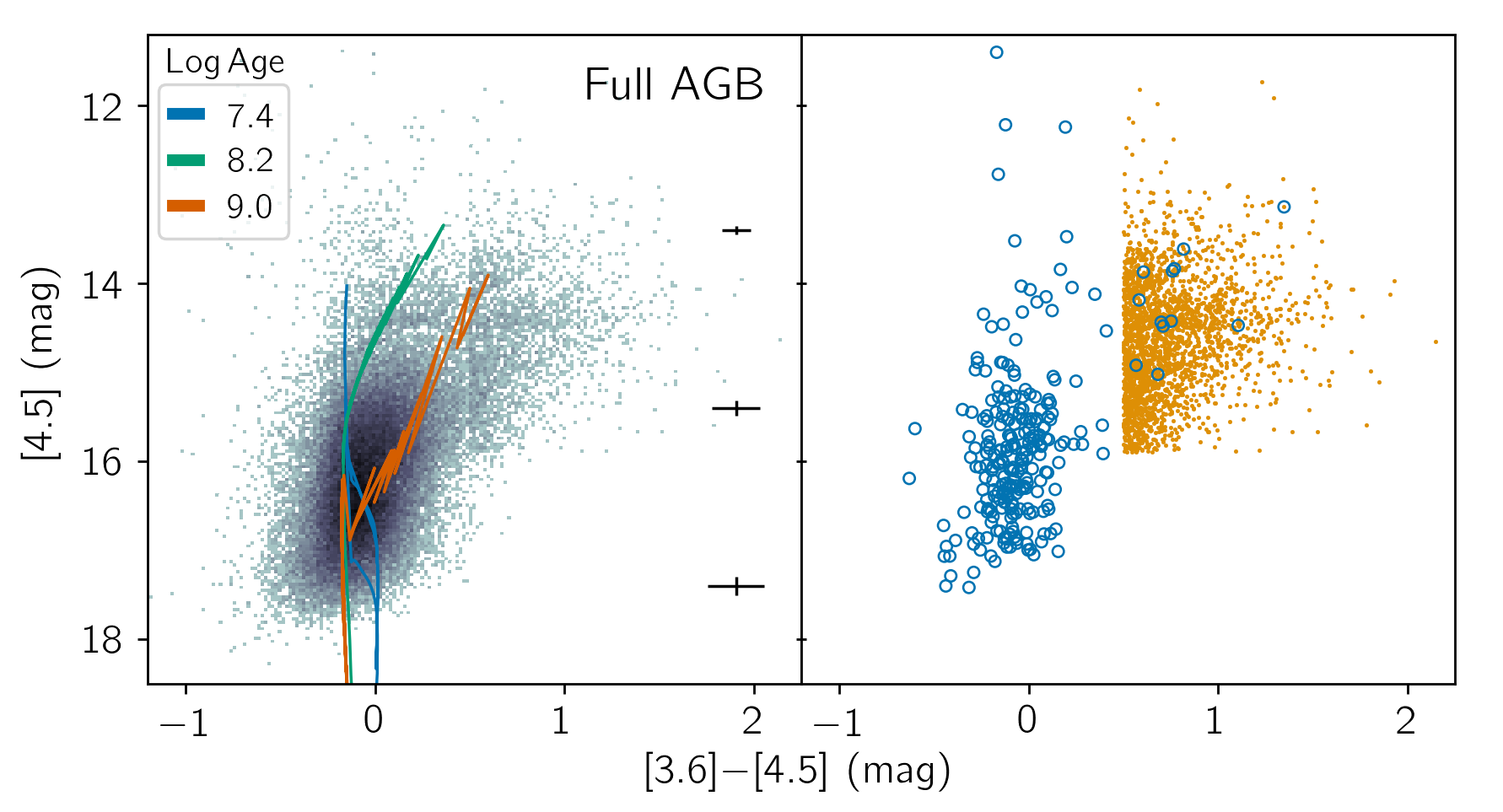}\\
 \includegraphics[width=0.48\linewidth]{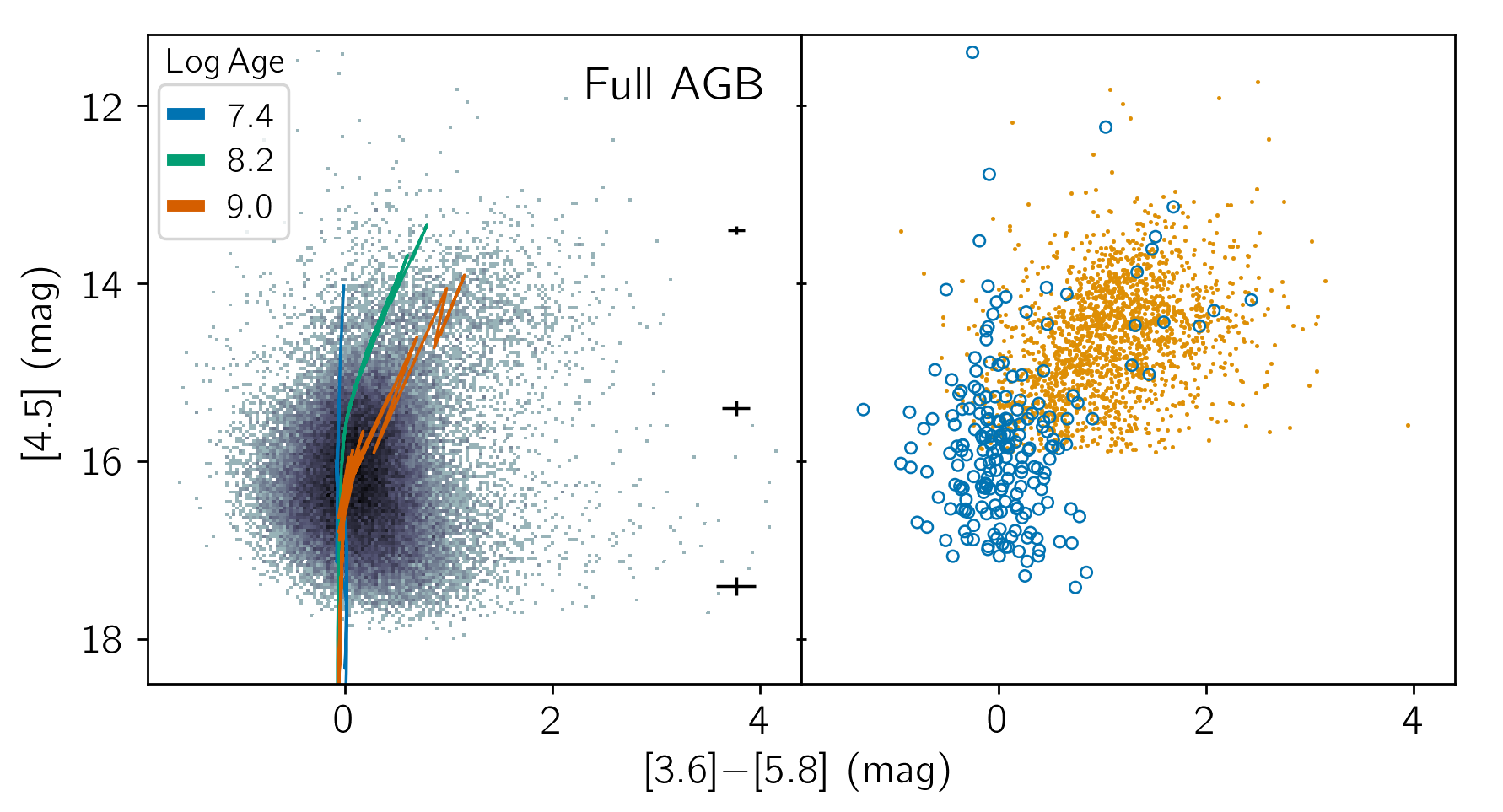}
 \includegraphics[width=0.48\linewidth]{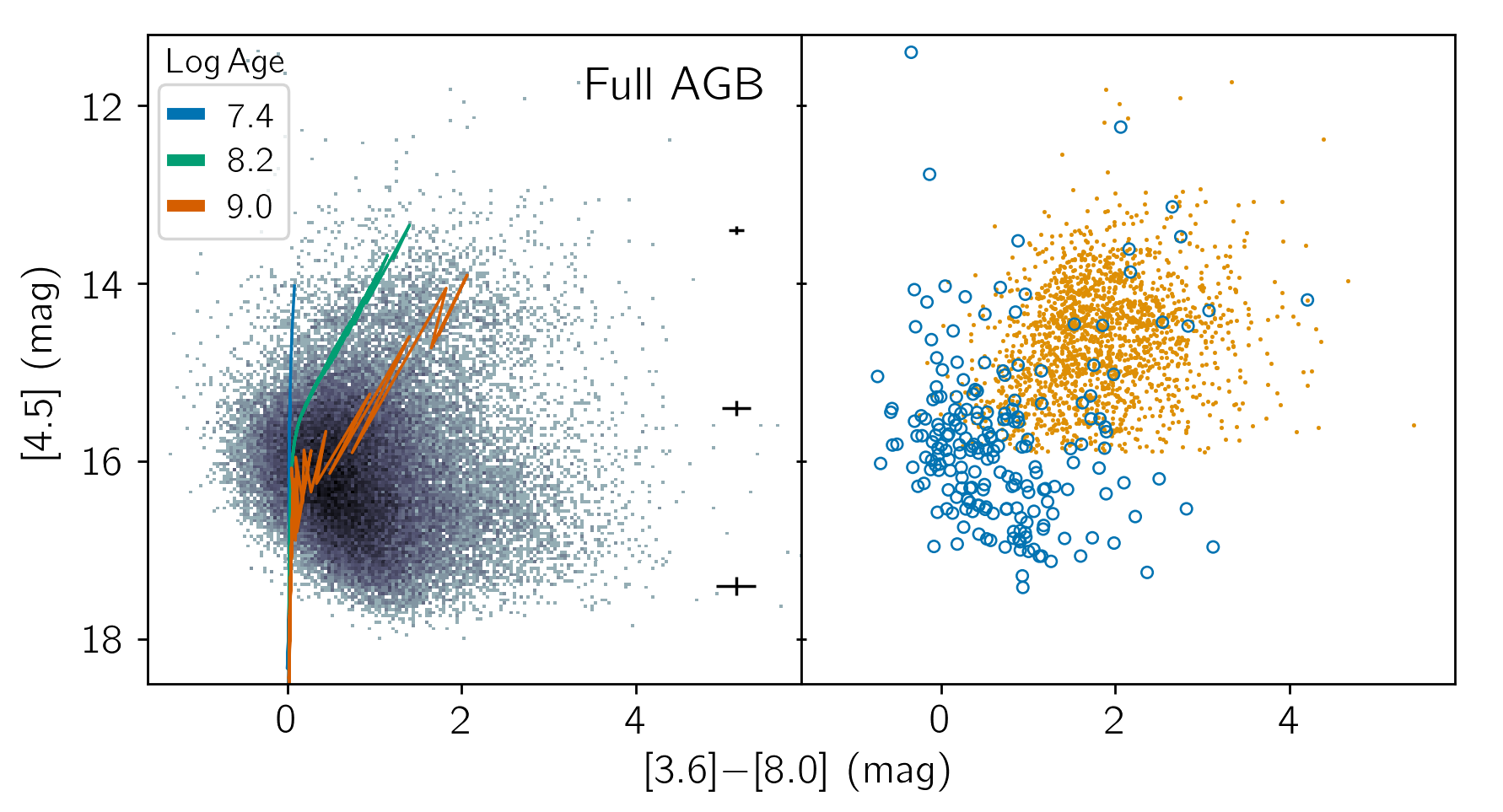}
 \caption{\spitz\ infrared CMDs of the M31 AGB candidates (black), AGB candidates classified with \spitz\ (orange), and cluster AGB candidates (blue). PARSEC isochrones \citep{Marigo2017} are shown to indicate the end stage of evolution before the onset of AGB mass loss; also shown are errorbars calculated using the average photometric uncertainty.  \\}
 \label{fig:cmds}
\end{figure*}

\begin{figure*}
\centering
    \includegraphics[width=\linewidth]{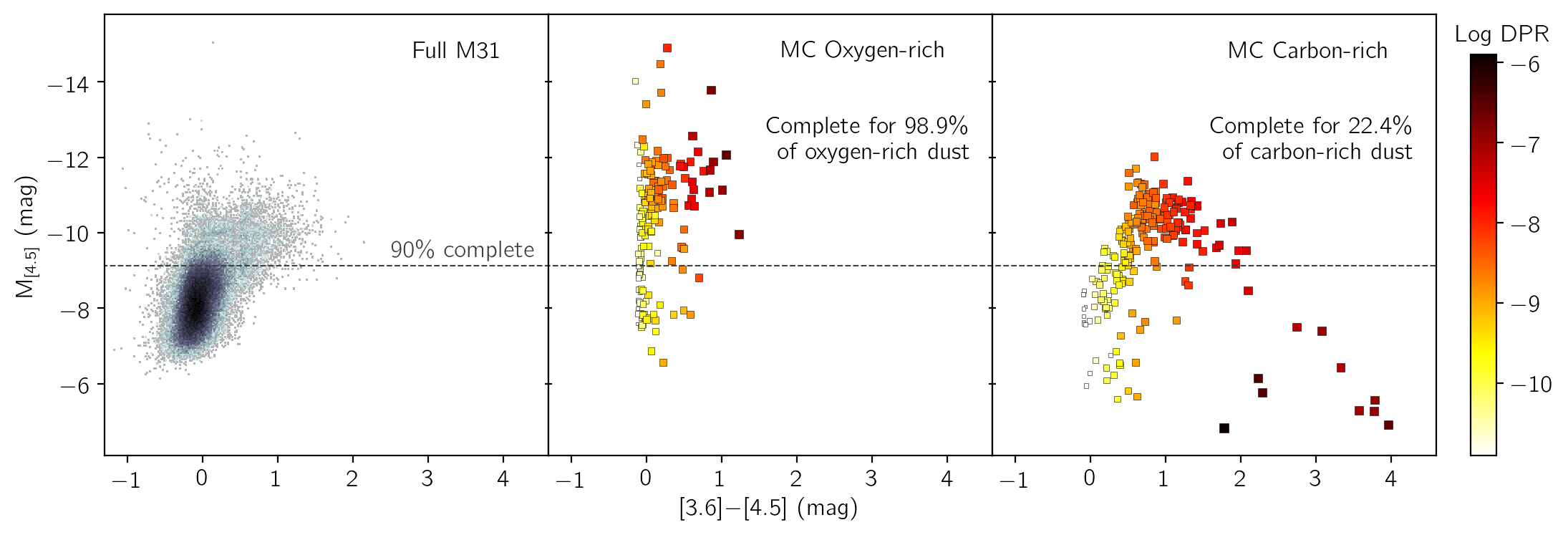}
    \caption{\spitz\ CMDs showing our M31 AGB sample (Left) along with the dustiest oxygen-rich (Center) and carbon-rich (Right) AGB stars found in the MCs \citep{Groenewegen2018}. The dust production rates (DPRs)} for the MC samples are shown in color (and size) and were measured using spectral energy distribution (SED) fitting. The threshold at which our \spitz\ data are 90\% complete  (M$_{[4.5]} \sim -9.12$ mag) is shown with a dashed line and the fraction of the dust injected by stars of that chemical subset above this limit shown as a percentage. \\
    \label{fig:missing_dusty}
\end{figure*}

Figure \ref{fig:cmds} shows infrared CMDs of our AGB candidates, our AGB candidates recovered with \spitz, and cluster AGB candidates. Comparing the shapes of the CMDs to those in the MCs, we see no clear indication of separate features related to changes in chemistry. The cluster AGB candidates seem to span the same color--magnitude space as the full AGB sample, indicating that they are also representative of the full AGB sample. We see that our AGB candidates classified with \spitz\ are highly reddened (falling to the right side of each panel), indicative of a high dust content. There is no noticeable change in the morphology of the CMDs with respect to deprojected radius except for sensitivity issues as a result of crowding.

\section{Catalog Properties}

\subsection{The x-AGB stars}\label{sec:x-AGB_stars}

AGB stars reach a point at which their mass-loss rate exceeds the nuclear-consumption rate, known as the ``superwind'' phase \citep{Renzini1981}. At this point, the timescale of evolution of the star is dictated by the mass-loss rate. Attempts to isolate stars in this short-lived phase led to the classification of ``extreme'' or x-AGB stars. While they represent a small fraction of the AGB populations ($\lesssim6\%$) in the MCs, they can account for up to $95\%$ of the dust \citep{Matsuura2009,Srinivasan2009,Boyer2012,Riebel2012}. We have therefore attempted to isolate this population in M31 using \spitz\ IRAC color and magnitude cuts to study their properties separately. We will use these x-AGB candidates to estimate the global dust injection of the AGB sample. The x-AGB candidates are selected using the following criteria: \\

Extreme AGB candidates
\begin{itemize}
    \setlength\itemsep{-0.25em}
    \item {[3.6]--[4.5] $>$ 0.25} \, and \, $[4.5] < 16.4$ mag  \vspace{0.2cm}
\end{itemize}

This is a sub-classification of our AGB candidate catalog, and is not a part of the AGB candidate criteria. It is similar, however, to the additional inclusion criteria listed previously ([3.6]--[4.5] $>$ 0.5), which includes sources in our AGB catalog irrespective of the \hst\ data.

The reason why we include AGB candidates with $[4.5] < 16.4$ mag and [3.6]--$[4.5]>0.5$ mag irrespective of their \hst\ data, but do not for colors between 0.25--0.5 mag is that we expect a high fraction of contamination from other dusty sources in this color range. If sources in this color range are also classified as AGB candidates based on the \hst\ data, we include them in the x-AGB subset. Of the \dustyagb\ sources we classify as x-AGB stars, \additionalinclusion\ of these sources are included with the additional inclusion criteria.

\begin{figure*}
    \centering
    \includegraphics[width=0.72\linewidth]{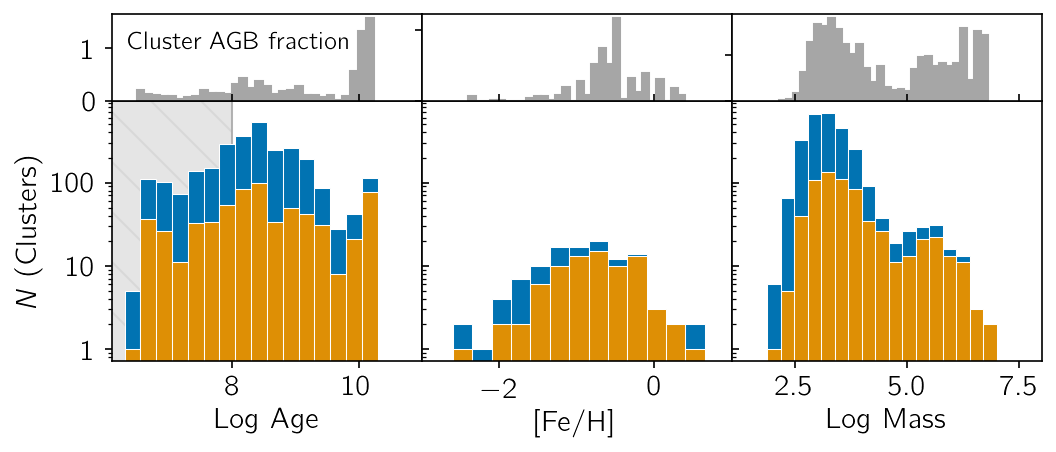}
    \caption{Histograms of the M31 clusters (blue) and their ages, masses, and metallicities, and the subset of those where we have detected AGB candidates (orange). Also shown are histograms of the fraction of individual sources within clusters, for different cluster properties. The age limit where cluster stars are not expected to have had enough time to evolve to the AGB phase is shown in the shaded region. \\}
    \label{fig:cluster_histograms}
\end{figure*}

We have identified \dustyagb\ x-AGB candidates in M31 using our IR photometry. Here we compare our sample with the AGB sample from the SAGE surveys in the MCs to study the differences in the sample properties and probe the limits of our data in isolating dusty AGB stars. Within M31, the x-AGB population makes up 1.3\% of the total AGB population. This is considerably lower than the values of 4.5\% and 6\% found in the LMC and SMC, respectively \citep{Riebel2012,Boyer2011}. We will examine several possible explanations for this low x-AGB fraction.

\subsubsection{x-AGB completeness}
The low fraction of x-AGB stars in M31, compared to the MCs, may suggest a real trend with metallicity, but may also be related to sensitivity issues, differences in selection criteria, and the star-formation history.

\paragraph{Sensitivity issues}
Our medium-band data suggests that we may be missing x-AGB stars due to completeness issues. In the near-IR (Figure \ref{fig:knee_plot}), we see that the x-AGB candidates that were chemically classified by \citet{Boyer2019} scatter in the direction of increased dust and extinction. While this confirms our dusty classification using \spitz, we lack sources that are so dusty that the dust significantly veils the water feature. With increasing dust, the veiling of the molecular features pushes these sources up and to the right of the color-color diagram (CCD) in the direction of the extinction vectors. These stars follow the extinction vectors until the features are completely veiled and the molecular signature is lost. At this point, they return to the juncture of the carbon- and oxygen-rich stars, near the location of the dusty sample from the LMC (open squares). In the IR (Figure \ref{fig:missing_dusty}), we also do not see dusty sources like the carbon stars in the MCs with [3.6]$-$[4.5] $>$ 2.75 mag and mass-loss rates $\sim 10^{-4}$\,M$_{\odot}$yr$^{-1}$ \citep{Groenewegen2018}. We expect M31 AGB stars to be producing predominantly oxygen-rich dust. While our sensitivity is only sufficient to have detected 22.4\% of the dustiest carbon-rich AGB stars in the MCs, this number is 98.9\% for the dustiest oxygen-rich AGB stars.

\paragraph{Star Formation History} A higher recent star formation would result in a higher fraction of massive dusty AGB stars due to their shorter evolutionary timescales. Given that the LMC has had a higher relative intensity of recent star formation than M31 and the SMC, but lies in between them in terms of the x-AGB fraction, this is unlikely to explain M31's lower x-AGB fraction.

\paragraph{Selection Criteria} The x-AGB classification aims to select those stars producing the bulk of the AGB dust. The boundaries of this classification, however, are not based on changes in stellar structure or evolution, but are defined empirically. The x-AGB classification criteria in the MCs requires a [3.6] magnitude above the TRGB and $J-[3.6] > 3.1$ mag. To avoid misclassification as a result of mismatches between the \hst\ and \spitz\ data, our x-AGB classification is based entirely on \spitz\ data. Our slightly different TRGB requirement in the [4.5] filter and color cut in the mid-IR ($[3.6]-[4.5]>0.25$ mag) may contribute to the lower x-AGB fraction in M31. The x-AGB fraction in 6 nearby dwarf galaxies was found to be between 2--6\%, as opposed to our 1.3\% x-AGB fraction, and these also used a \spitz-only x-AGB criteria \citep{Boyer2017}.

\subsection{Cluster statistics}
\label{sec:cluster_statistics}

Cluster AGB candidates were recently identified by \citet{Girardi2020}. Here, we re-identify AGB candidates in clusters using our AGB classification criteria, which are slightly different from that used by \citet{Girardi2020}. The cluster properties \citep[age, mass, and metallicity; ][]{Johnson2015} of sources that lie within the estimated radii are listed in Table \ref{table:photometry} and are shown in Figure \ref{fig:cluster_histograms}. Metallicities were measured spectroscopically \citep{Caldwell2011}, and are primarily limited to the older, more luminous clusters.

We identify \clusteragbs\ cluster AGB candidates, including 96\% (672 / 697) of the cluster AGB candidates identified in \citet{Girardi2020}. Of our additional cluster AGB candidates, 74\% (456 / 616) are located in Bricks 1 or 3 \citep[ignored in ][]{Girardi2020}, and the majority of the remainder meet our F160W magnitude criteria (F160W $<$ 18.28 mag) but not that of the \citet{Girardi2020} catalog (F160W $<$ 18.14 mag). A table showing the differences between our selection and that of \citet{Girardi2020} is shown Appendix A and the results are discussed in \S\ref{sec:agb_criteria_effectiveness}.

\citet{Girardi2020} showed that M31 cluster sources identified with the PHAT data are also modestly affected by crowding. The crowding parameter was calculated for each of the cluster sources in each filter, which calculates the amount of additional flux had nearby stars not bit fit simultaneously during the photometry process. The crowding parameter in the near-IR filters was found to be overwhelmingly below 0.2 mag, indicating a small impact as a result of crowding.

With our chemically-classified sample and by folding in the \spitz\ photometry, we are also able to identify a handful of chemically-classified ($N$=\clusterchem) and x-AGB candidates ($N$=\dustyclusteragb) within clusters. We show \hst\ image cutouts of the x-AGB cluster candidates, as well as the source SEDs fit with radiative transfer models using the Dusty Evolved Star Kit \citep[{\asciifamily DESK};][]{Goldman2020} in Appendix B. These sources are likely biased toward larger fluxes in the IR as a result of the lower spatial resolution of the \spitz\ data, and crowding near the cluster centers. Follow-up observations of these sources in the IR are needed to confirm that the IR excess is in fact associated with the AGB candidates.

A fraction of the cluster AGB candidates that we identify may in fact be foreground contamination or M31 field AGB stars. \citet{Girardi2020} presented an in-depth analysis of the M31 cluster AGB sample selected from PHAT. They show that, by extrapolating the field star density function down to the centers of the M31 clusters, around half of the potential AGB cluster members are likely spatially coincident field stars. They also argued that clusters with log age $<$ 8 should not host AGB stars, as their intermediate- and low-mass stars would still be in a previous evolutionary stage. Out of our \clusteragbs\ cluster AGB candidates, \clustersourcesyoungerthaneight\ cluster sources are found coincident with these younger clusters. Additionally, while contamination from RSGs is expected to be low in our global AGB sample, the fraction of RSGs in young clusters is expected to be considerably higher.

While the M31 cluster sample shows much promise for calibrating stellar evolutionary models, a careful assessment of the candidates is required to identify true AGB cluster stars. \\

\begin{figure}[b]
    \centering
    \includegraphics[width=0.85\linewidth]{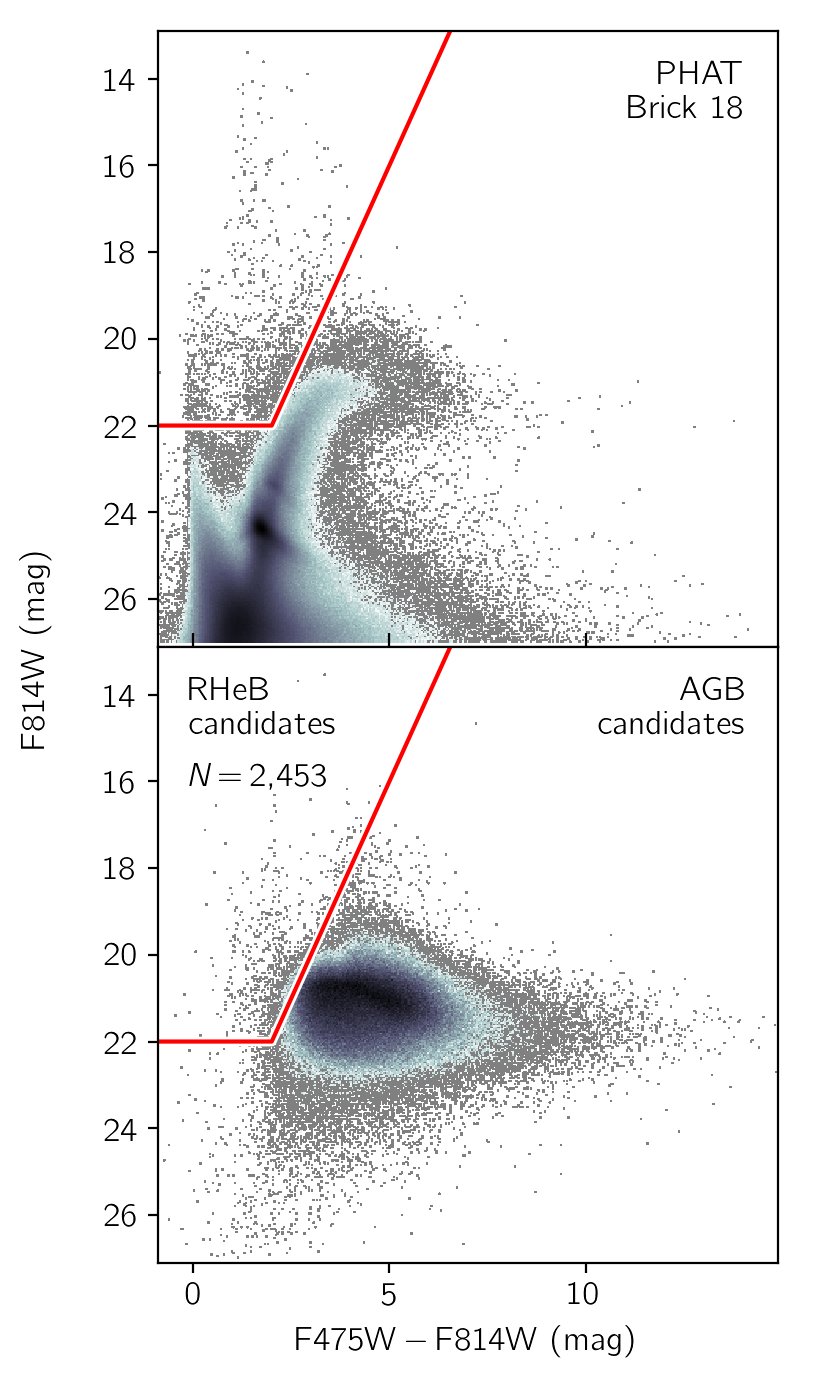}
    \caption{Optical CMDs showing the RHeB candidates and how they are selected. We used the PHAT photometry in Brick 18, where crowding is minimal, to refine our cut. These sources are flagged but included in our AGB catalog. \\}
    \label{fig:RHeB_cmd}
\end{figure}

\subsection{RHeB Stars}\label{sec:RHeB_stars}
Within our AGB candidates we likely have a small fraction of contamination from warmer giants and supergiants. Evolved stars fusing helium within their core (i.e., blue loop, and red supergiants) are often studied as a single population as they can be observationally indistinguishable. We have attempted to isolate these red helium burning (RHeB) stars using color and magnitude cuts in the optical. While we suspect that they may be in a different evolutionary phase, we include them in our AGB catalog as they are still expected to contribute to the dust budgets of galaxies.

To isolate the RHeB candidates we have used cuts in magnitude and color space (Figure \ref{fig:RHeB_cmd}). We have determined these cuts visually, trying to isolate the warmer and more luminous stars. We use an optical CMD of PHAT Brick 18 to determine our cuts, as sources in the outskirts of the galaxy are less affected by crowding and completeness issues. In total we flag \RHeB\ sources as potential RHeB candidates using the following criteria: \\

RHeB candidate (both of the following):
\begin{itemize}
    \setlength\itemsep{-0.25em}
    \item F814W$ < 22$ mag
    \item Above the line $(F814W) = -2 \times (F475W-F814W) + 26$ mag
\end{itemize}

Using a hand-picked and incomplete subset of 184 cluster RHeB/RSG stars in M31, 65 (35\%) were included in the \citet{Girardi2020} AGB selection. Within our sample, this number drops to 14 (8\%) as a result of our F814W--F160W color cut, 8 of which are flagged as RHeB candidates. This gives us confidence that we have reduced the contamination from these more massive stars.

Within the RHeB candidates are sources also classified in our other sub-classifications. The RHeB sample includes 53 red supergiant candidates from \citet{Ren2021}, 53 cluster AGB candidates, 88 chemically-classified AGB candidates, and 90 AGB candidates included with \spitz.

\subsection{Carbon-star luminosity distribution}\label{sec:carbon-star_luminosity_distribution}
Within the small regions where AGB chemical types have been determined from medium-band \hst\ imaging \citep{Boyer2019}, we can use our additional photometry, measurements, and classifications, to study the properties of metal-rich carbon stars in greater detail.

Current models disagree on the predicted mass limit for forming carbon stars. The ranges are expected to vary dramatically with metallicity with the largest uncertainties stemming from the unclear effect of metallicity on the efficiency of the HBB and TDU processes. Models suggest that the carbon star mass range can vary from 1.75--7 M$_{\odot}$ in metal-poor environments (Z=0.007), to 2--4.5 M$_{\odot}$ at solar metallicity, and  3.25--4.0 M$_{\odot}$ in metal-rich environments \citep{Karakas2014}. Other metal-rich models predict a more narrow carbon star mass range of 2.5--3.5 M$_{\odot}$ \citep{Ventura2020}, and observations have suggested a lower and even more narrow mass range. In the Milky Way, \citet{Marigo2020} discovered a kink in the CO white dwarf mass function attributed to a low mass boundary of carbon stars at $1.8-1.9$\,M$_{\odot}$. We will use our more metal-rich chemically-classified subsample to provide additional constraints on these critical chemical transitions.

\begin{figure}
    \centering
    \includegraphics[width=\linewidth]{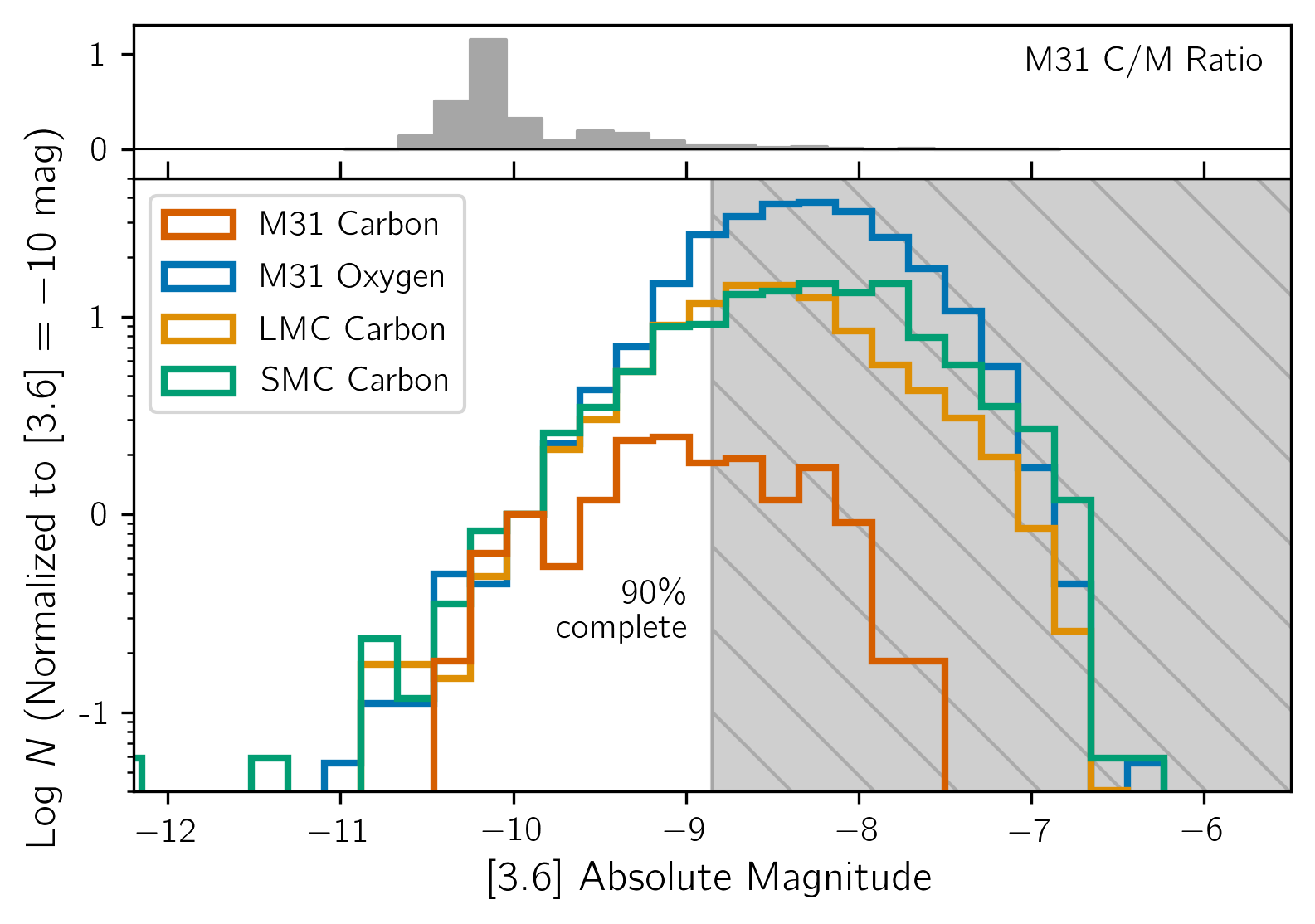}
    \caption{The distribution of IRAC [3.6] absolute magnitudes for the carbon-rich samples within M31 and the MCs \citep{Boyer2019,Boyer2011}. Also shown is the luminosity function for the M31 oxygen-rich sample showing that our \spitz\ data is able to detect oxygen-rich AGB candidates fainter than our faintest carbon stars.\\}
    \label{fig:chem_lum_func}
\end{figure}

\begin{figure*}[]
    \hspace{0.cm}
    \includegraphics[width=0.29\linewidth]{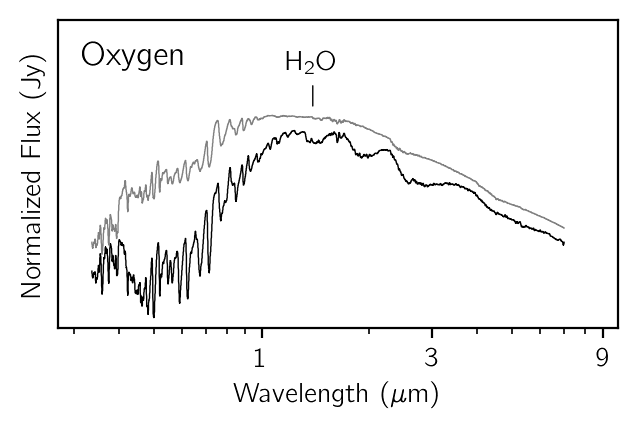}\hspace{0.33cm}
    \includegraphics[width=0.29\linewidth]{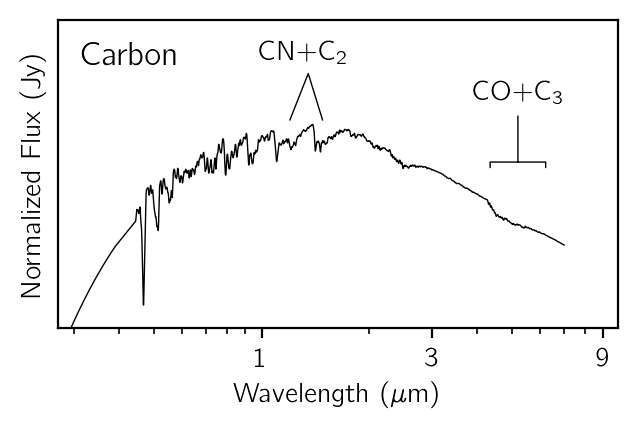}\hspace{0.33cm}
    \includegraphics[width=0.29\linewidth]{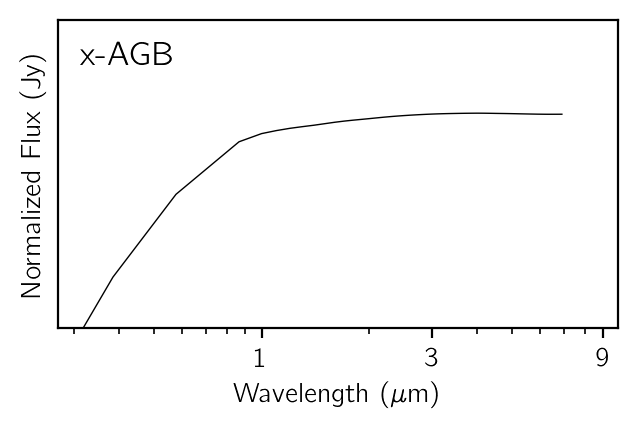}\hspace{0.3cm}
    \includegraphics[width=0.32\linewidth]{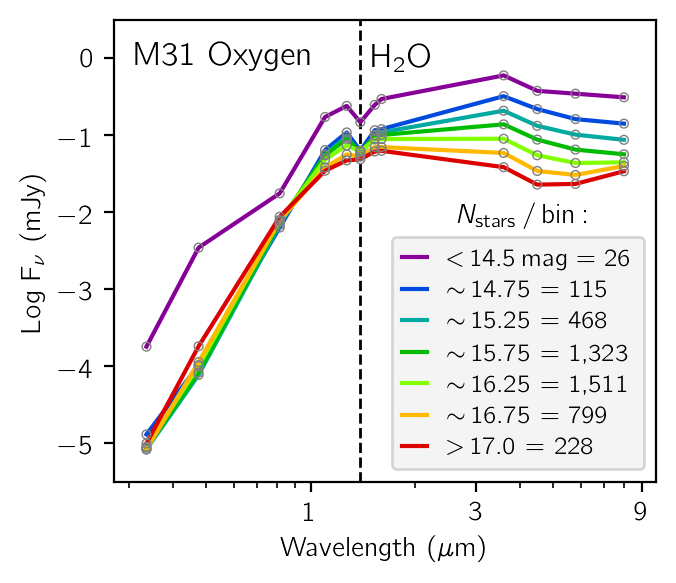}
    \includegraphics[width=0.32\linewidth]{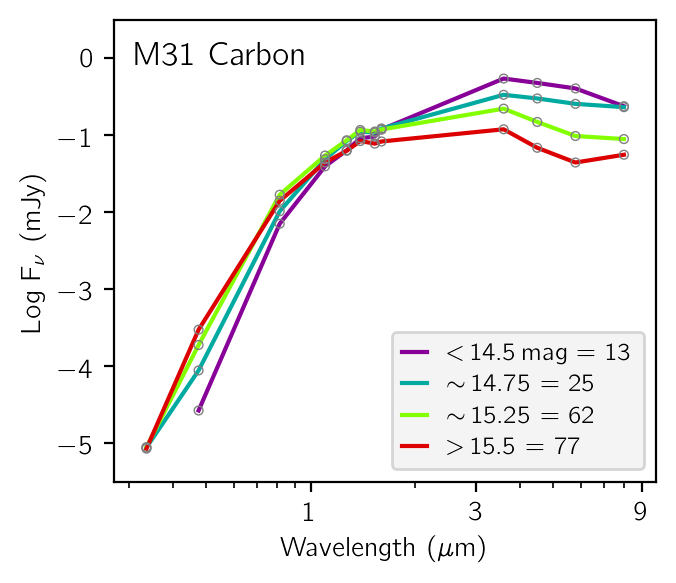}
    \includegraphics[width=0.32\linewidth]{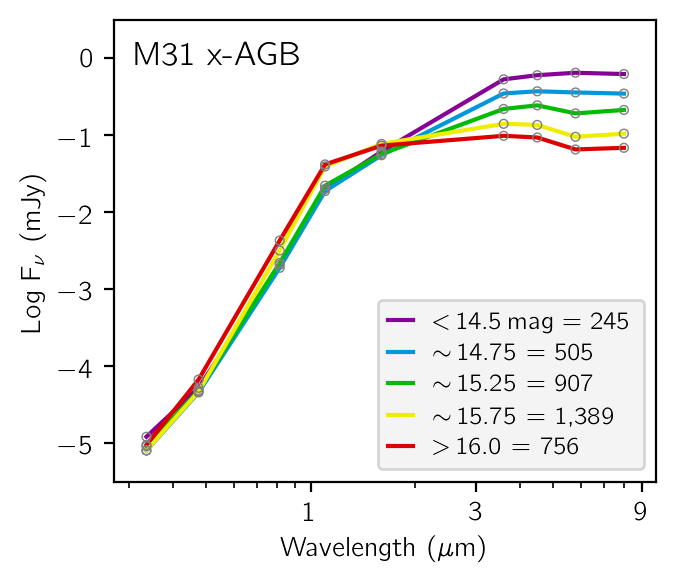}
    \includegraphics[width=0.32\linewidth]{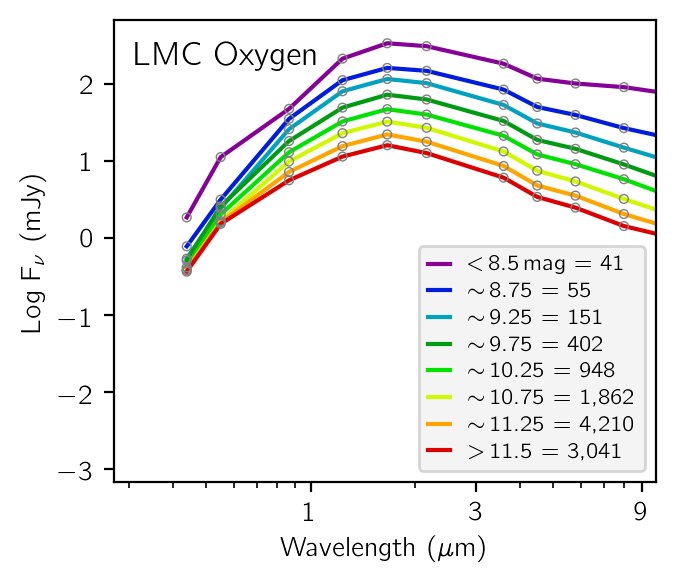}
    \includegraphics[width=0.32\linewidth]{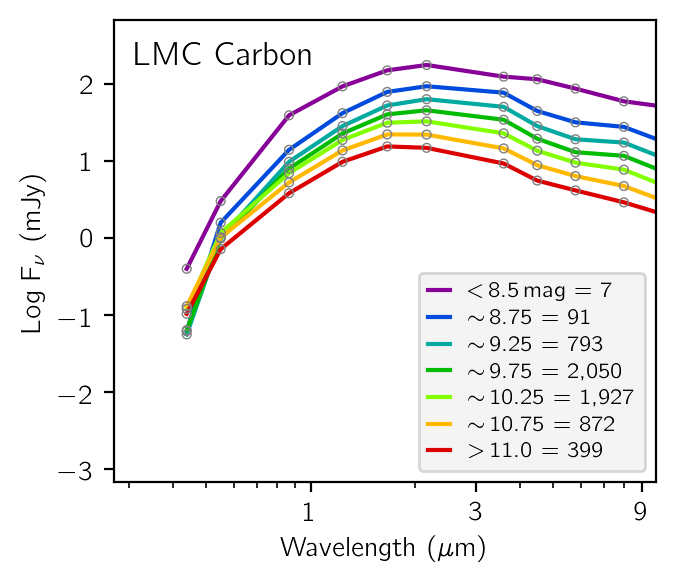}
    \includegraphics[width=0.32\linewidth]{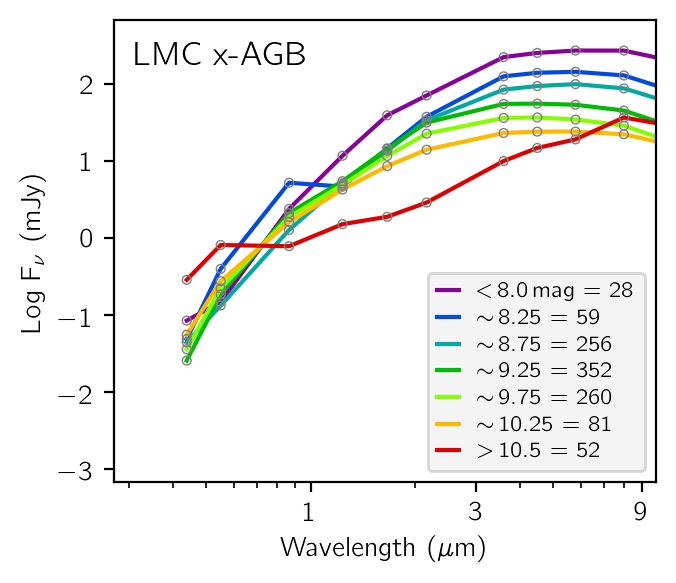}
    \includegraphics[width=0.32\linewidth]{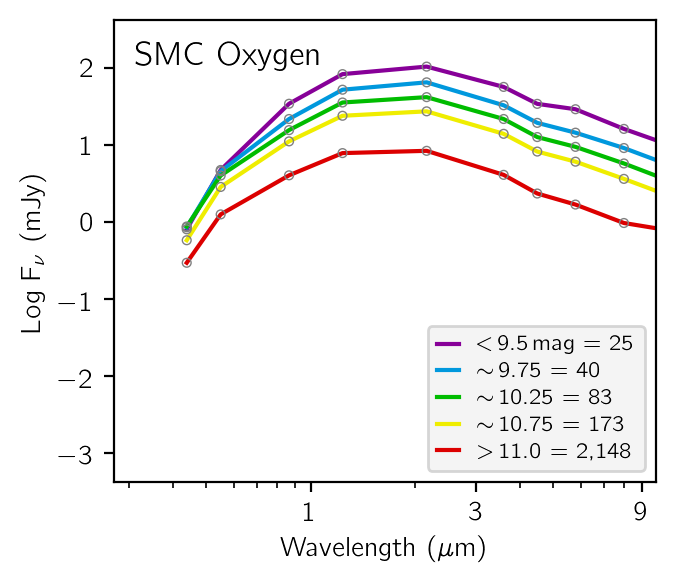}\hspace{0.3cm}
    \includegraphics[width=0.32\linewidth]{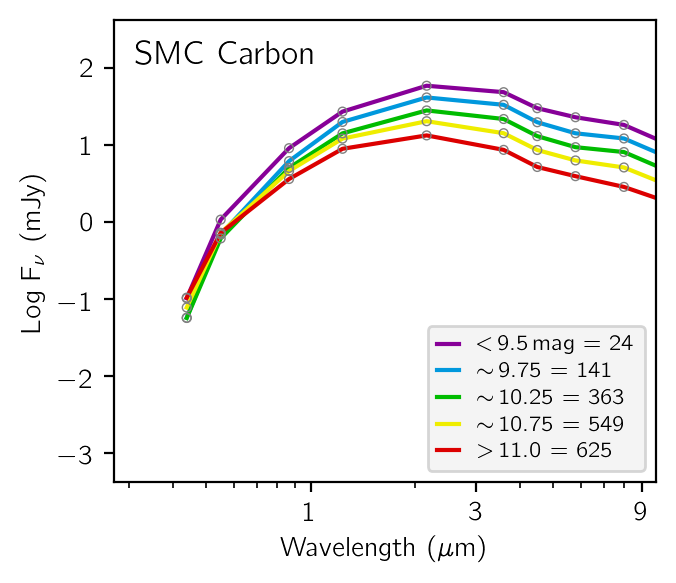}\hspace{0.3cm}
    \includegraphics[width=0.32\linewidth]{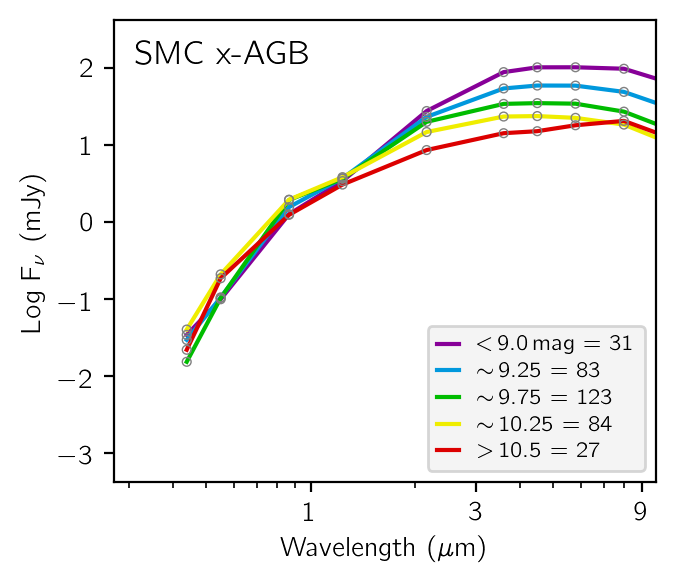}
    \caption{Median SEDs of the oxygen-rich, carbon-rich, and x-AGB candidates within the PHAT footprint of M31. We include only sources with IRAC 3.6\,$\mu$m data and those outside of Brick 1, the region closest to the galaxy center. The data are binned by their absolute IRAC [3.6] magnitude, with a bin size of 0.5 mag, and the number of stars in each bin represented in corresponding color on the right of each figure. We have excluded the MIPS 24\,$\mu$m magnitudes due to the lower spatial resolution of the observations; we have also removed the medium-band \hst\ data for the x-AGB median SED due to the low number of x-AGB sources with medium-band data. Example model spectra from the COMARCS \citep{Aringer2009,Aringer2016} code are shown at the top for reference, with two oxygen-rich models at 2600 K (black) and 3300 K (gray), as well as another carbon-rich model at 3400 K, all of which assume no dust. We also show a dusty model created using the {\asciifamily DUSTY} code \citep{Elitzur2001}. \\}
    \label{fig:median_SEDs}
\end{figure*}

We estimate the luminosity boundaries of the carbon-star mass function using the luminosity function of our carbon-rich AGB candidates. Previous attempts to compare the carbon-star luminosity functions across galaxies have been hampered by the use of different filters. Here, our IRAC photometry is directly comparable to the IRAC photometry for the MCs from the SAGE surveys. Using this data we can finally make a direct empirical comparison.

The carbon-rich AGB candidates span a similar, yet more limited range in infrared magnitudes and colors (Figure \ref{fig:ir_criteria_cmd}). The luminosity function of the data shows a peak in the distribution around [3.6] $\sim$ $-$9 mag, where the \spitz\ data begin to lose sensitivity (Figure \ref{fig:chem_lum_func}). The photometry are complete at brighter magnitudes, suggesting a real upper limit, slightly fainter that those of the MCs. The carbon star luminosity function also shows a sharp drop at M$_{3.6} \sim -8$ mag. As our oxygen-rich stars (shown in blue) can be seen reaching magnitudes fainter that this lower limit, this suggests a real lower luminosity/mass limit unrelated to sensitivity issues. This also continues the trend of a more limited luminosity range with decreasing average metallicity across the three galaxies. This more limited luminosity range of the M31 carbon stars (as opposed to the MCs) is consistent with the expectation and observational evidence of decreased TDU efficiency in more metal-rich environments. 

In-falling dwarf galaxies and streams, and binary interactions (extrinsic carbon stars) are capable of producing metal-poor M31 carbon stars whose properties are not tied to TDU or these mass limits \citep{Majewski2003,Huxor2015,Escala2021}. \citet{Boyer2019} show, however, that the C/M ratio in M31 is lower than predictions based on other Local Group galaxies. Furthermore, they find a decrease in the C/M ratio from the disk to the bulge as expected for a metallicity gradient across the disk. They also find no evidence of spatial over-densities that would be associated with interloping carbon stars. This indicates that the M31 AGB population, including the carbon stars, is more metal-rich than those in the MCs.

\subsection{Median SEDs}\label{sec:median_seds}
To better understand AGB stars in more metal-rich environment, we must study how energy at different wavelengths is transported through their atmosphere and circumstellar envelope. The SEDs of our AGB candidates peak in the mid-IR with different shapes for the carbon- and oxygen-rich candidates, due to the different molecular and dust species that they produce. Using our sub-samples of AGB candidates, we have binned the SEDs of our oxygen-rich, carbon-rich, and x-AGB candidates by their IRAC [3.6] magnitudes. We show these binned median SEDs as well as those classified in the MCs \citep{Riebel2010,Boyer2011} in Figure \ref{fig:median_SEDs}.

\subsubsection{M31 median SEDs}
For our oxygen-rich AGB candidates we can clearly see the water feature at 1.39\,$\mu$m (labeled as H$_2$O), increasing in absorption with increasing 3.6\,$\mu$m flux. As far as we know, this is the first time that this has been seen photometrically. This is seen most dramatically in the most luminous oxygen-rich SEDs, which contains some foreground sources, and one possible background galaxy. As AGB stars evolve, their envelopes become more extended. This allows for cooler molecules like water to form in abundance. At the same time, material is levitated out to large radii through pulsations, where it cools and condenses into dust. Hydrostatic models have shown that with decreasing temperatures, as well as increasing mass, oxygen-rich evolved stars show stronger absorption of molecules like water, in the IR \citep{Aringer2016}. We suspect that the increase in the strength of the water feature and the increase at 3.6\,$\mu$m is likely a reflection of these stars being more massive and having cooler dustier envelopes. For our carbon-rich median SED, we see evidence for the CO + C$_3$ feature around 4--8\,$\mu$m in our fainter bins. These features are from photospheric molecular transitions and have been detected spectroscopically in nearby carbon stars \citep{Jorgensen2000,Zijlstra2006}. In our more luminous bins, the SED shapes starts to resemble the x-AGB median SEDs, which may indicate that these are dusty carbon stars.

We expect M31's x-AGB population to be dominated by oxygen-rich chemistry, unlike in the more metal-poor MCs, which are carbon-rich; we expect this for two reasons. First, the decreased efficiency of TDU events at high metallicity shrinks the mass range over which TP-AGB stars produce carbon-rich atmospheres, and consequently the size of the carbon star population \citep{Karakas2010,Marigo2017}. Secondly, oxygen-rich dust formation is expected to rely on the initial metal content to act as seed nuclei (unlike carbon stars), and we expect a high natal abundance of oxygen in metal-rich environments. As a result, we expect oxygen-rich dust to form more easily and in larger quantities in M31 as it is metal-rich. With that expectation, the fact that the x-AGB SEDs closely resembles the most-luminous carbon-rich SEDs may suggest a bias in the empirically determined x-AGB classification scheme.

\subsubsection{Median SED differences in M31 and the MCs}
A key difference that we see between the median SEDs in M31 and the MCs is that M31 IR fluxes are generally higher. This is likely related to our bias against fainter stars in the IR, below our detection threshold. It may also be affected by the lower spatial resolution with increasing wavelength, or the limited calibration of the longer-wavelength data from \citet{Khan2017}.

Comparing the carbon-rich AGB candidates in M31 with those of the MCs, we see that in M31, the CO + C$_3$ feature seems to increase in strength toward lower luminosity. We also see the CO + C$_3$ feature in fainter bins. This may reflect the cleaner separation of chemical types using \hst\ medium-band filters, as opposed to \emph{J} and \emph{K}$_{\rm S}$ in the MCs. It is unlikely that this broad feature is related to differences in our GSC and longer-wavelength catalog from \citet{Khan2017}, as the feature is also seen in our [4.5] GSC data.

\begin{figure}
    \centering
    \includegraphics[height=3.5cm, trim=4 4 4 4,clip]{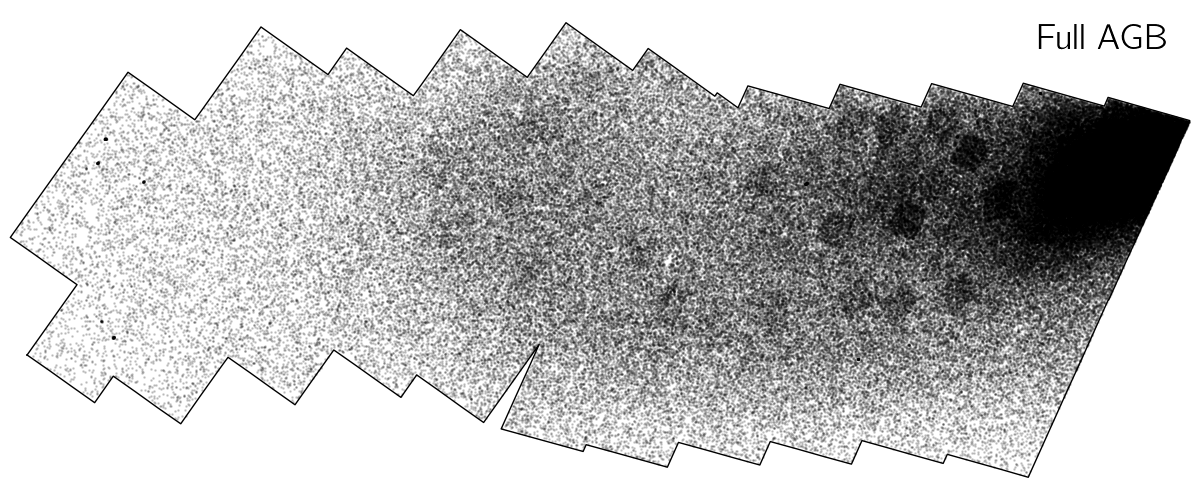}
    \includegraphics[height=3.5cm, trim=4 4 4 4,clip]{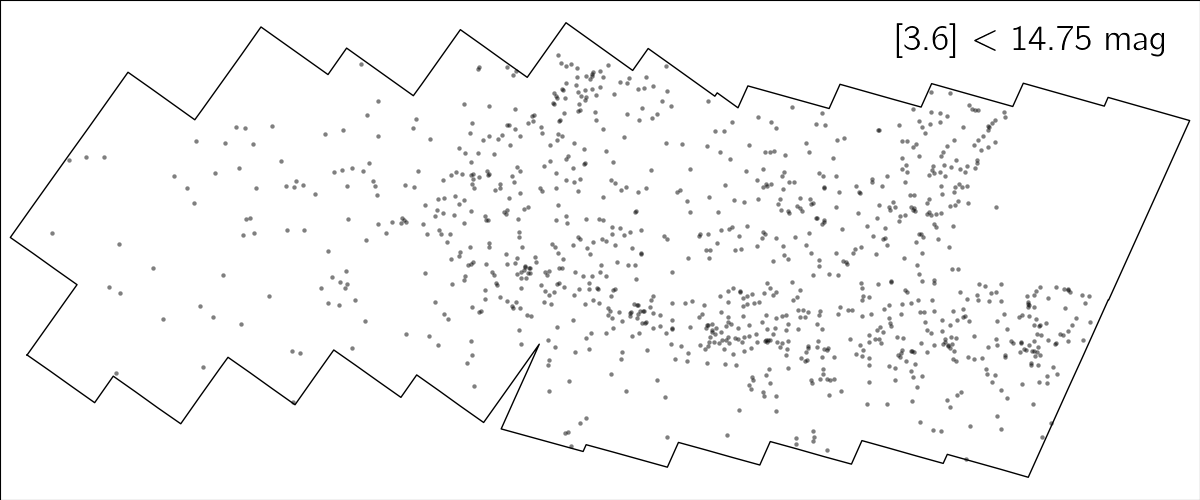}
    \includegraphics[height=3.5cm, trim=4 4 4 4,clip]{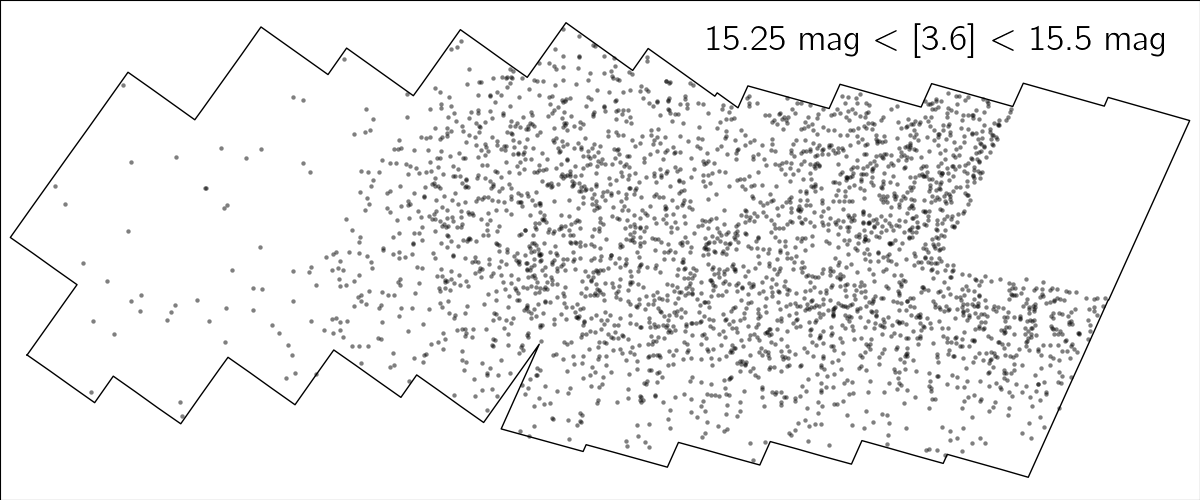}
    \includegraphics[height=3.5cm, trim=4 4 4 4,clip]{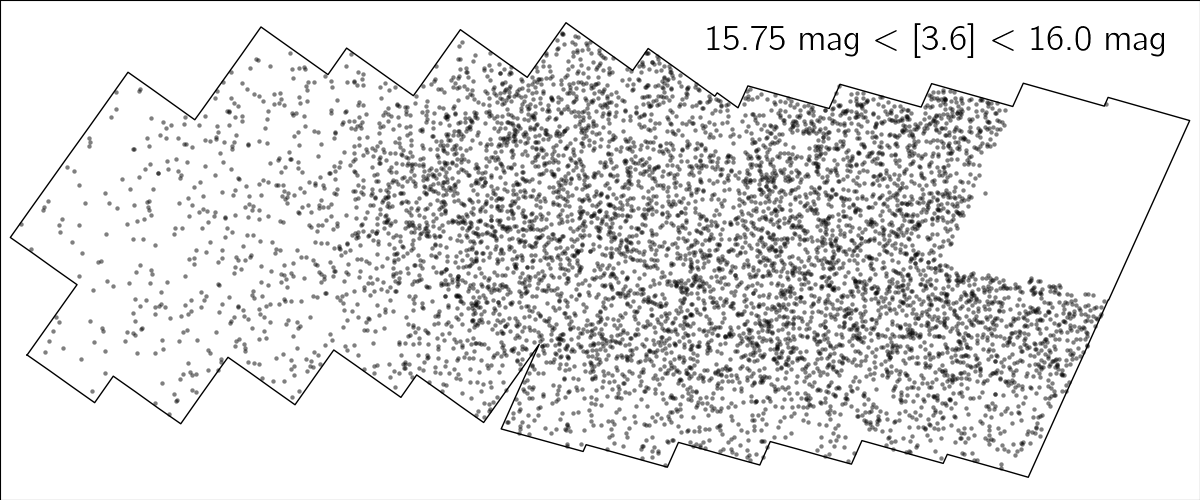}
    \includegraphics[height=3.5cm, trim=4 4 4 4,clip]{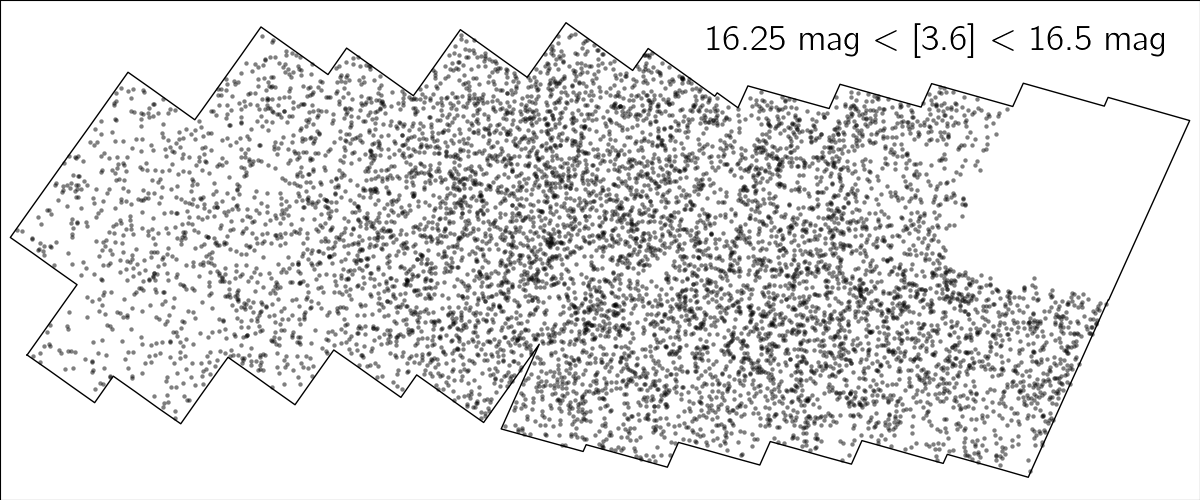}
    \caption{The spatial distribution of the full AGB candidates and distributions of AGB candidate subsets of differing IR magnitude; we omit sources within Bricks 1 and 3, which are significantly affected by crowding. Our most luminous sources are primarily located near the 10\,kpc ring, a region of recent star formation. Sources classified in \citet{Boyer2019}, which have additional medium-band \hst\ photometry, are included in the AGB catalog. As a result, a higher density of AGB candidates can be seen in these regions in the first panel, shown with white squares in Figure \ref{fig:spatial_distribution}. \\}
    \label{fig:bright_distribution}
\end{figure}

\begin{figure}
    \centering
    \includegraphics[height=3.5 cm, trim=4 4 4 4,clip]{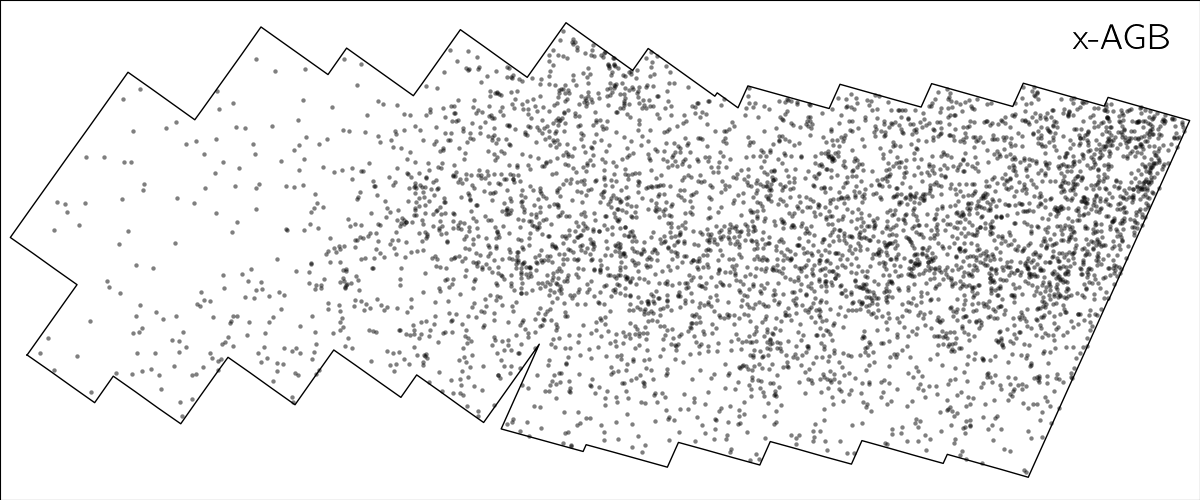}
    \caption{The spatial distribution of sources classified as x-AGB stars. \\}
    \label{fig:dusty_distribution}
\end{figure}

\subsection{AGB spatial distribution}\label{sec:agb_spatial_distribution}

The ages, metallicities, and stellar density of stars across M31 has been well characterized by PHAT. We can leverage this information to interpret the spatial distribution of AGB stars and their subtype and luminosity classes. M31 hosts a metal-rich bulge with a decrease in density and metallicity with radius \citep{Dalcanton2012, Gregersen2015}. Three ring features are visible within the PHAT footprint. Two faint rings at 5 and 15\,kpc, and a more clear ring at 10\,kpc. Star formation has been shown to vary across the disk with the most clear variation in the star forming in the 10\,kpc ring, where star formation has been ongoing for 500\,Myr \citep{Lewis2015}. Most of the star formation, however, occurred prior to 8\,Gyr \citep{Williams2017}. We will look at the spatial distribution of the AGB candidates to determine if they show a similar picture as these studies of the galaxy's star formation.

In Figure \ref{fig:spatial_distribution} the AGB sample in the PHAT footprint shows a concentration toward M31's bulge, and smooth drop-off in the spatial distribution towards the outer disk. The edge of the 10\,kpc ring is visible within the AGB sample indicating less star-formation in the outer disk. There is also a higher density of sources in the regions previously observed by \citet{Boyer2019} due to differences in the selection criteria; this difference is discussed further in \S \ref{sec:agb_criteria_effectiveness} and Appendix A.

In Figure \ref{fig:bright_distribution} we compare the full AGB sample distribution to subsets of different IR brightness. The most luminous AGB candidates in the IR seem to follow the shape of the 10\,kpc ring. This is where star formation has been occurring relatively recently, and where the star formation history (SFH) is consistent with the presence of massive AGB stars and RSGs. The global fraction of RSGs in M31 is expected to be much lower than that of AGB stars. As a result, we generally expect little contamination from RSG stars, but may have some in our most luminous bins. This distribution also indicates that these sources are unlikely to be foreground stars as those would follow a more uniform distribution. Moving to the fainter spatial distributions, we see a gradual increase in the smoothness of the distribution. This seems to be illustrating the migration of stars from their formation sites primarily within the 10\,kpc ring, with the youngest AGB stars still near the 10\,kpc ring, and the oldest AGB stars fully dispersed throughout the disk. We can also look at the spatial distribution of the x-AGB sample to better understand where these stars are producing dust.

The spatial distribution of the x-AGB candidates shows an clear enhancement at 10\,kpc. However, while our x-AGB sample is biased towards brighter IR fluxes, we see an abundance of sources at $r<10$\,kpc (Figure \ref{fig:dusty_distribution}). This indicates that the dust production is not limited to a subset of stellar masses (e.g. massive AGB stars) and points to a relatively smooth injection of AGB dust within the interior of the 10\,kpc ring. The inclination of M31's disk may lead to an increase in the internal interstellar extinction for sources in the lower half of our spatial distribution, or near side of the galaxy. As a result, we may expect more reddened sources in the lower half of Figure \ref{fig:dusty_distribution}. While there may be a hint of this, it remains unclear if the x-AGB distribution is affected by the galaxy's inclination.

\subsubsection{Probing age with AGB stars}
Comparing the spatial distribution of M31's AGB sample with its younger populations allows us to better understand the galaxy's features, SFH, and evolution. We will compare the AGB sample to a subset of M31's RGB stars to further probe the effect of age on the spatial distribution of these samples.

Previously, \citet{Dalcanton2012} found an over-density of RGB stars affiliated with the 10\,kpc ring, suggesting it is a longer-lived feature. Maps of M31's SFH show the 10\,kpc ring is also broader for older populations \citep{Lewis2015}, suggesting a scenario in which stars form within molecular clouds, the natal clouds disperse, and most stars diffuse from loose associations into the surrounding environment \citep{Harris1999,Bastian2009}. This scenario is consistent with the binned spatial distributions of our AGB sample shown in Figure \ref{fig:bright_distribution}. Given these results, and measurements of the SFH, we expect the RGB spatial distribution to be similarly dispersed as our faintest/oldest AGB candidates.

We have defined the RGB population using a F110W--F160W color between 0.5 and 1.2 mag, and reaching one magnitude below the TRGB in either of these filters. We also remove RGB stars in Bricks 1 and 3. While not a complete sample of the RGB stars within each region, we use the samples for a relative age comparison, and avoid completeness issues by restricting the magnitudes of the RGB sample (\emph{N}\,$\sim$\,\rgbs) to well above the completeness limits in PHAT. While this ratio has not been calibrated on an absolute scale, it provides us with a relative age probe.

\begin{figure}[]
    \centering
    \includegraphics[height=3.1cm, trim=4 4 4 4,clip]{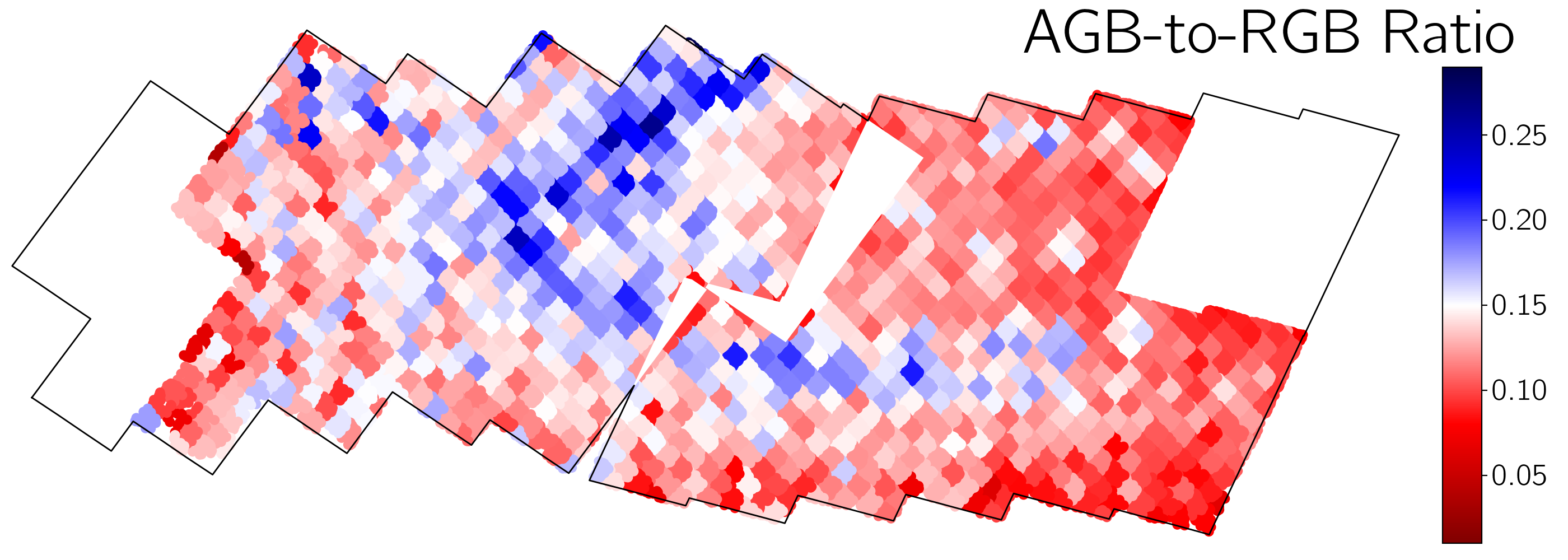} \hspace{0.1cm} \\
    \vspace{0.2cm}\includegraphics[height=3.1cm]{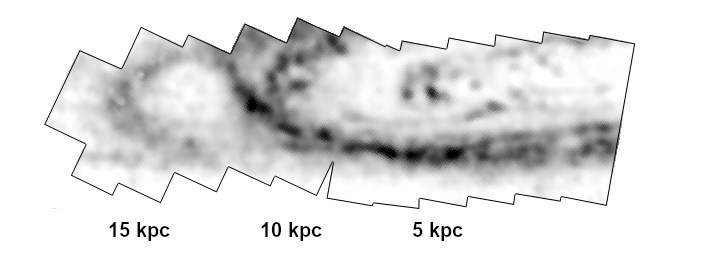}\\

    \caption{Top: A map of the ratio of AGB-to-RGB stars in the PHAT footprint. The red areas represent older regions dominated by RGB stars, where more massive stars have already evolved through the AGB phase. The ratio of AGB-to-RGB stars beyond the 10\,kpc ring is noisier due to low source density. We have excluded Bricks 1 and 3 due to confusion, Bricks 22 and 23 due to lower source counts, and the overlapping Brick regions due to miscounted/duplicated RGB stars. Bottom: An example dust map within the PHAT footprint region created using the dust maps by \citet{Utomo2019}, shown here to illustrate the locations of the star-forming rings. \\}
    \label{fig:agb_rgb_map}
\end{figure}

We have used the ratio of AGB stars to RGB stars in spatial bins as a diagnostic for changes in the ages of populations (Figure \ref{fig:agb_rgb_map}); a similar analysis has been done in smaller regions in M31 \citep{Hamren2015,Boyer2019}. We have mapped the AGB-to-RGB ratio by spatially-binning the populations in 50 bins within the PHAT footprint. Comparing this map with the M31 dust maps by \citet{Utomo2019}, we see that the AGB enhancement traces the dustiest regions within the 10\,kpc ring. In this view we also see the magnitude of the full AGB sample's enhancement at 10\,kpc in relation to the fall off in density of the RGB sample toward larger radii.

\begin{figure}
    \centering
    \includegraphics[width=\linewidth]{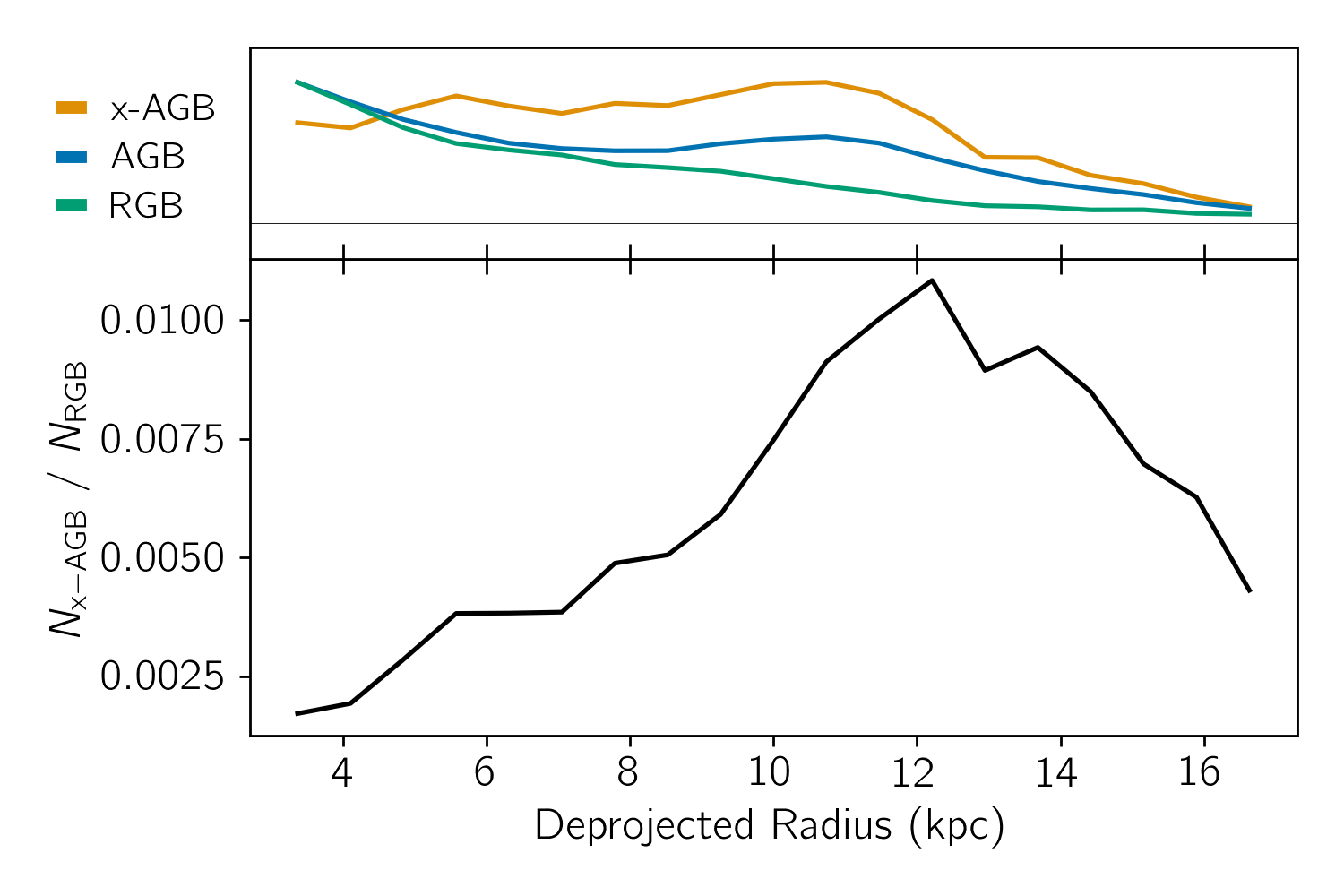}
    \caption{The ratio of x-AGB stars to RGB stars shown in Figure \ref{fig:agb_rgb_map} with respect to deprojected radius. The normalized RGB, AGB, and x-AGB sample distributions are shown above in color. The x-AGB fraction peaks around 12\,kpc, near the location of recent star formation within the 10\,kpc ring. \\}
    \label{fig:agb_rgb_ratio}
\end{figure}

\begin{figure*}
    \centering
    \includegraphics[width=0.49\linewidth]{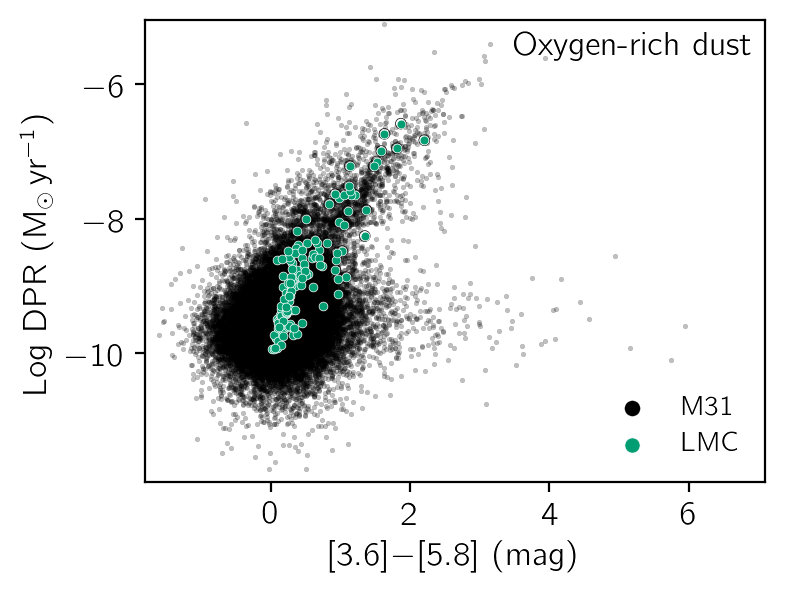}
    \includegraphics[width=0.49\linewidth]{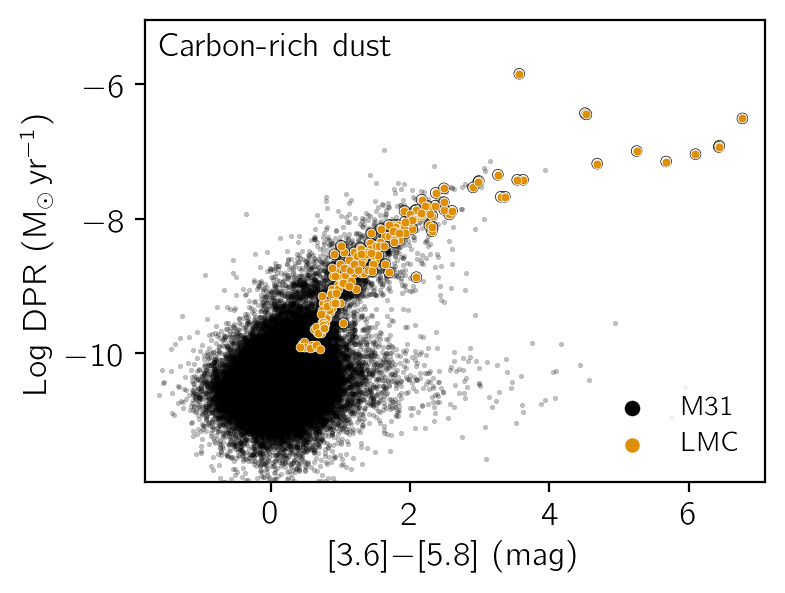}
    \caption{Preliminary DPRs for the M31 AGB candidates assuming oxygen- (left) and carbon-rich chemistry (right). The DPRs are estimated by fitting a quadratic to the [3.6]--[4.5] color vs. Log\,DPR data of the dustiest AGB stars from the SAGE-LMC and SAGE-SMC samples \citep{Groenewegen2018}, and applying the relation to our sample. We apply the color--mass-loss relations to all M31 AGB candidates with IRAC 3.6 and 4.5\,$\mu$m data for the oxygen-rich and carbon-rich case.\\ }
    \label{fig:mdot_color_relation}
\end{figure*}

We have also looked at the radial distribution of the x-AGB sample to better understand where dust is being produced. Figure \ref{fig:agb_rgb_ratio} shows the ratio of x-AGB to RGB stars with respect to deprojected radius. The distribution of this ratio is similar to that of the full AGB sample and RGB sample, with a clear enhancement centered around 12\,kpc. While we see a consistent fraction of x-AGB candidates in the interior of the ring, the peak at slightly larger radii may suggest a small general outward migration from the 10\,kpc ring.

In order to use stellar samples to probe M31, we have to consider the completeness issues associated with the RGB, AGB, and x-AGB samples and the selection criteria used. We expect that the \spitz\ data, and likewise the x-AGB sample, is more severely crowding-limited near the galaxy center. We also expect that as the RGB sample was selected using \hst\ photometry, and the x-AGB sample was selected using \spitz\ photometry, we may be missing some of the RGB sample obscured by dust within M31's 10\,kpc ring. The fact that we see a lower x-AGB fraction at larger radii where both issues of completeness are mitigated, however, may suggest a real x-AGB over-density associated with the exterior side of the 10\,kpc ring. \\

\section{AGB dust production}\label{sec:agb_dust_production}

\subsection{Color--DPR relations}\label{sec:color-dpr_relation}

To determine the level of dust injection by the M31 AGB sample we fit and apply color--dust-production-rate relations using data from the more-complete SAGE sample. We have used the dustiest (log DPR $>-10$ M$_{\odot}$\,yr$^{-1}$) carbon- and oxygen-rich LMC AGB stars from \citet{Groenewegen2018} and fitted these samples separately. We fitted their [3.6]--[4.5] colors and DPRs\footnote{\citet{Groenewegen2018} present total mass-loss rates calculated by measuring DPRs from SED-fitting and scaling them by a gas-to-dust ratio of 200; we use the un-scaled DPRs here.} with quadratic functions of the form $ax^2+bx+c$; we show the best fit results for the oxygen-rich sample ($a=-0.310$, $b=2.634 $, $c=-9.335$) and carbon-rich sample ($a=-0.407$, $b=2.396$, $c=-10.256$) in Figure \ref{fig:mdot_color_relation}. Using these relations and assuming that all stars belong to a single spectral type (carbon- or oxygen-rich), we estimate the total AGB dust injection for all of the AGB candidates in the PHAT footprint with data in the [3.6] and [4.5] filters and not in Bricks 1 or 3; we measure the total DPR in the PHAT footprint as 1.19$ \times 10^{-5}$ M$_{\odot}$ yr$^{-1}$ and 2.13$ \times 10^{-4}$ M$_{\odot}$ yr$^{-1}$, for the carbon- and oxygen-rich relations, respectively.

We expect the M31 AGB population to be producing both carbon- and oxygen-rich dust. The fraction of either type, however, is unclear. We can estimate the chemical composition of the AGB dust being produced in M31 by using our subset of medium-band data where we have chemically-classified x-AGB candidates. While our chemically-classified x-AGB sample ($N=329$) is 86\% oxygen-rich, applying our color relations to the corresponding chemically-classified x-AGB samples yields 97.8\% of the dust as oxygen-rich. This is in stark contrast to the LMC where the x-AGB sample is composed of 3\% oxygen-rich stars, that produce 13\% of the x-AGB dust \citep{Riebel2012}.

Compared to the MCs \citep{Riebel2012, Srinivasan2016}, we expect more oxygen-rich dust production from M31 AGB stars, as a result of their more metal-rich environment.
With M31's more-narrow carbon-star mass range, we also expect more dusty oxygen-rich AGB stars of moderate mass. These are likely contributing to the increase in the fraction of oxygen-rich dust. As a side-note, this also has the effect of a lower average mass (and mass-loss rate) for oxygen-rich x-AGB stars in M31. To truly understand the oxygen-rich fraction of the dust, however, we need to have the sensitivity to detect any highly obscured and dusty carbon star below the detection threshold of our IR observations (see \S \ref{sec:x-AGB_stars}).

\subsection{Dust-Injection Rate}\label{sec:dust-injection_rate}

The total dust mass in M31 has been measured by modeling both its extinction and emission (see Table \ref{table:dust_budgets}). More recently, \citet{Draine2014} estimated the total dust in M31 using \spitz/IRAC data to be 5.4$ \times 10^{7}$\,M$_{\odot}$. \citet{Dalcanton2015}, however, found that the dust models used in this work over-predict the extinction by a factor of $\sim$\,2.5, which suggests that the true dust mass is slightly lower. We can compare these measurements to DPRs of the AGB population to assess the impact of the AGB stars on the dust budget of M31.


\begin{deluxetable}{lcr}
\tablewidth{\columnwidth}
\tabletypesize{\small}
\tablecolumns{3}
\tablecaption{Previous measurements of M31's dust mass, as well as AGB dust injection fractions calculated for the MCs and now M31.  \label{table:dust_budgets}}

\tablehead{
\colhead{M31 Dust Mass (10$^{7}$\,M$_{\odot}$)}
& &
}

\startdata
\citet{Haas1998} & & 3.8 \\
    \citet{Schmidtobreick2000} & & 1.3 \\
    \citet{Montalto2009} & & 7.6 \\
    \citet{Draine2014} & & 5.4 \\
    \citet{Dalcanton2015} & & $5.4 \div 2.5$ (2.2\rlap{)}\\
    \hline
    \\
    AGB \rlap{Dust Injection Fraction} & &\\
    \hline
    \citet{Matsuura2009} &LMC& \llap{$\sim$}1.6\,\% \\
    \citet{Boyer2012} &SMC& $\sim$2.1\,\% \\
    \citet{Gordon2014} &LMC/SMC& $\times 10$ ($\sim$20\,\%\rlap{)} \\
    \citet{Srinivasan2016} &SMC& $\times 2$ ($\sim$40\,\%\rlap{)}\\
    \citet{Nanni2018} &SMC& $\times 5$ ($\sim$100\,\%\rlap{)}\\
    {\it This work} &M31&  0.9\,--\,35.5\,\%\\
    \hline\\
\enddata
\tablenotetext{}{{\bf Note.} Dust injection fractions are highly uncertain. This uncertainty stems primarily from the poorly-constrained dust grain lifetimes. These estimates also make assumptions for expansion velocities, drift velocities, gas-to-dust ratios, geometry, and optical constants. Additionally, in smaller dwarf galaxies like the SMC, only a handful of stars can dominate the dust injection, with the possibility of incomplete sampling resulting in dramatically different results.}
\end{deluxetable}

Previous analyses of the SEDs of evolved stars in the MCs have estimated the fraction of dust created by AGB stars (Table \ref{table:dust_budgets}). These methods, however, differ, from color--mass-loss relations \citep{Matsuura2009}, to infrared excess--mass-loss relations \citep{Boyer2012}, and DPRs calculated using SED-fitting \citep{Riebel2012, Srinivasan2016}. Additionally there are differences in the assumed properties of the circumstellar envelope and the life-cycle of dust.

We can use our AGB candidates to estimate the impact of their dust injection in M31. We are limited to around one third of M31's disk and may be biased against the dustiest carbon stars. That being said, we have the sensitivity to detect the bulk of the oxygen-rich dust producers, expected to dominate the AGB dust injection. Scaling our PHAT DPR by the relative size of M31's full disk ($\sim$\,3) we get a global DPR = 6.39\,$\times$\,$10^{-4}$\,M$_{\odot}$ yr$^{-1}$. To estimate the fraction of M31's ISM dust produced by AGB stars we must consider the lifetimes of dust grains within the ISM.

\subsection{Dust Budget}\label{sec:dust_budget}

\citet{Jones2011} have argued that current estimates of global lifetimes of dust grains in the ISM are likely too uncertain to provide useful constraints. We have, however, provided estimates for the fraction of dust with a circumstellar origin using several scenarios for dust lifetimes. We assume a DPR that has been constant since the formation of the oldest stars within M31 \citep[12\,Gyr;][]{Williams2017}. We also assume a current dust budget of 2.16$ \times 10^{7}$\,M$_{\odot}$, 2.5 times lower than the value measured by \citet{Draine2014}, in accordance with the findings of \citet{Dalcanton2012}.

\begin{enumerate}
    \item Assuming no dust destruction, the AGB population is capable of producing 35.5\% (7.67\,$\times$\,$10^{6}$\,M$_{\odot}$) of the measured dust in M31.
    \item Assuming the much shorter dust grain lifetimes estimated for the Milky Way of 300\,Myr \citep{Draine1979,Jones1994} would indicate that 0.9\% (1.92,$\times$\,$10^{5}$\,M$_{\odot}$) of M31's dust is unprocessed AGB dust.
    \end{enumerate}

Oxygen-rich silicates are expected to live longer in the ISM than carbon-rich dust species (e.g. graphite). Grain-grain collisions are also expected to shatter larger dust grains creating smaller fragments, that could be preserved as cores capable of re-growing mantles in the ISM. The lifetimes of the cores, however, are only expected to be 3--4 times longer than the unfragmented grains \citep{Jones1994}. For our second dust lifetime scenario, this extends the upper limit to $\sim 3\%$ for any AGB dust, or the remnants of AGB dust acting as a core of the dust grains.

Pre-solar dust grains have shown a non-negligible fraction of dust with circumstellar signatures from multiple stars \citep{Gail2009}. Whether M31 dust grains are capable of retaining any characteristics of their circumstelllar origin remains unclear. In order to make more realistic estimates of the dust budget, we need a more  sensitive survey of the galaxy in the mid-IR. More work on the dust-destruction rate is also needed to constrain the life-cycle of dust and estimate the fraction of dust from AGB stars.

\section{Conclusions}

We have completed the first comprehensive census of the AGB population in a metal-rich galaxy, and used the results to estimate the contribution of AGB stars to the dust budget of M31. \smallskip

\begin{itemize}
    \item Using near-IR data from the PHAT survey, we have identified \agb\ AGB candidates across one third of M31's disk. These results were then combined with mid-IR data from \spitz\ to isolate \dustyagb\ AGB stars expected to be producing the bulk of the AGB dust in M31. Using clusters identified in M31, we identified \clusteragbs\ AGB candidates within clusters, some of which have measured cluster ages and metallicities.

    \item We match our data to AGB stars previously chemically-classified (carbon or oxygen), and compare them to the more complete samples in the MCs. The M31 AGB sample is shown to be dominated by oxygen-rich AGB stars. In the small footprints where we have chemical classifications, we find that 97.8\% of the dust being produced by the largest dust producers is oxygen-rich.

    \item Increasingly luminous and dusty AGB stars are found more frequently associated with the 10\,kpc ring, a site of recent star formation. For the older AGB candidates faint in the IR, we see a uniform distribution across the PHAT footprint, consistent with a scenario of increased mixing with age.

    \item We have used a color--dust-production-rate relation based on the oxygen-rich x-AGB sample from the LMC to estimate the M31 AGB dust injection. Using different scenarios for the dust lifetimes, we estimate that AGB stars account for 0.9--35.5\% of M31's global dust budget. More constraints on dust grain lifetimes are needed to provide more realistic estimates of the dust budget.
\end{itemize}

\acknowledgements{We thank the anonymous referee for their helpful and constructive comments. We would like to thank Cliff Johnson for the helpful discussions about the cluster sources. We would like to thank Grace Telford for the help with calculating integrated fluxes (that ultimately, were not used). This work was supported by NASA via Astrophysics Data Analysis award 80NSSC19K0529. L. Girardi acknowledges support from the ERC Consolidator Grant funding scheme (project STARKEY, G.A. n. 615604). {S. Srinivasan acknowledges support from UNAM-PAPIIT Programme IA104820.} This research is based on archival data obtained with the NASA/ESA Hubble Space Telescope obtained from the Space Telescope Science Institute, which is operated by the Association of Universities for Research in Astronomy, Inc., under NASA contract NAS 5–26555. This work is based in part on archival data obtained with the Spitzer Space Telescope, which was operated by the Jet Propulsion Laboratory, California Institute of Technology under a contract with NASA. Some of the data presented in this paper were obtained from the Mikulski Archive for Space Telescopes (MAST) at the Space Telescope Science Institute. The specific observations analyzed can be accessed via \dataset[10.17909/t9-eahb-c508]{https://doi.org/10.17909/t9-eahb-c508}.}

\facilities{\hst, \spitz}
\software{DESK \citep{Goldman2020}, DOLPHOT \citep{dolphin2000}, Astropy \citep{astropy:2013, astropy:2018}, Matplotlib \citep{matplotlib2007}, Scipy \citep{scipy2020}, Numpy \citep{numpy2011,numpy2020} IPython \citep{ipython2007}.}

\clearpage

\bibliography{references_2020}{}
\bibliographystyle{aasjournal}

\appendix
\section{AGB criteria comparison}
Here, we compare our AGB selection criteria with two recent AGB catalogs in the PHAT footprint. \citet{Boyer2019} surveyed 20 small footprints (Figure \ref{fig:spatial_distribution}), and \citet{Girardi2020} classified AGB stars in PHAT clusters. Here, we classify stars across the entire PHAT footprint using both the \hst\ data and \spitz\ data. Table \ref{table:cut_comparison} summarizes the classification criteria for each of these surveys. Some of photometry are required to meet the sharpness and crowding criteria (GST), outlined in \S2. Our classification criteria recovers \leopercentagerecovered\% of the \citet{Girardi2020} cluster AGB sample, and \boyerpercentagerecovered\% of the \citet{Boyer2019} chemically-classified AGB sample. Of the \citet{Girardi2020} AGB candidates not recovered, most did not meet our F814W--F160W color requirement. This criteria was added to mitigate contamination from bluer stars (supergiants, foreground stars, etc). For the \citet{Boyer2019} AGB candidates that we did not recover, most did not meet our magnitude criteria. The \citet{Boyer2019} classification criteria required photometry above the TRGB in any of the five near-IR bands, as opposed to our two near-IR bands. This resulted in more of that sample falling below the TRGB in F110W and F160W, and being missed in our classification.

\begin{table}[h]
\caption{The cuts used for classifying AGB stars in \citet{Boyer2019}, \citet{Girardi2020}, and this work\label{table:cut_comparison}. \medskip}
\begin{tabular}{|c|ccc|}
\hline
Requirements &
Boyer+19 &
\multicolumn{1}{|c|}{Girardi+20} &
\multicolumn{1}{c|}{\it This Work} \\
\hline
            & F127M $<$ 18.80 mag & \multicolumn{1}{|c|}{}                      &              \\
            & F139M $<$ 18.66 mag & \multicolumn{1}{|c|}{}                      & {F110W $<$ 19.28 mag} \\
Magnitude   & F153M $<$ 18.27 mag & \multicolumn{1}{|c|}{F160W $<$ 18.14 mag}    & {or}   \\
            & F110M $<$ 19.28 mag & \multicolumn{1}{|c|}{}                      &   {F160W $<$ 18.28 mag} \\
            & F160M $<$ 18.28 mag & \multicolumn{1}{|c|}{}                      &    \\
            &  (any)              & \multicolumn{1}{|c|}{}                    &    \\
\hline
            & F127M $-$ F139M $<$ 0.3 mag   & \multicolumn{1}{|c|}{}                                    & {F110W $-$ F160W $>$ 0.88 mag$^{*}$}\\
Color       & and                           & \multicolumn{1}{|c|}{F110W $-$ F160W $>$ 0.88 mag}  & and\\
            & F139M $-$ F153M $<$ 0.3 mag   & \multicolumn{1}{|c|}{}                                    & {F814W $-$ F160W $>$ 2.4 mag$^{*}$}\\
\hline
            &                           & \multicolumn{2}{|c|}{}\\
GST         & All 5 filters    & \multicolumn{2}{|c|}{F110W \& F160W}\\
            &                           & \multicolumn{2}{|c|}{}\\
\hline
                        &                               & \multicolumn{1}{|c|}{}            & {[3.6]$-$[4.5] $>$ 0.5 mag}\\
Additional inclusion    & \multirow{2}{*}{\textemdash}  & \multicolumn{1}{|c|}{\multirow{2}{*}{\textemdash}} & {and} \\
criteria                &                               & \multicolumn{1}{|c|}{}            & {$[4.5] < 16.4$ mag}\\
                        &                               & \multicolumn{1}{|c|}{}            & (or inclusion in Boyer+19)\\
\hline
                        &                               & \multicolumn{1}{|c|}{}            & {[3.6]$-$[4.5] $>$ 1.9 mag}\\
Galaxy Removal          & \textemdash                   & \multicolumn{1}{|c|}{\textemdash} & {and} \\
                        &                               & \multicolumn{1}{|c|}{}            & {$[4.5] < 13.4$ mag}\\
\hline
                        & F127M $-$ F153M $<$ 0 mag     & \multicolumn{1}{|c|}{}            & {[3.6]$-$[4.5] $>$ 0.25 mag}\\
Dusty Classification    & and                           & \multicolumn{1}{|c|}{\textemdash} & {and} \\
                        & 18 mag $<$ F153M $<$ 20 mag   & \multicolumn{1}{|c|}{}            & {$[4.5] < 16.4$ mag}\\
                        \hline
\end{tabular}
\begin{itemize}[label={}]
 \small
 \medskip
 \item $^*$If data in both filters.
 \item {\bf Note.} The GST criteria for \hst\ and \spitz\ is further explained in \citet{Williams2014} and \S\ref{sec:spitzer}, respectively. The IRAC photometric catalog only includes sources that meet the GST criteria. For sources classified only in the IR, the $[3.6]$ and $[4.5]$ filters would also be included in the GST requirements. We additionally removed any sources deemed duplicates or image artifacts.
 \end{itemize}
\end{table}

\newpage

\section{Cluster AGB candidates with additional classifications}
We have identified \dustyclusteragb\ x-AGB candidates that reside within the expected radius of clusters with measured properties \citep{Johnson2015}. These clusters give us pockets of sources with uniform metallicities and ages, quantities that are difficult to measure in individual stars. Studying these sources will allow us to isolate the effects of initial mass and metallicity on the dust production and evolution of AGB stars, critical for constraining stellar and chemical evolutionary models. Some of our x-AGB cluster sources (7/17; Table \ref{table:cluster_ages}) have cluster ages likely too young to host AGB stars (Log Age\,[yr] $<$ 8.0). This likely indicates that they are red supergiants, or field AGB candidates not associated with the cluster.

Figure \ref{fig:d_cluster} shows \hst\ F160W image cutouts of the M31 clusters where we have identified x-AGB candidates. Also shown are the SEDs of these x-AGB candidates fit with radiative transfer models. The dimensions of the cutout images are four times the apparent cluster radius (e.g. AP\,552: 14.2\arcsec $\times$ 14.2\arcsec). We fit the SEDs with the Dusty Evolved Star Kit \citep[{\asciifamily DESK};][]{Goldman2020} using a grid of oxygen-rich radiative transfer models assuming dust grains from \citet{Ossenkopf1992} and a distance of 776\,kpc. The \hst\ and \spitz\ data were taken at different times, and assumed variability of these sources is likely to affect the observed shape of the SED. Crowding in the well-populated clusters may also affect the IR fluxes. For all but two of the sources (shown in Figure \ref{fig:dusty_cluster_mismatch}), the models fit the data well. The well-fit sources have luminosities generally $\sim$\,10,000 L$_{\odot}$ and gas mass loss rates $\sim$\,7 $\times 10^{-6}$ M$_{\odot}$yr$^{-1}$. \\

\begin{figure}
    \centering
    \includegraphics[height=\dclusterfigsize]{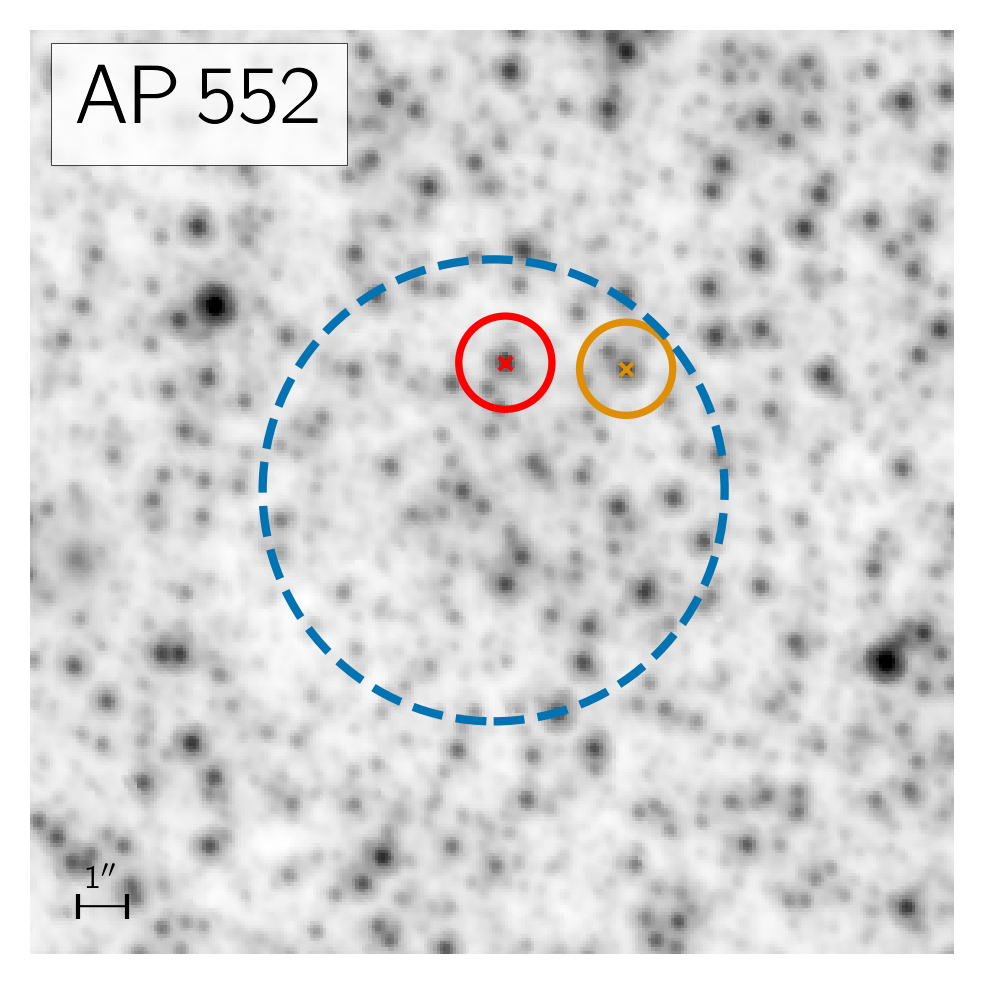}
    \includegraphics[height=\dclusterfigsize]{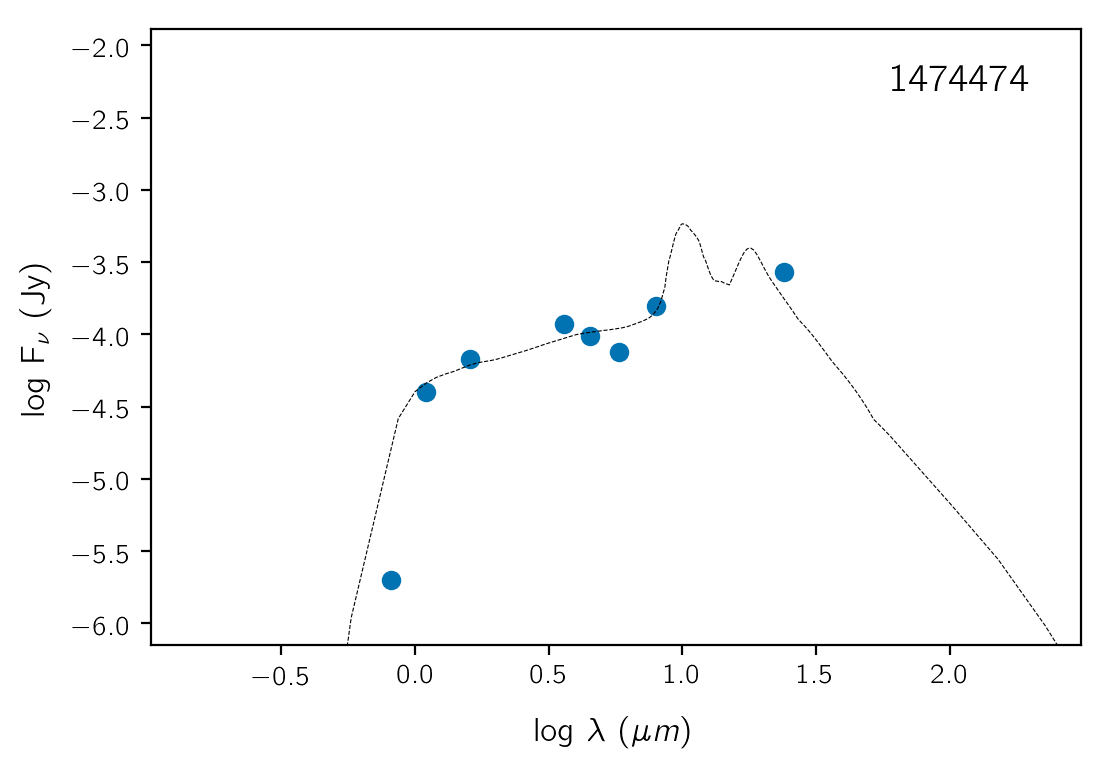} \hspace{\dclustergap}
    \includegraphics[height=\dclusterfigsize]{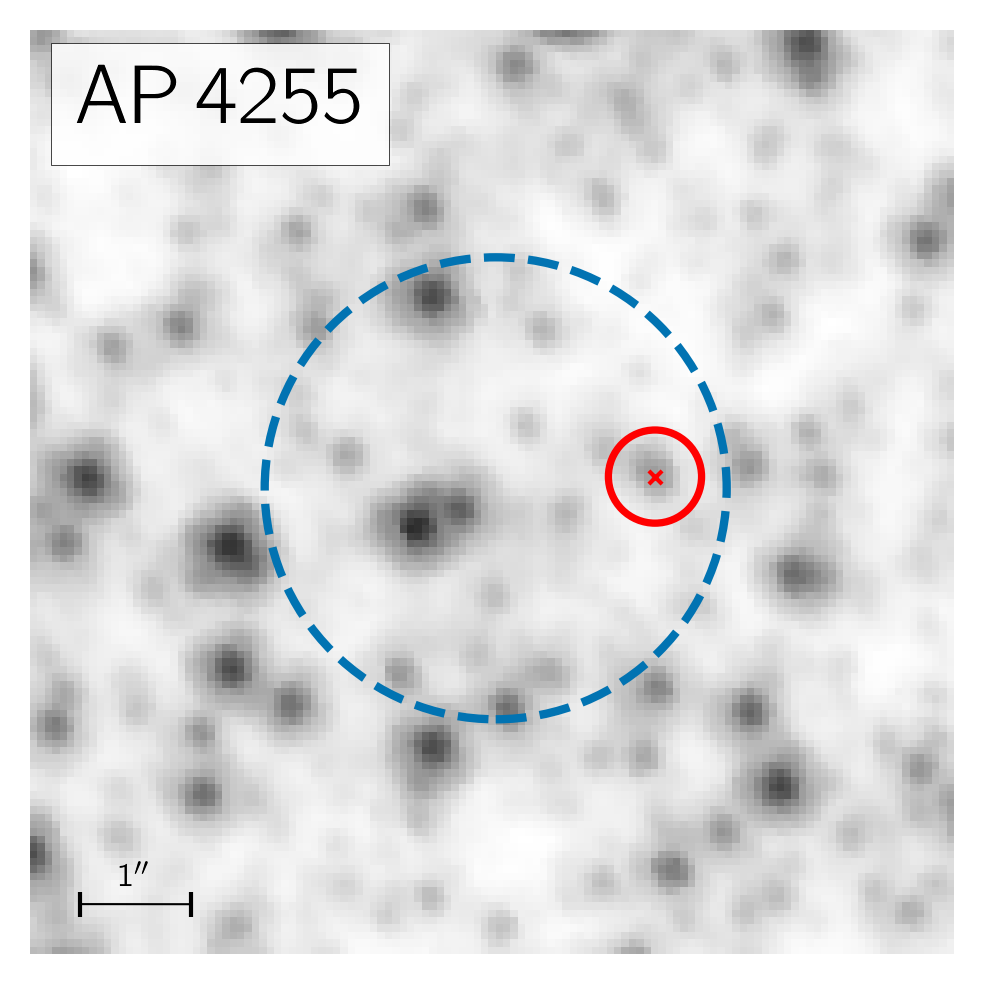}
    \includegraphics[height=\dclusterfigsize]{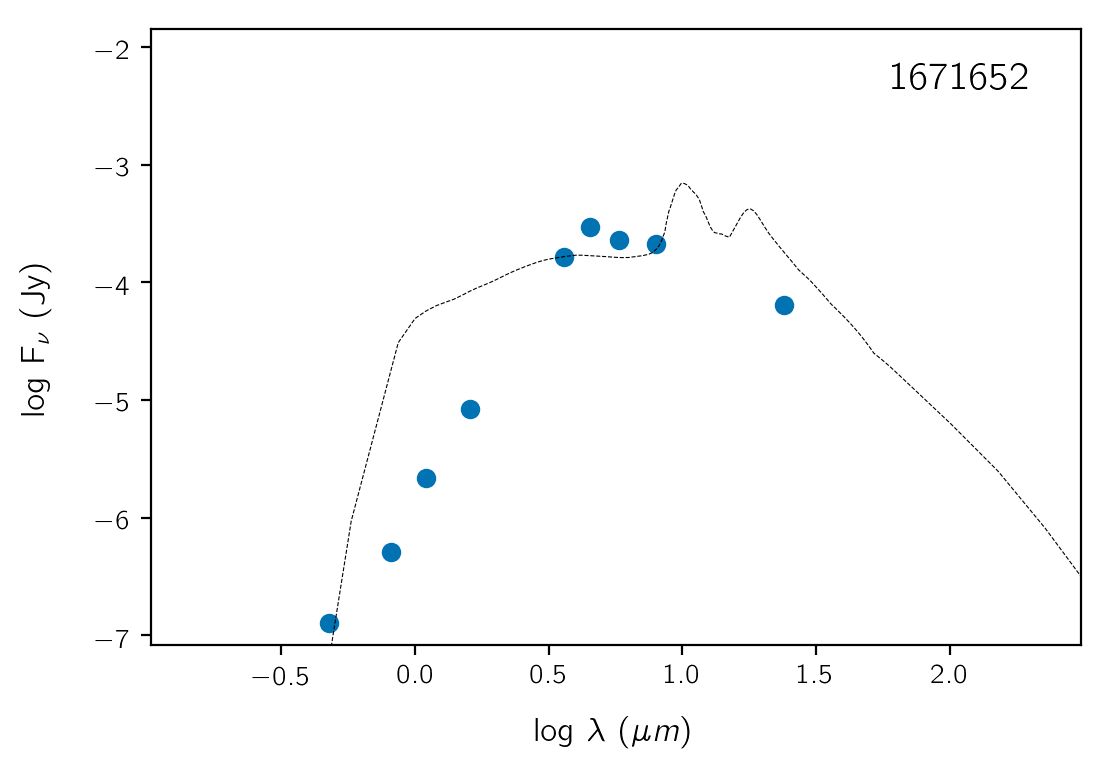}\\
    \includegraphics[height=\dclusterfigsize]{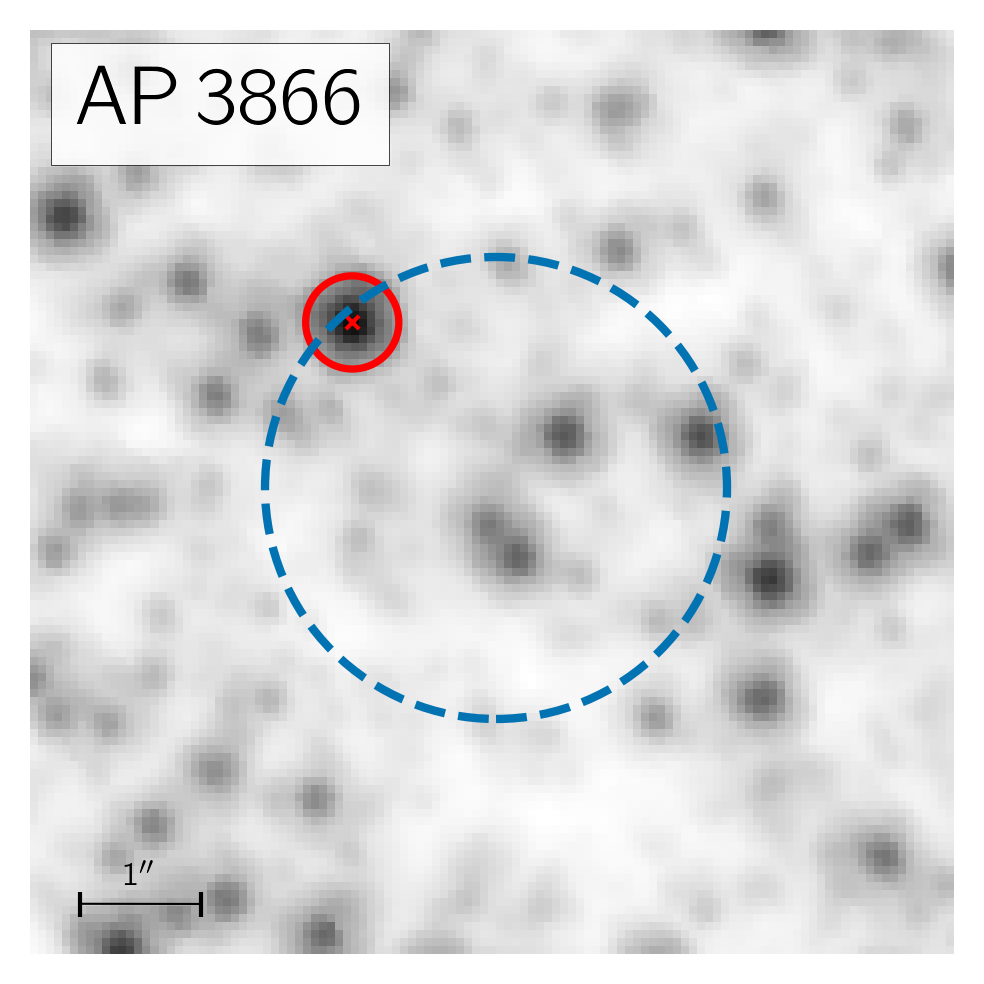}
    \includegraphics[height=\dclusterfigsize]{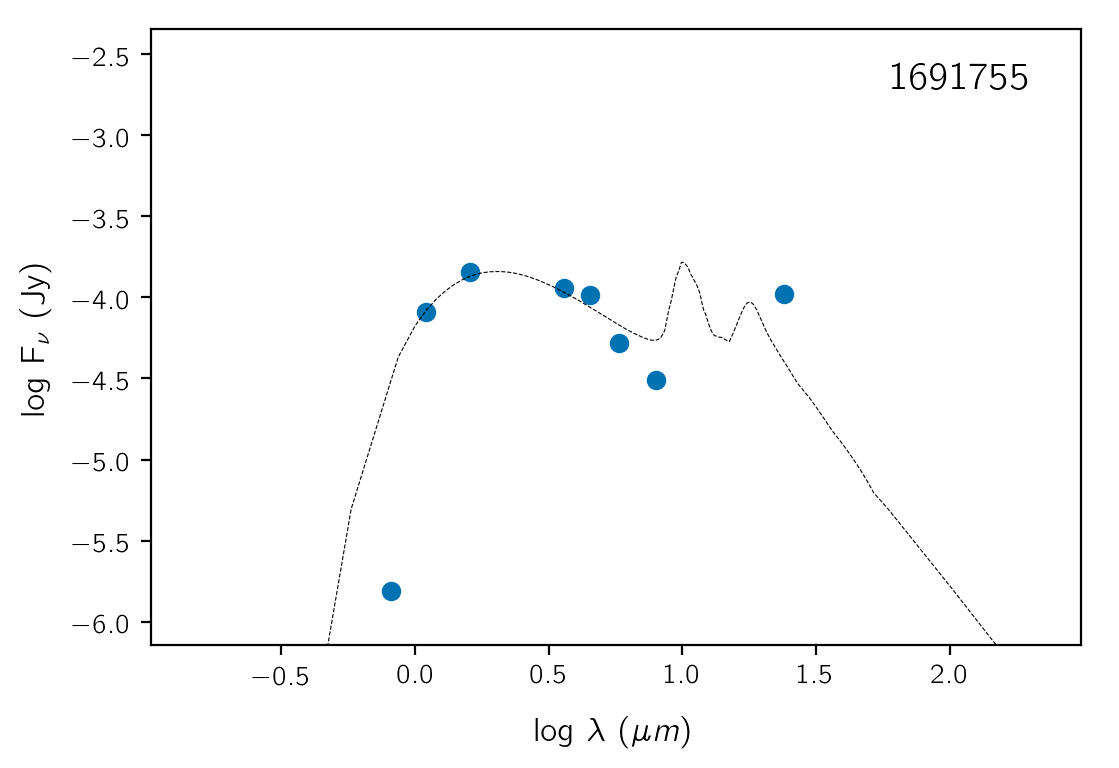} \hspace{\dclustergap}
    \includegraphics[height=\dclusterfigsize]{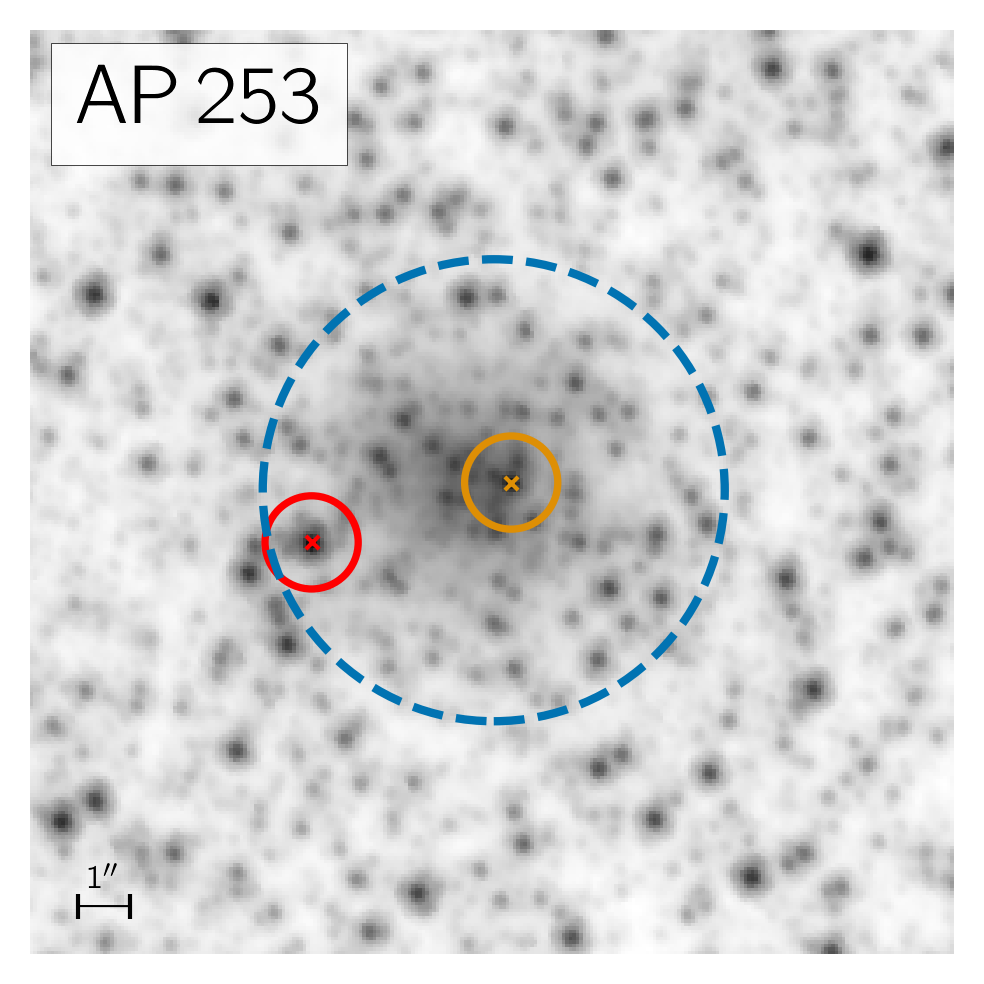}
    \includegraphics[height=\dclusterfigsize]{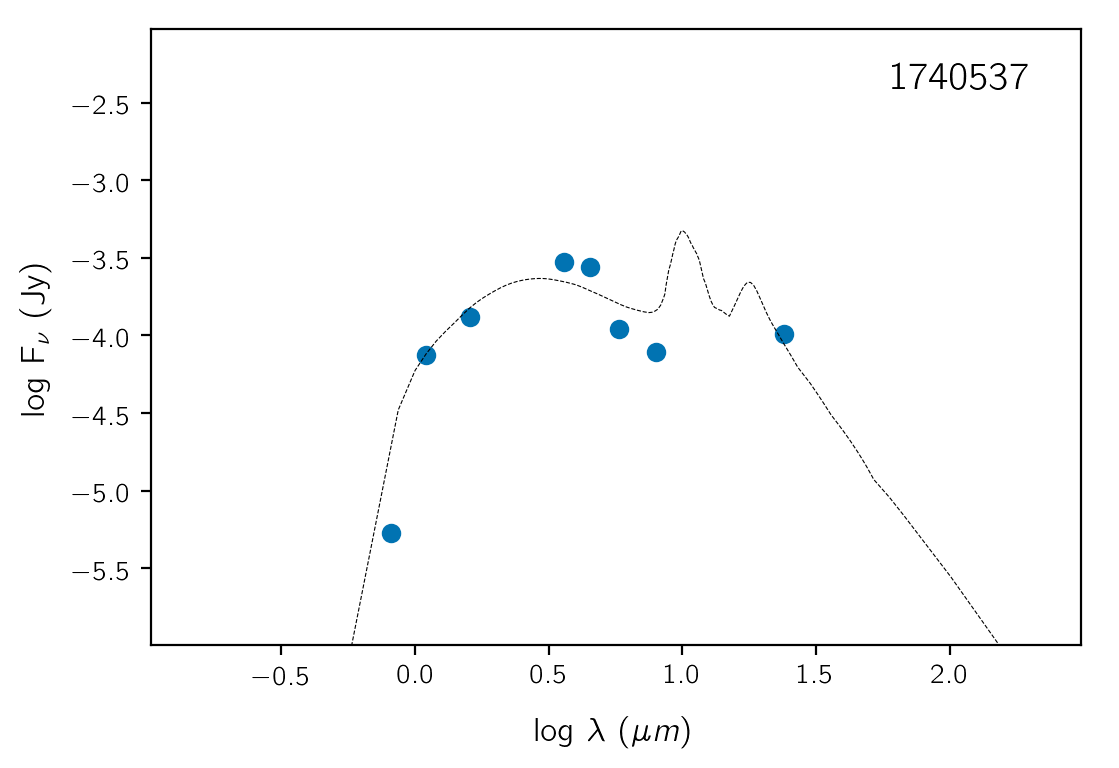}\\
    \includegraphics[height=\dclusterfigsize]{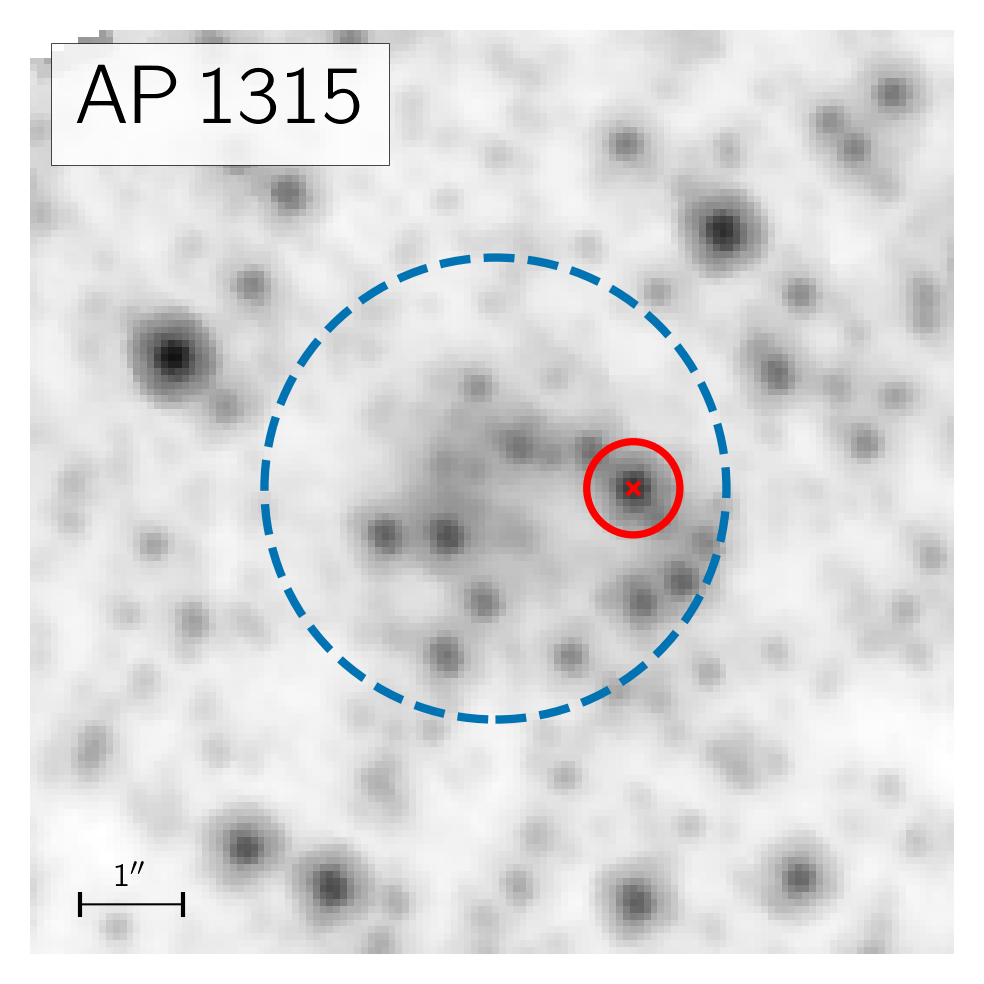}
    \includegraphics[height=\dclusterfigsize]{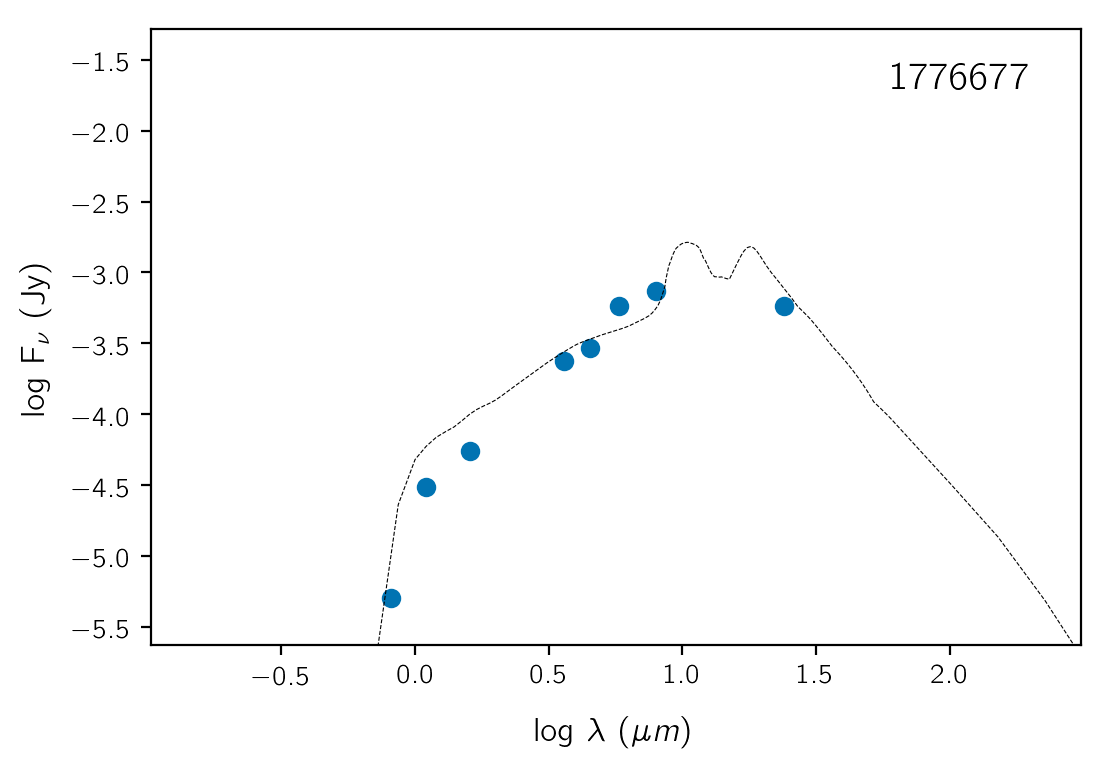} \hspace{\dclustergap}
    \includegraphics[height=\dclusterfigsize]{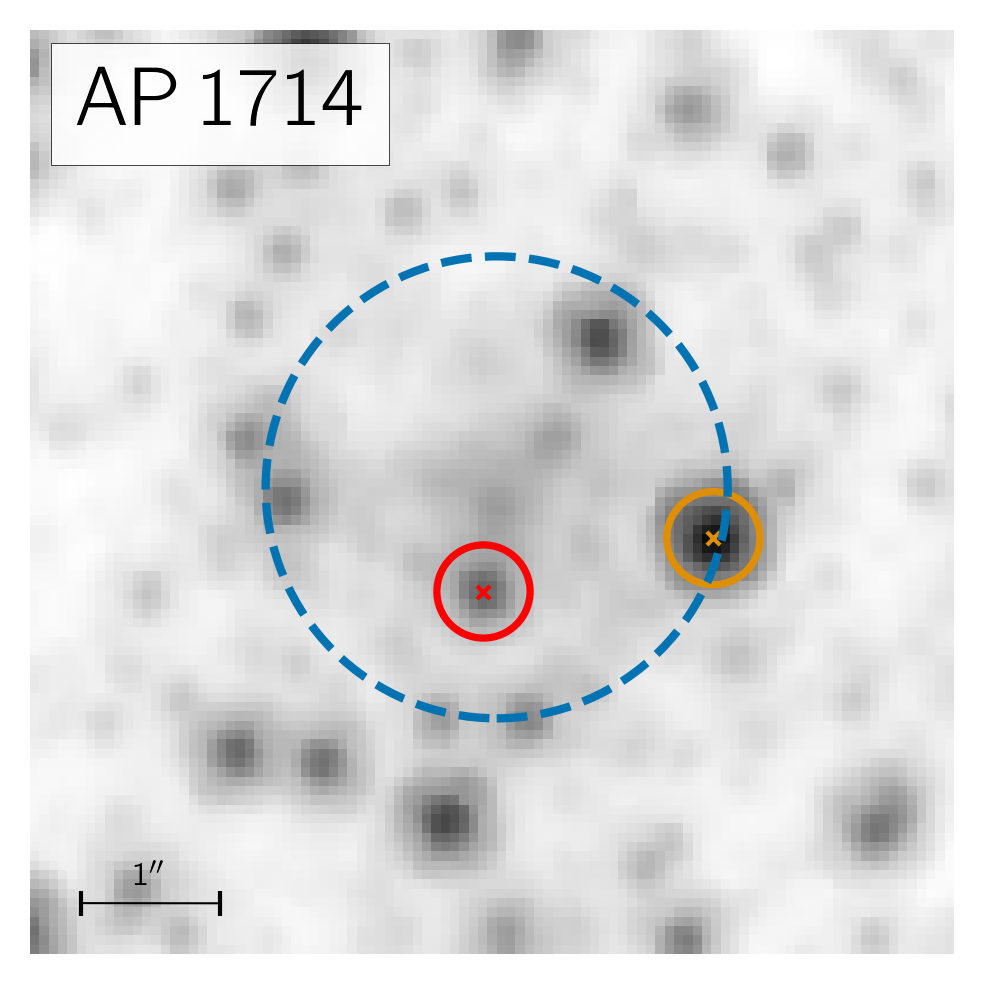}
    \includegraphics[height=\dclusterfigsize]{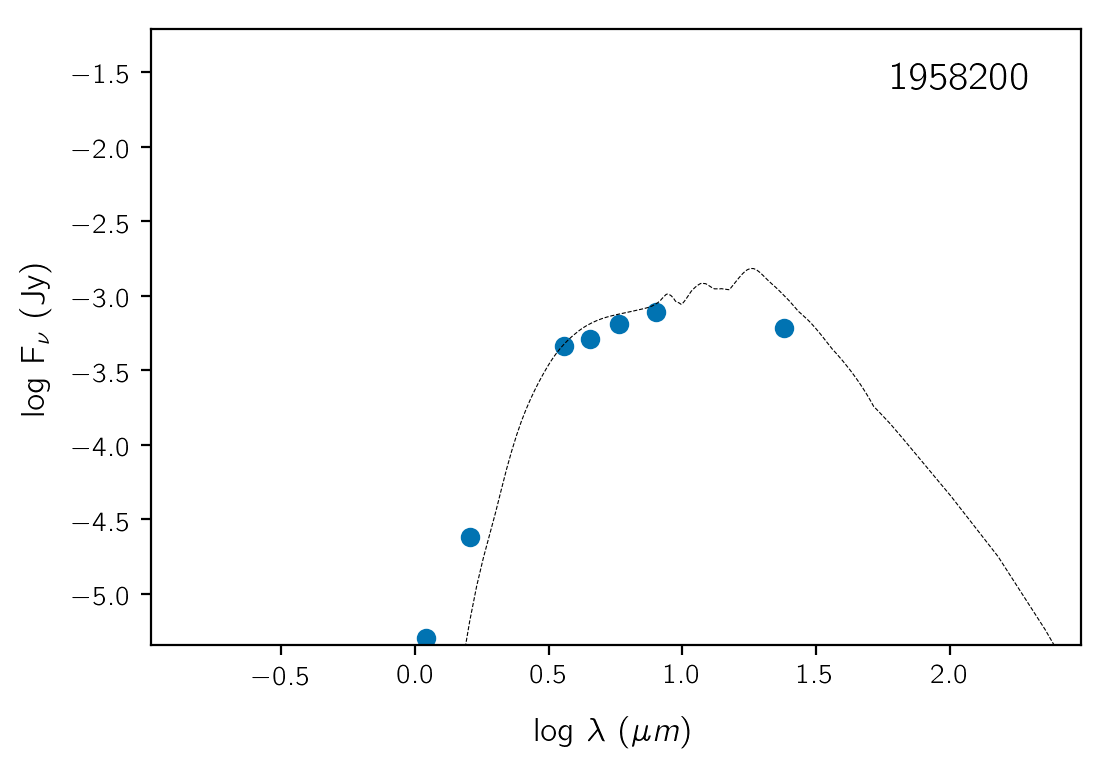}\\
    \includegraphics[height=\dclusterfigsize]{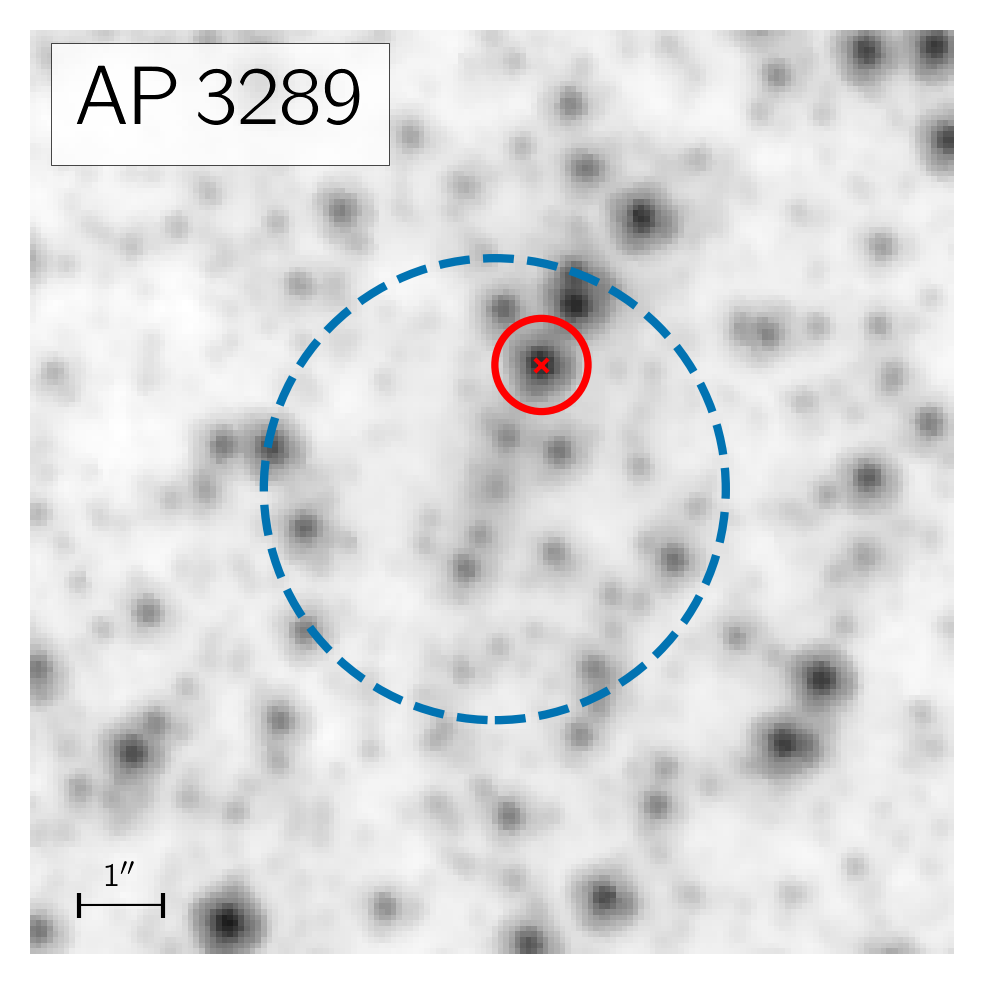}
    \includegraphics[height=\dclusterfigsize]{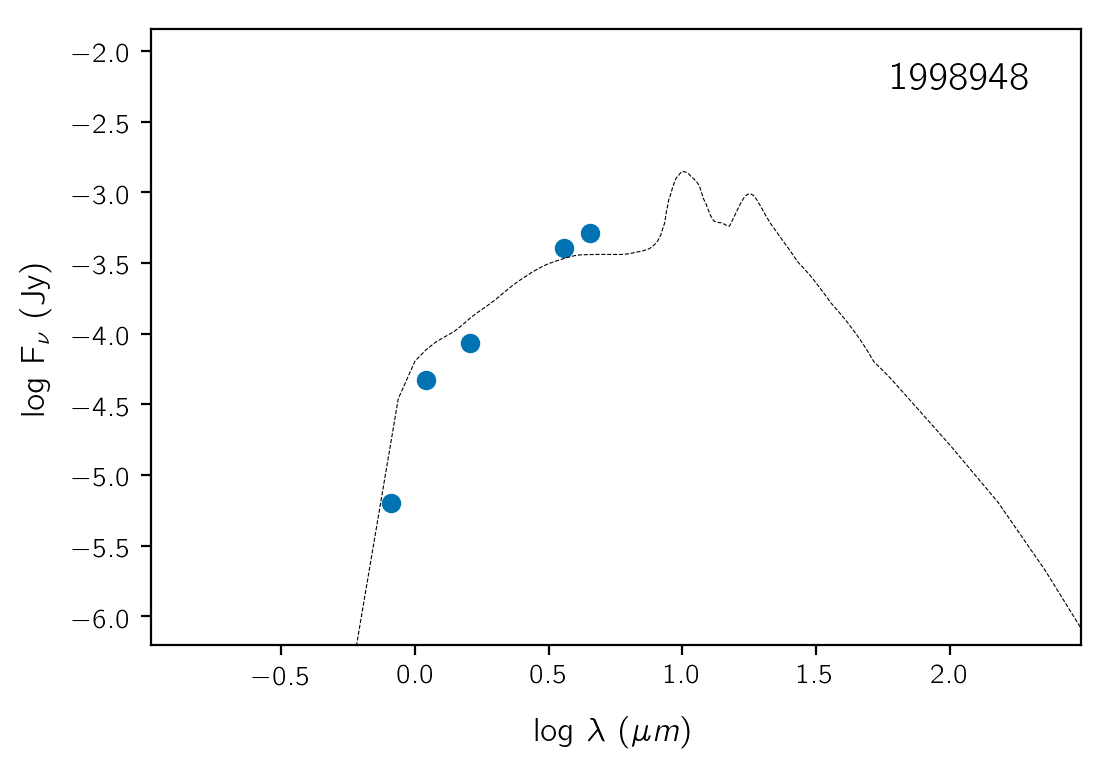} \hspace{\dclustergap}
    \includegraphics[height=\dclusterfigsize]{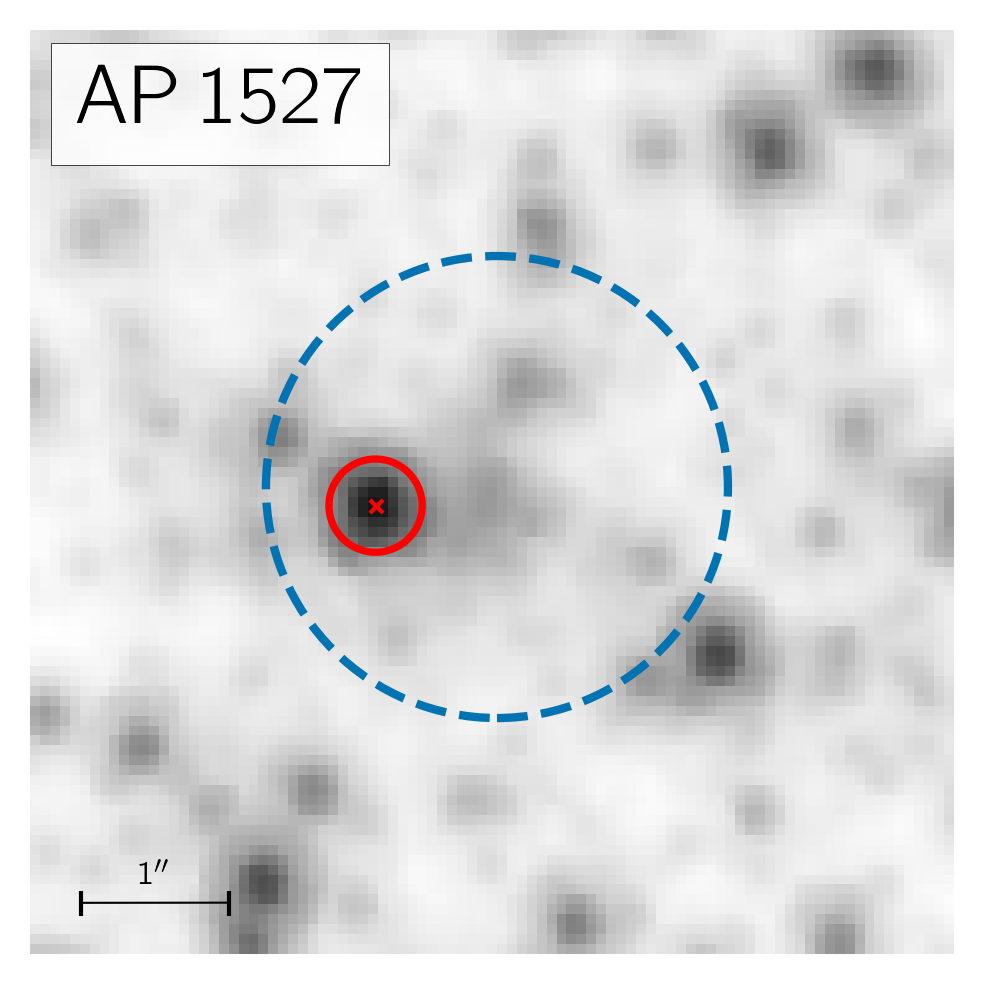}
    \includegraphics[height=\dclusterfigsize]{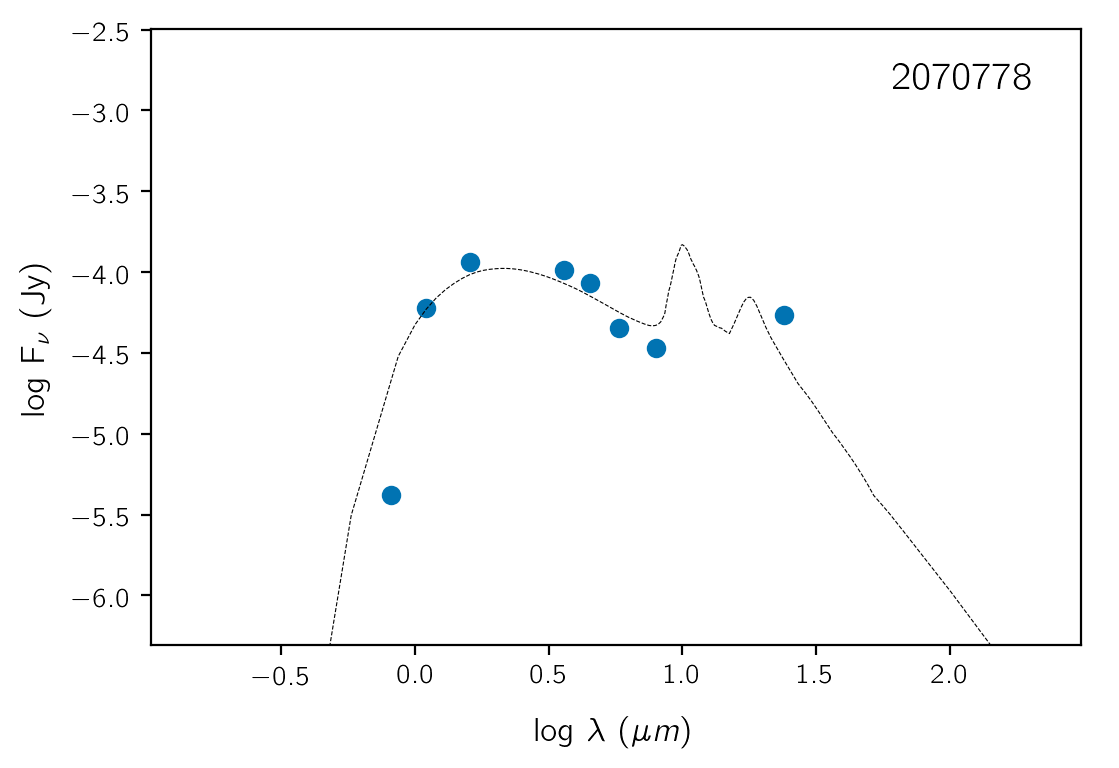}\\
    \includegraphics[height=\dclusterfigsize]{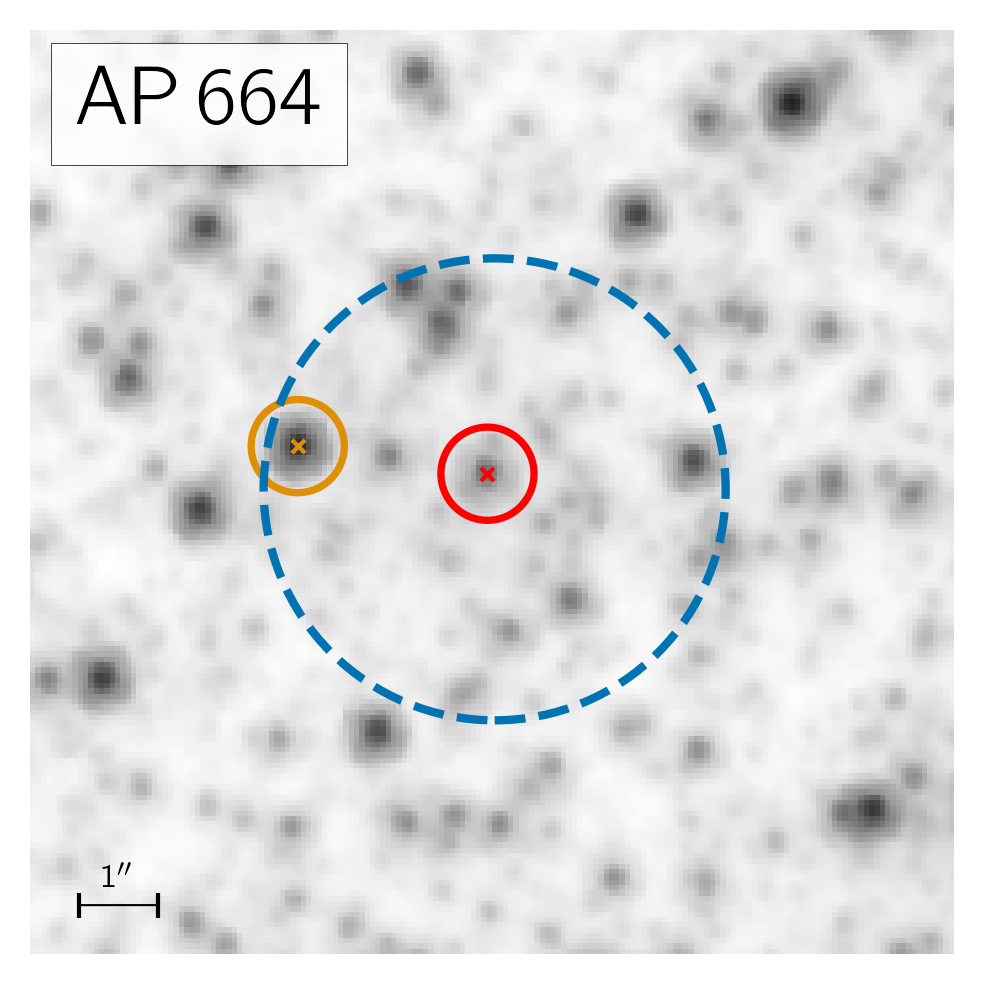}
    \includegraphics[height=\dclusterfigsize]{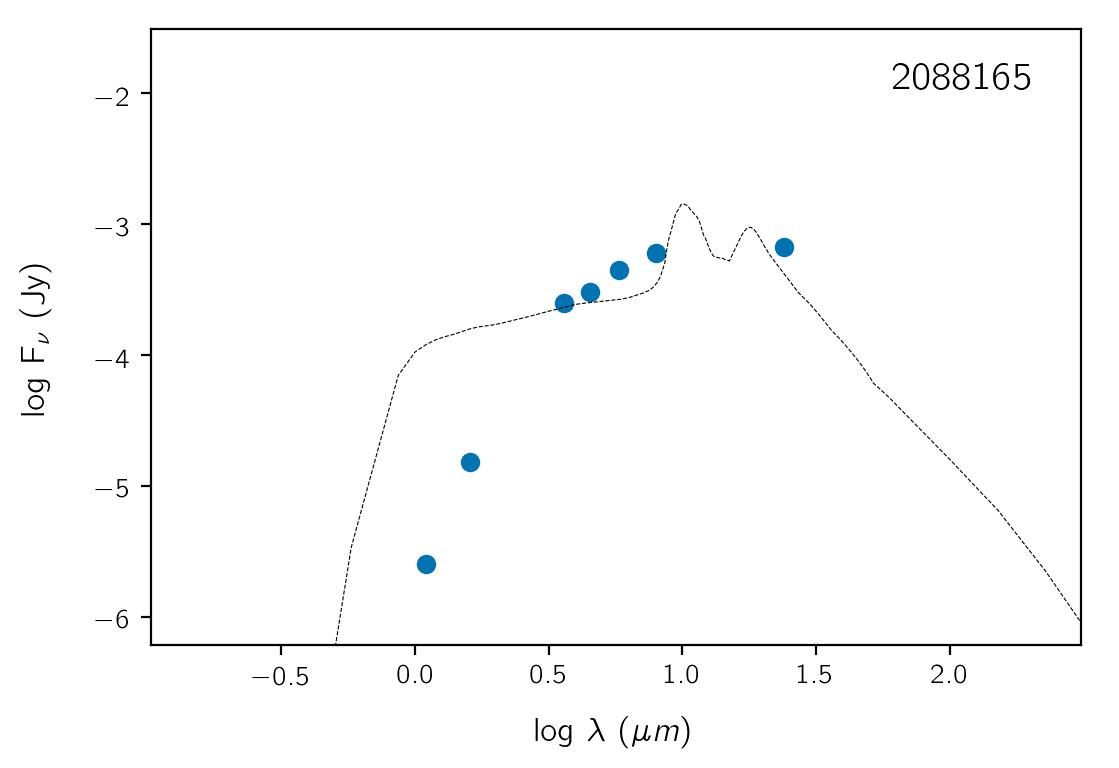} \hspace{\dclustergap}
    \includegraphics[height=\dclusterfigsize]{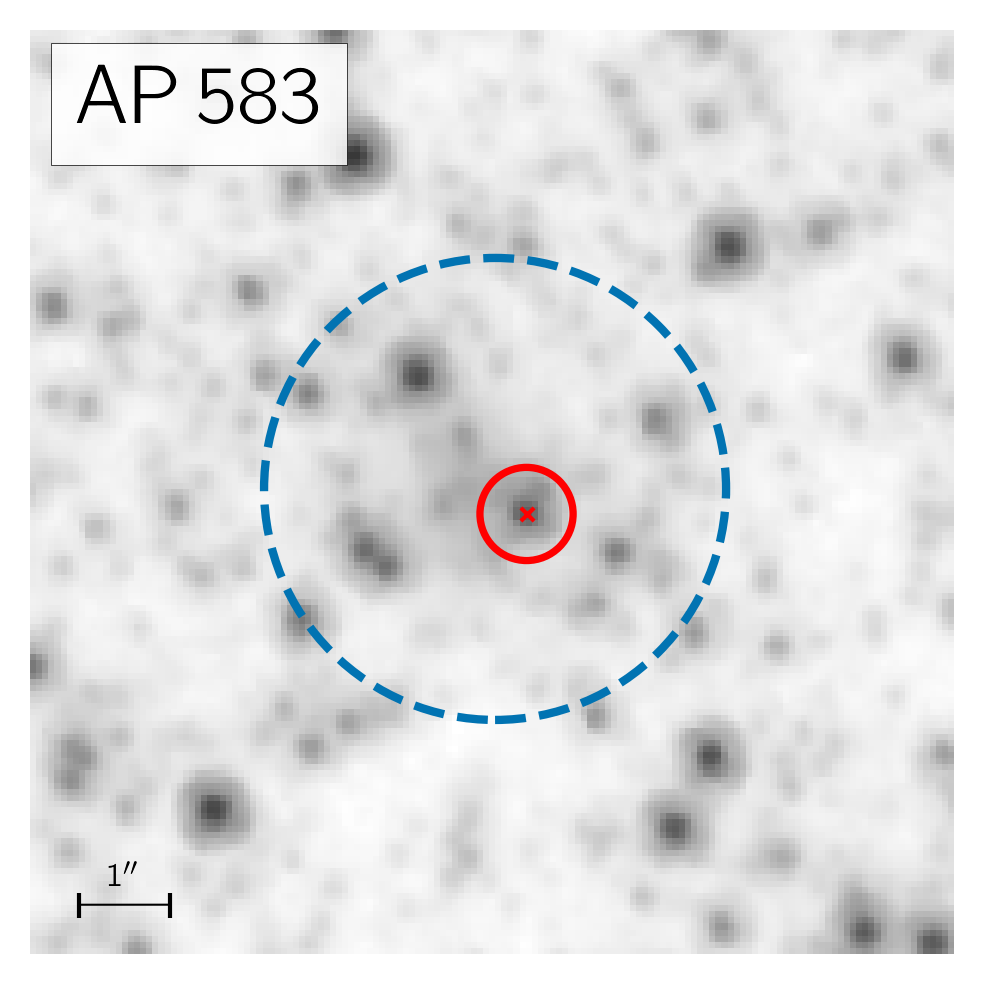}
    \includegraphics[height=\dclusterfigsize]{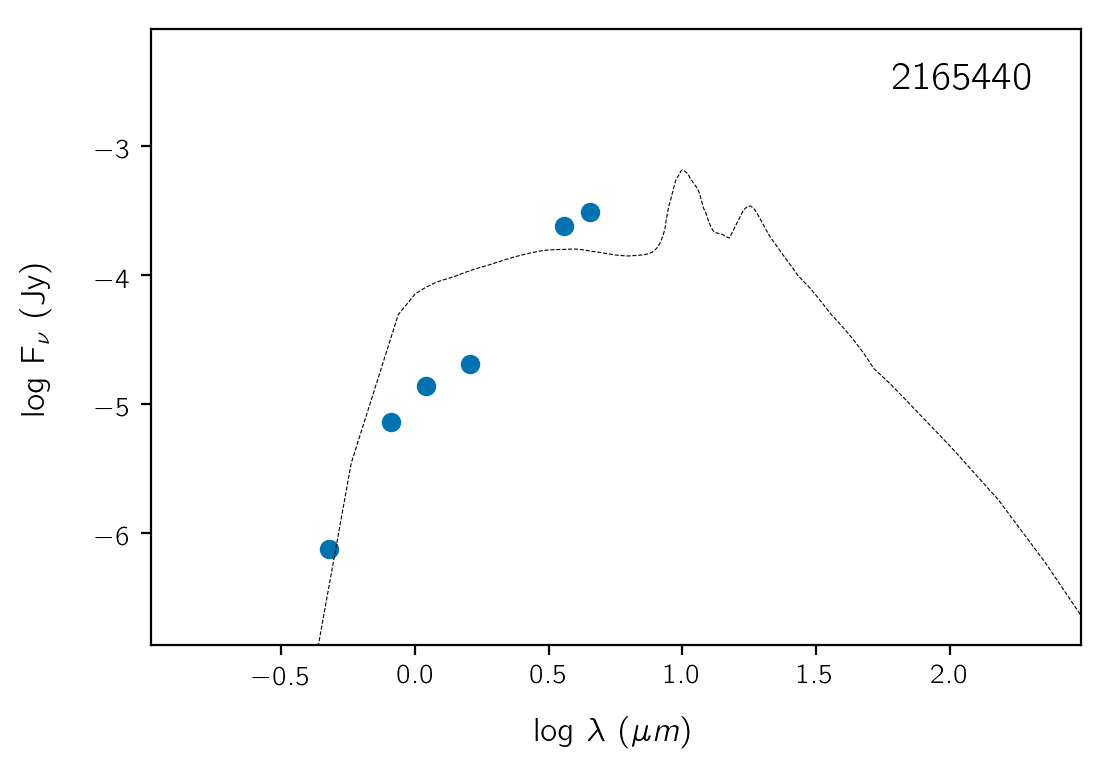}
    \caption{M31 clusters identified by \citet{Johnson2015} where we have also identified x-AGB candidates (red circles). On the left of each panel is an \hst\ F110W image cut-out with dimensions of four times the cluster apparent radius (APRAD; dashed blue circle). Additional AGB candidates that we have identified in these clusters are shown with orange circles. The right panel shows the SEDs of the dusty cluster sources (in red on the left) fit with a grid of oxygen-rich radiative transfer models using the Dusty Evolved Star Kit \citep[\desk;][]{Goldman2020}. \\}
    \label{fig:d_cluster}
\end{figure}

\setcounter{figure}{22}

\begin{figure}
    \centering
    \includegraphics[height=\dclusterfigsize]{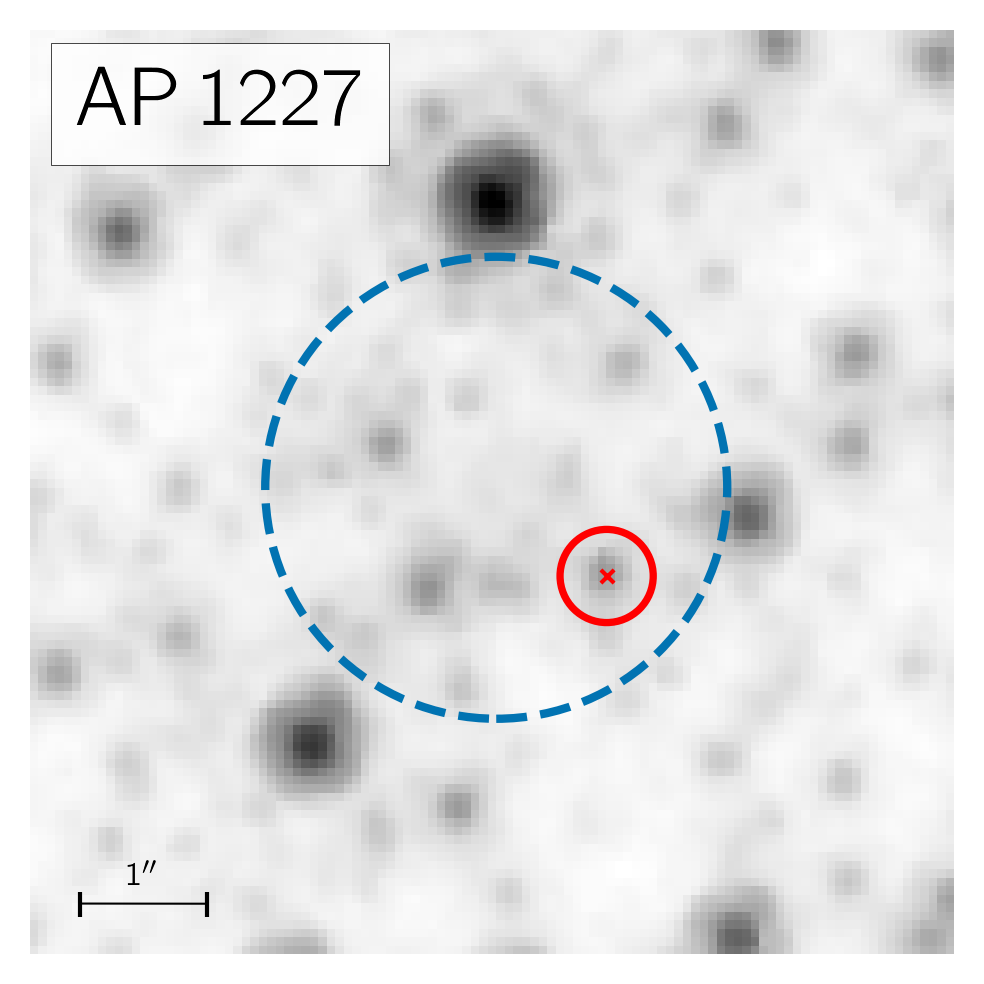}
    \includegraphics[height=\dclusterfigsize]{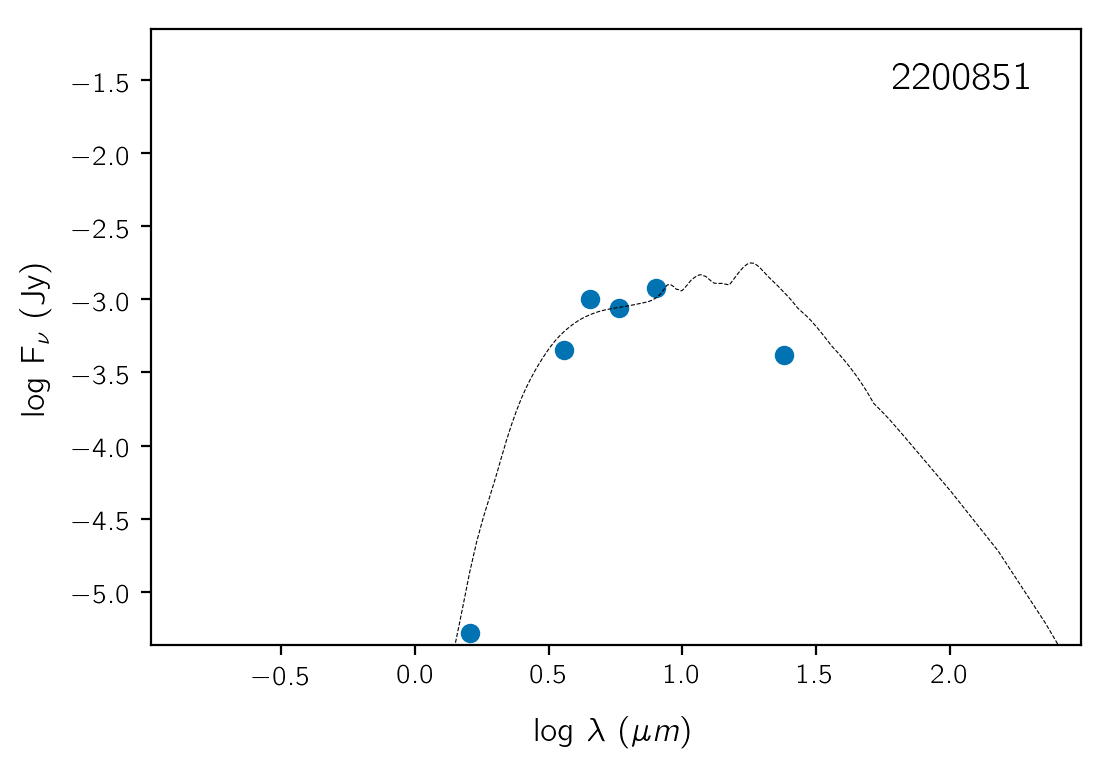} \hspace{\dclustergap}
    \includegraphics[height=\dclusterfigsize]{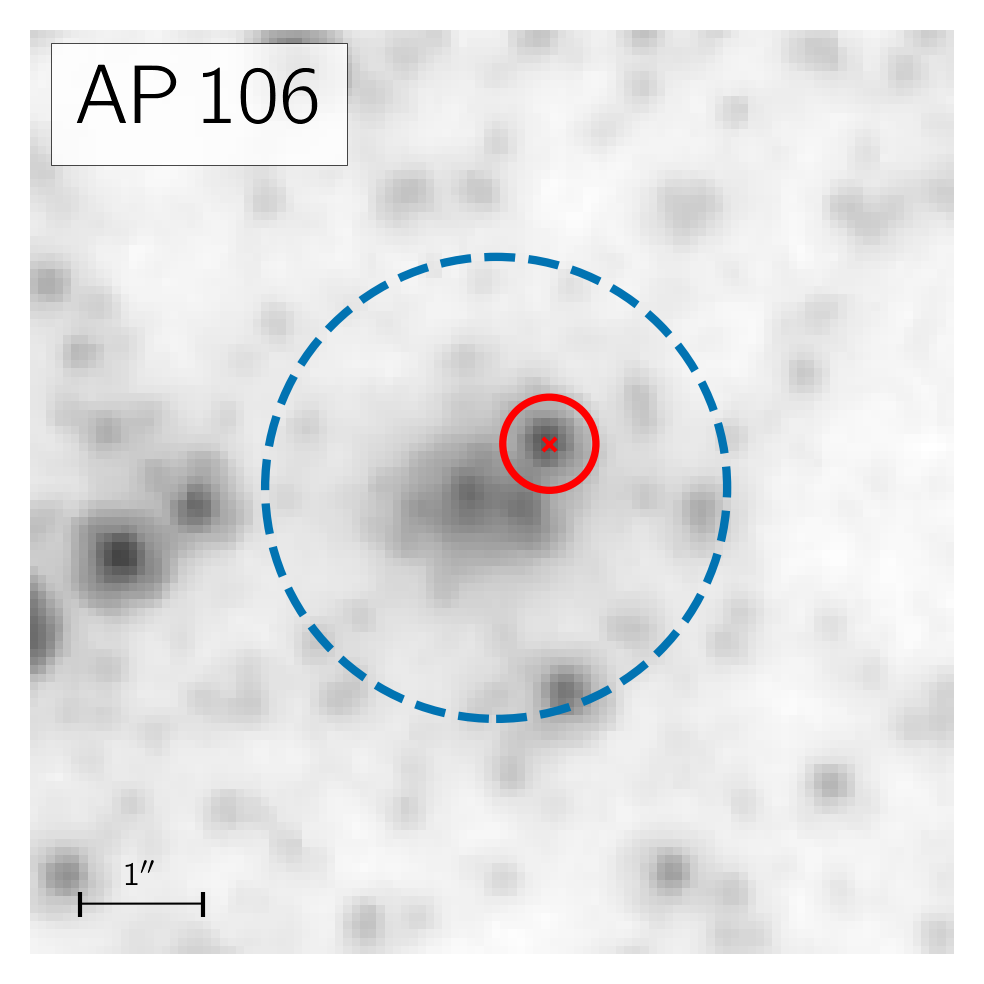}
    \includegraphics[height=\dclusterfigsize]{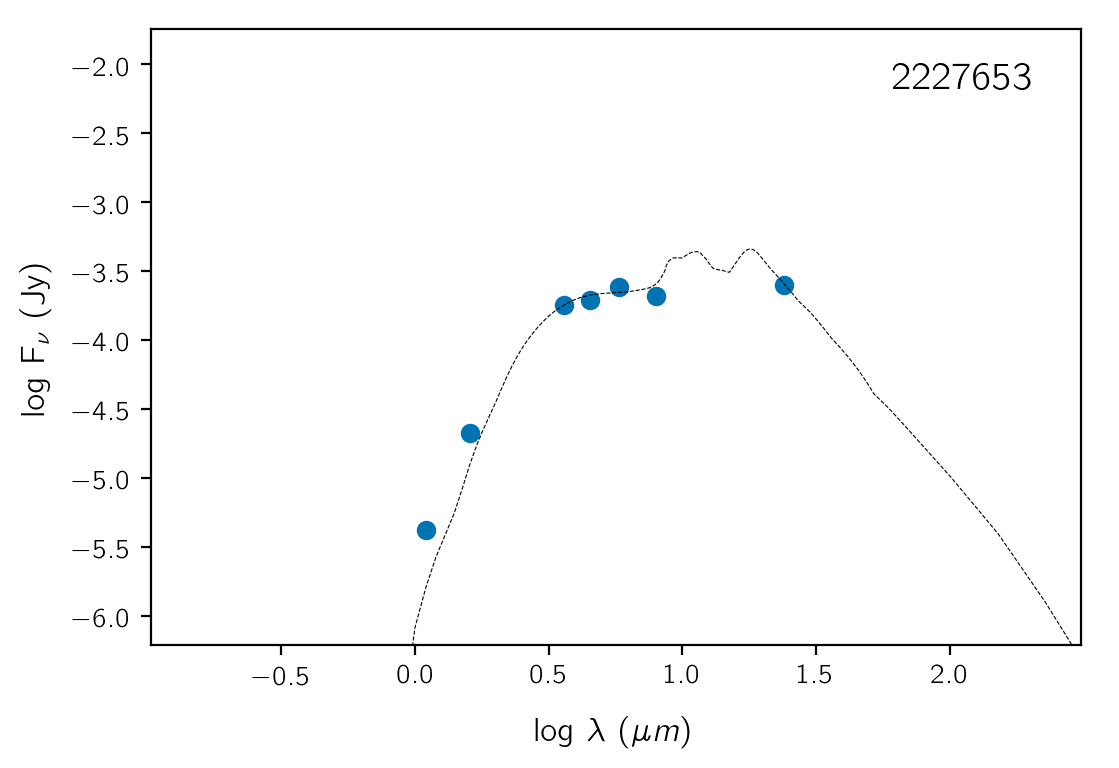} \\
    \includegraphics[height=\dclusterfigsize]{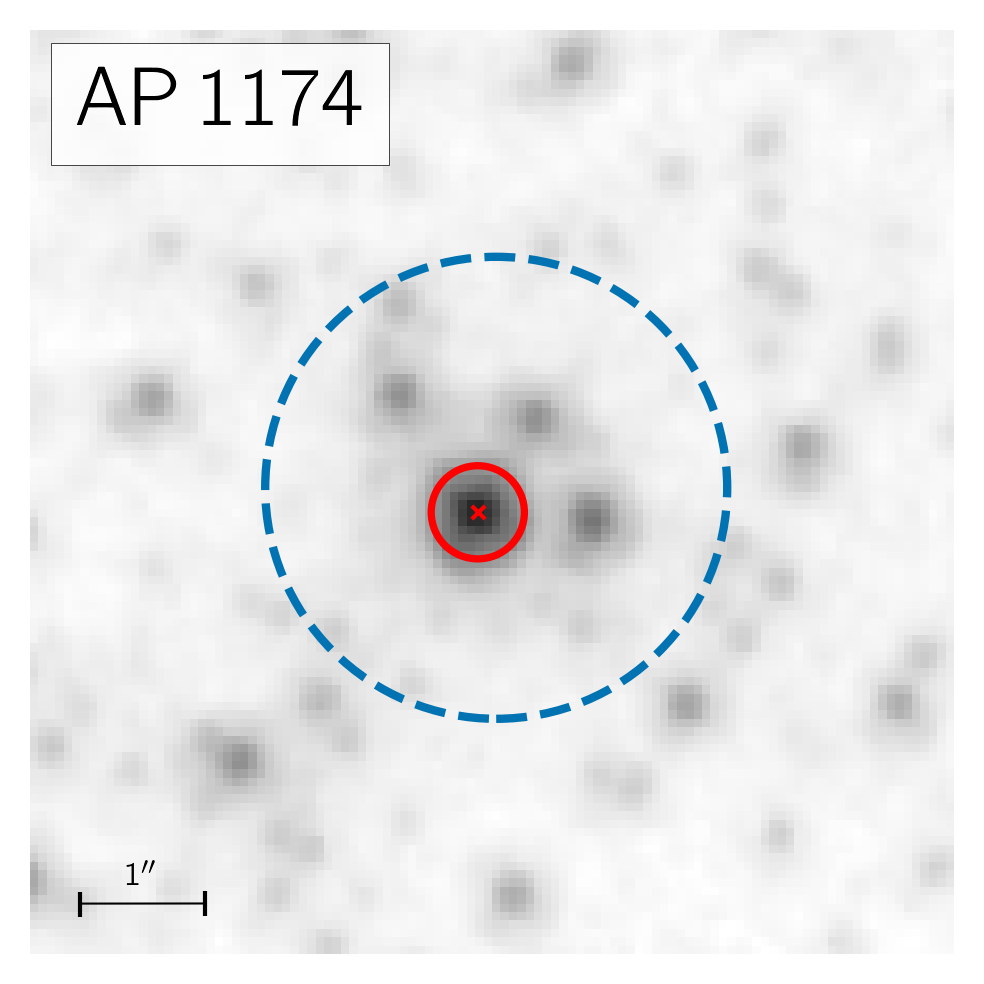}
    \includegraphics[height=\dclusterfigsize]{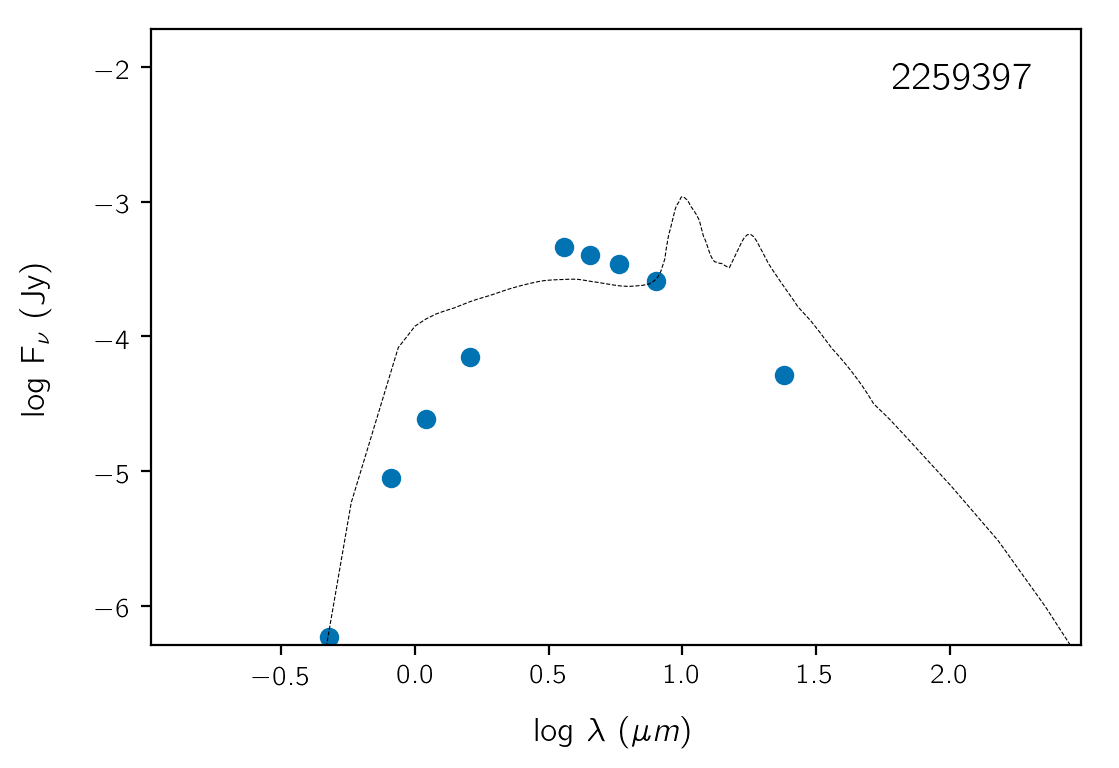} \hspace{\dclustergap}
    \includegraphics[height=\dclusterfigsize]{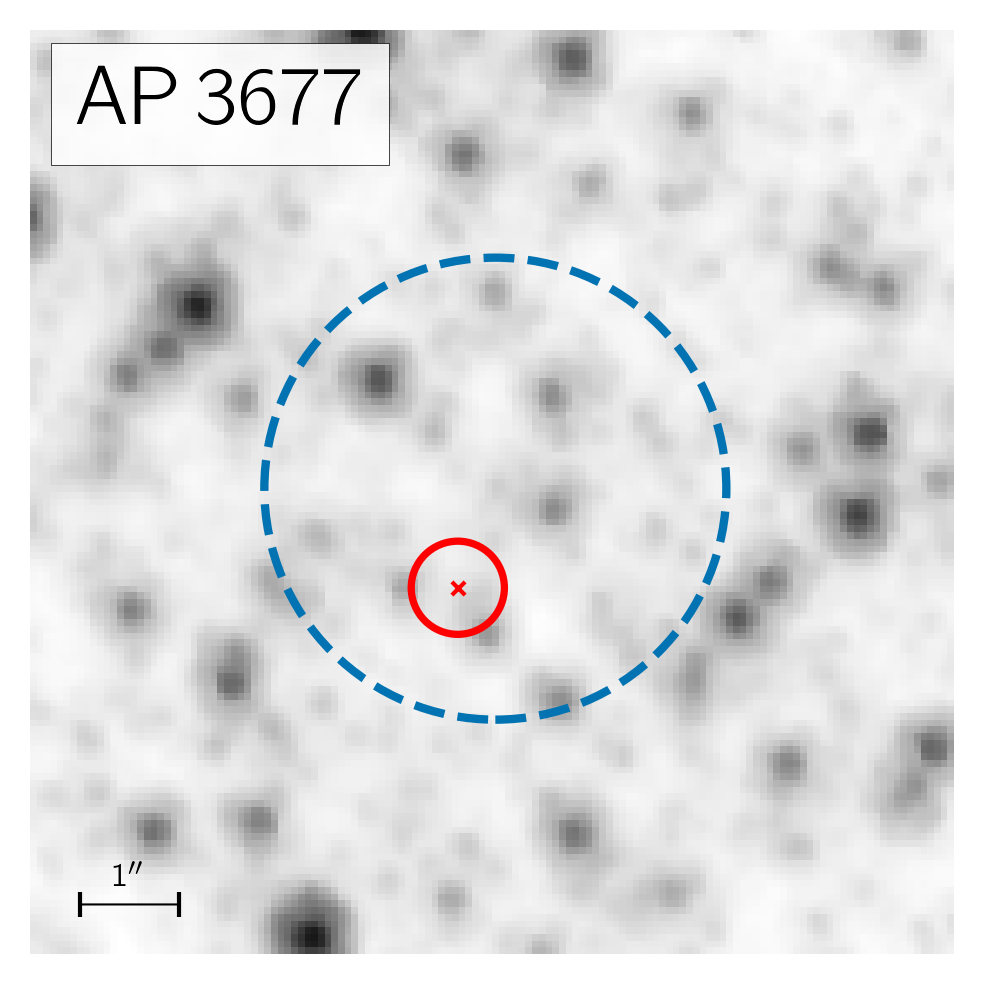}
    \includegraphics[height=\dclusterfigsize]{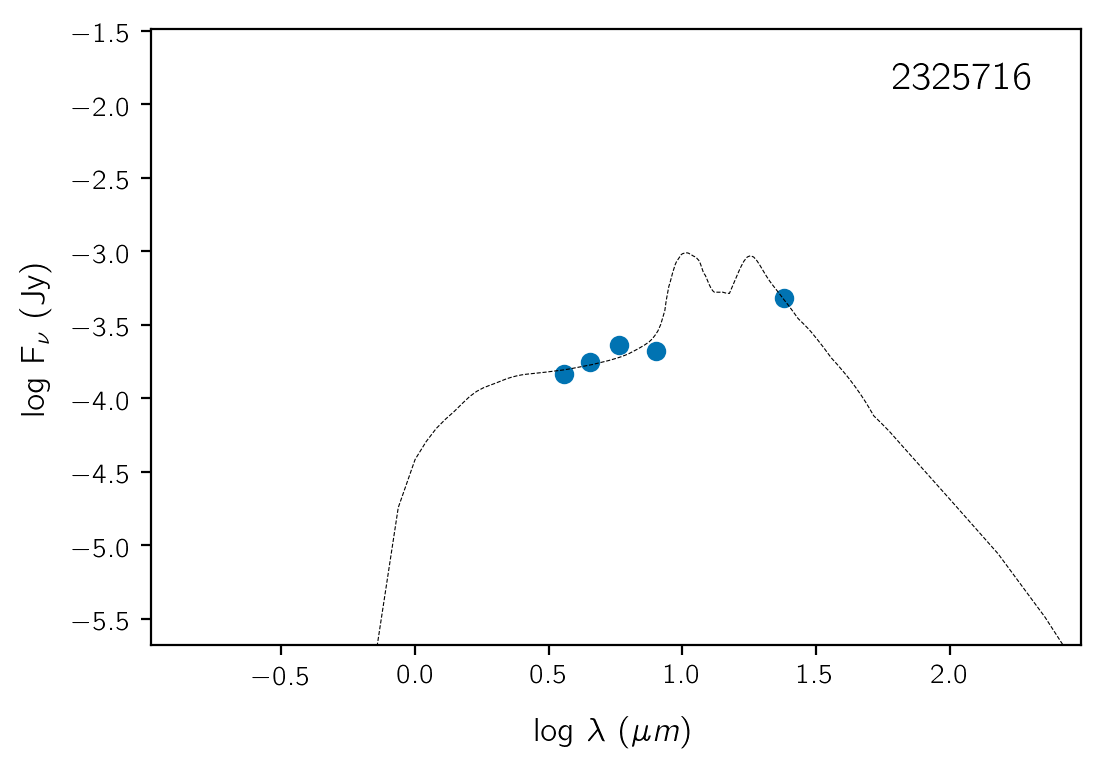} \\
    \includegraphics[height=\dclusterfigsize]{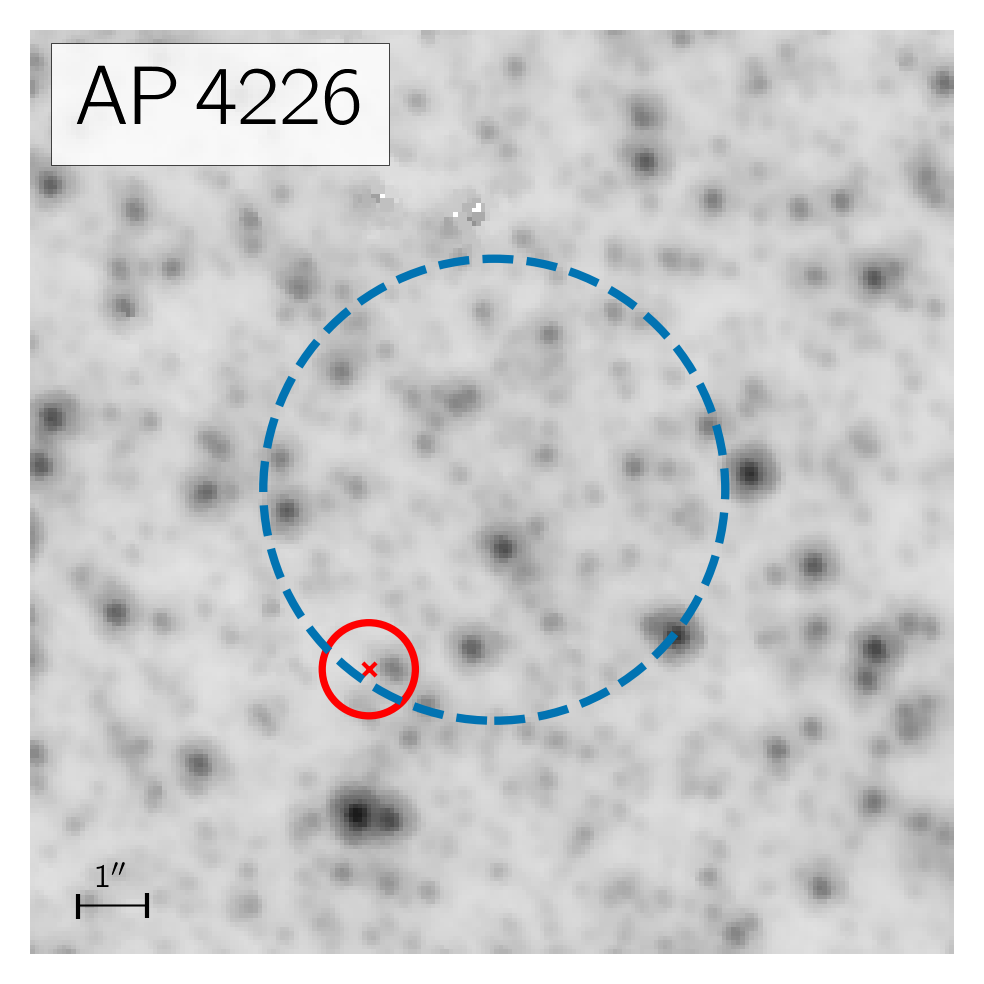}
    \includegraphics[height=\dclusterfigsize]{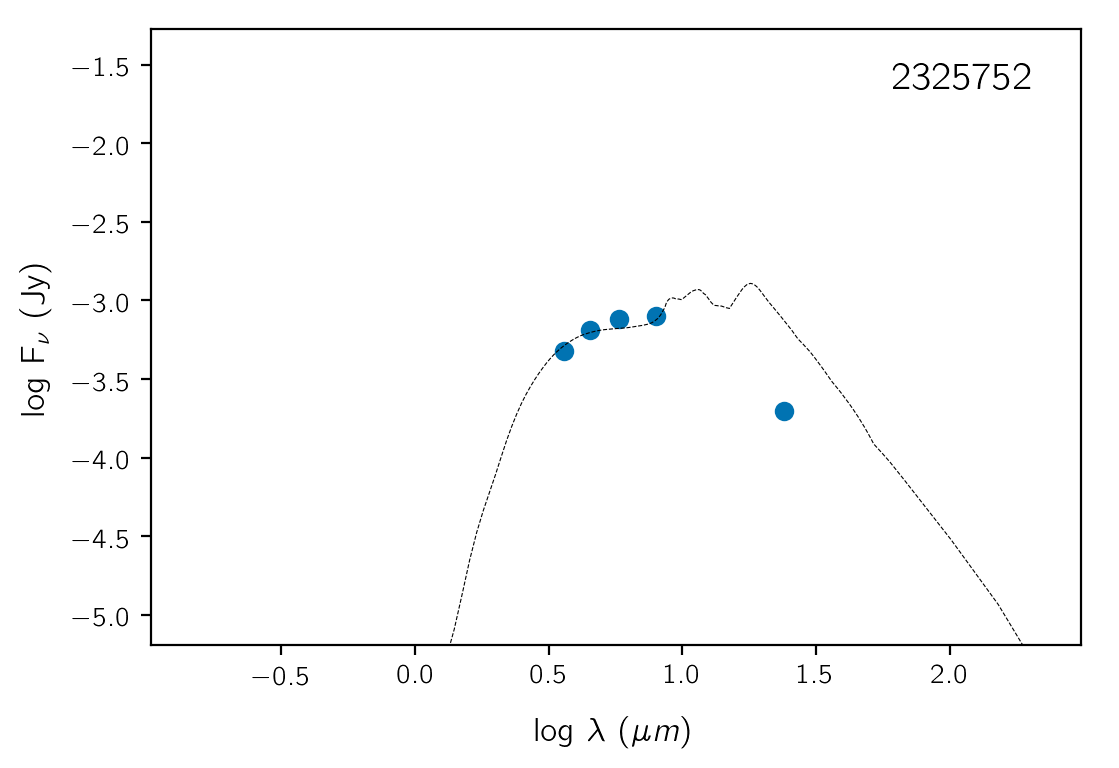}
    \caption{continued.}
\end{figure}

\begin{figure}
    \centering
    \includegraphics[height=\dclusterfigsize]{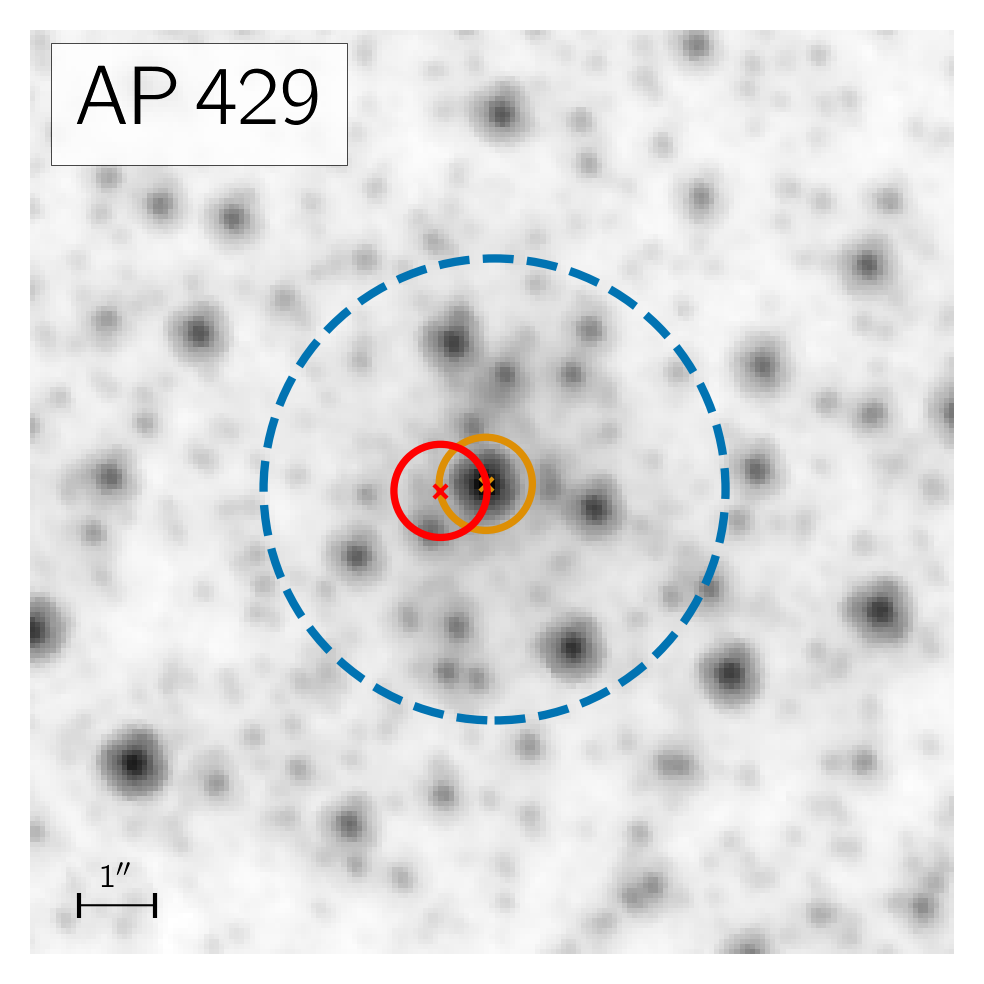}
    \includegraphics[height=\dclusterfigsize]{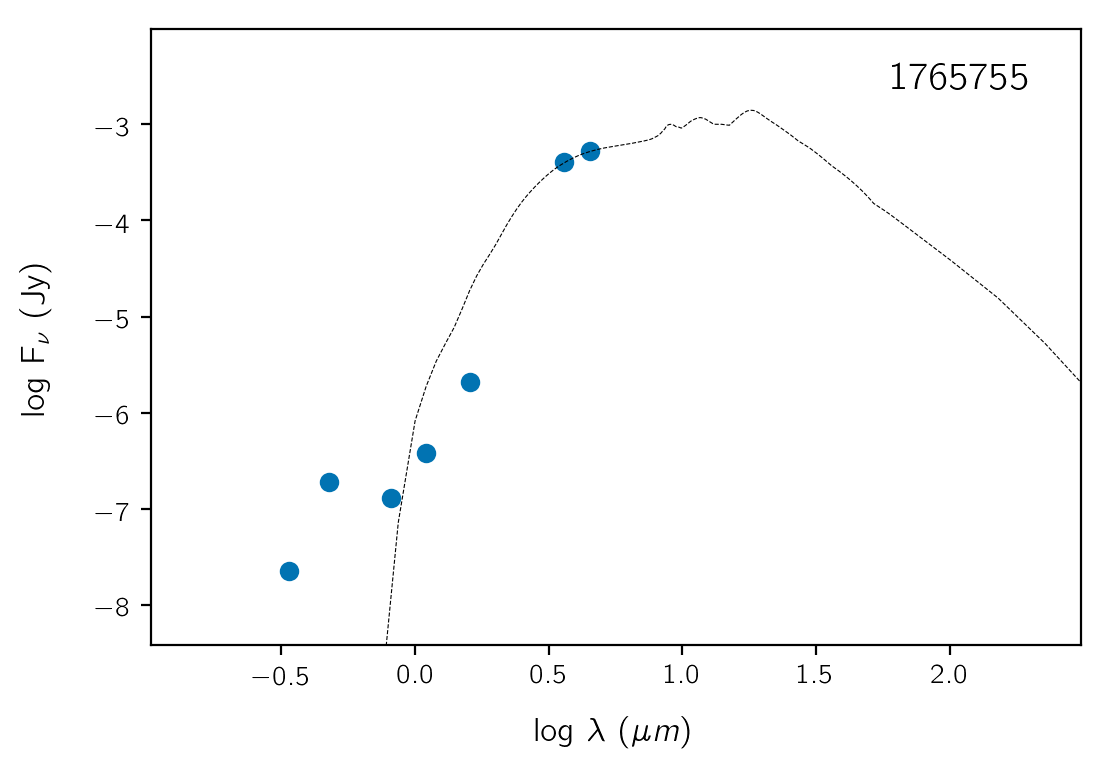}
    \includegraphics[height=\dclusterfigsize]{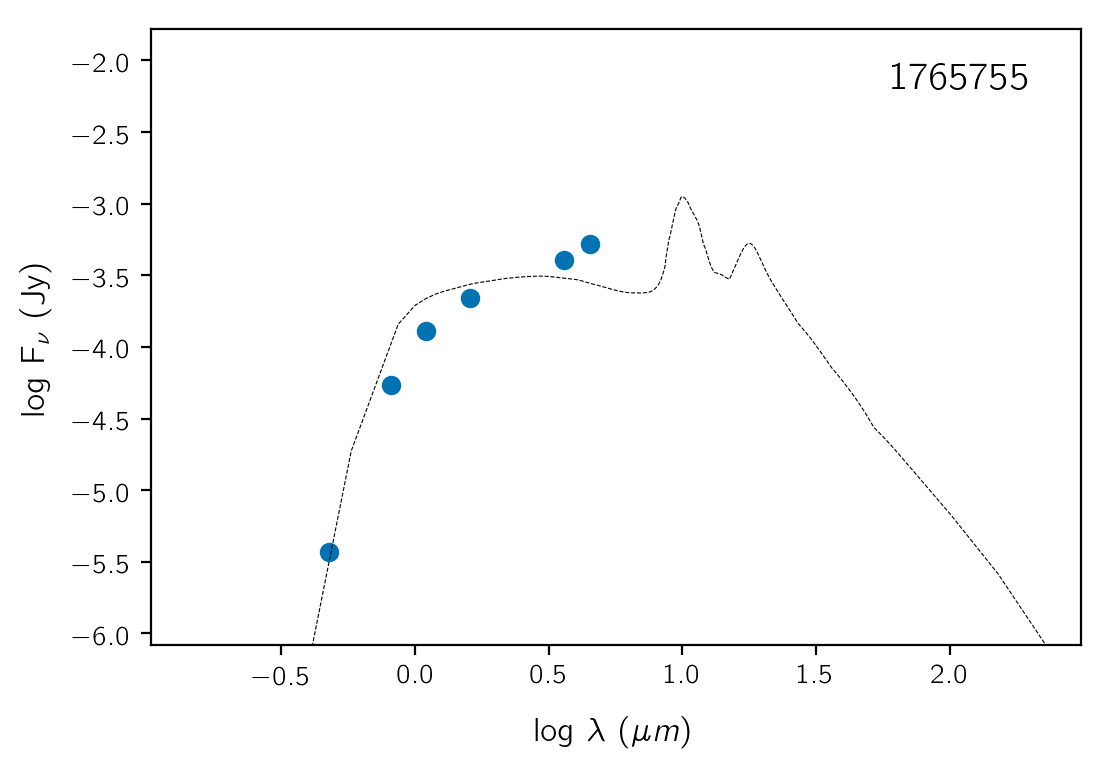} \\
    \includegraphics[height=\dclusterfigsize]{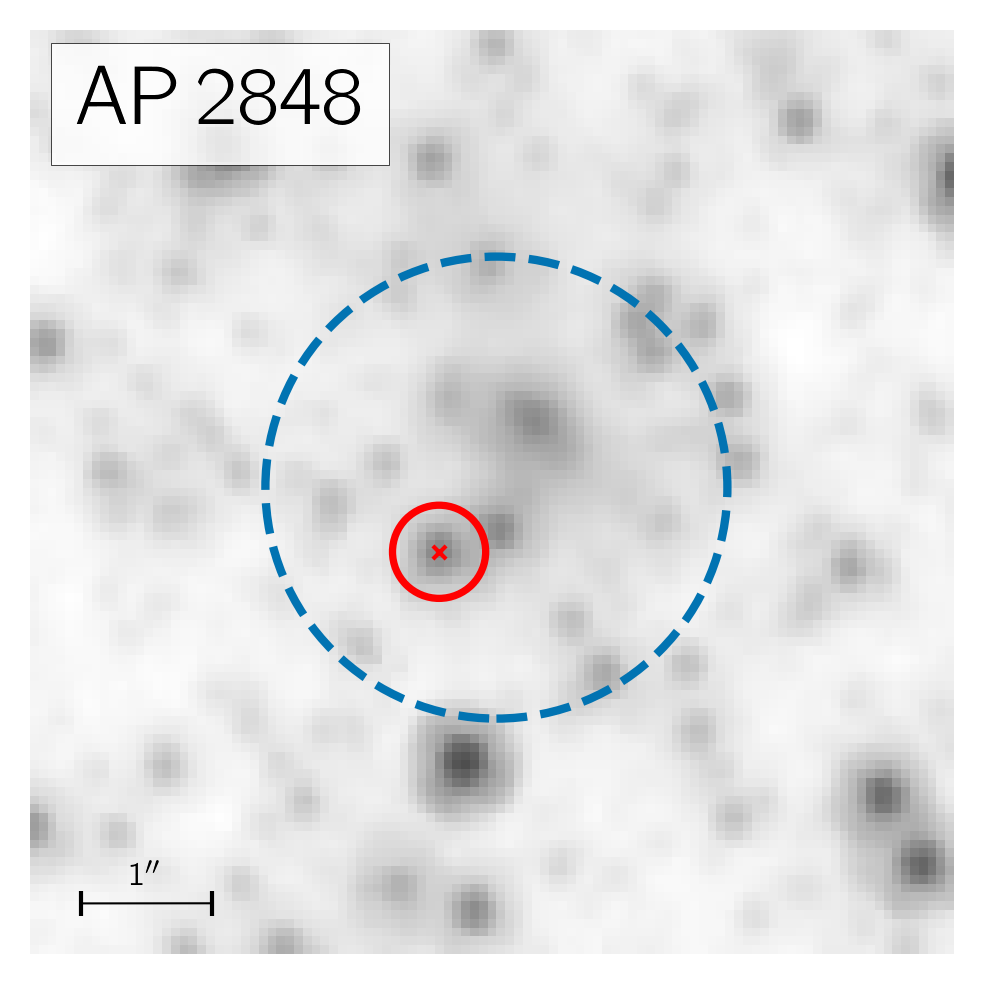}
    \includegraphics[height=\dclusterfigsize]{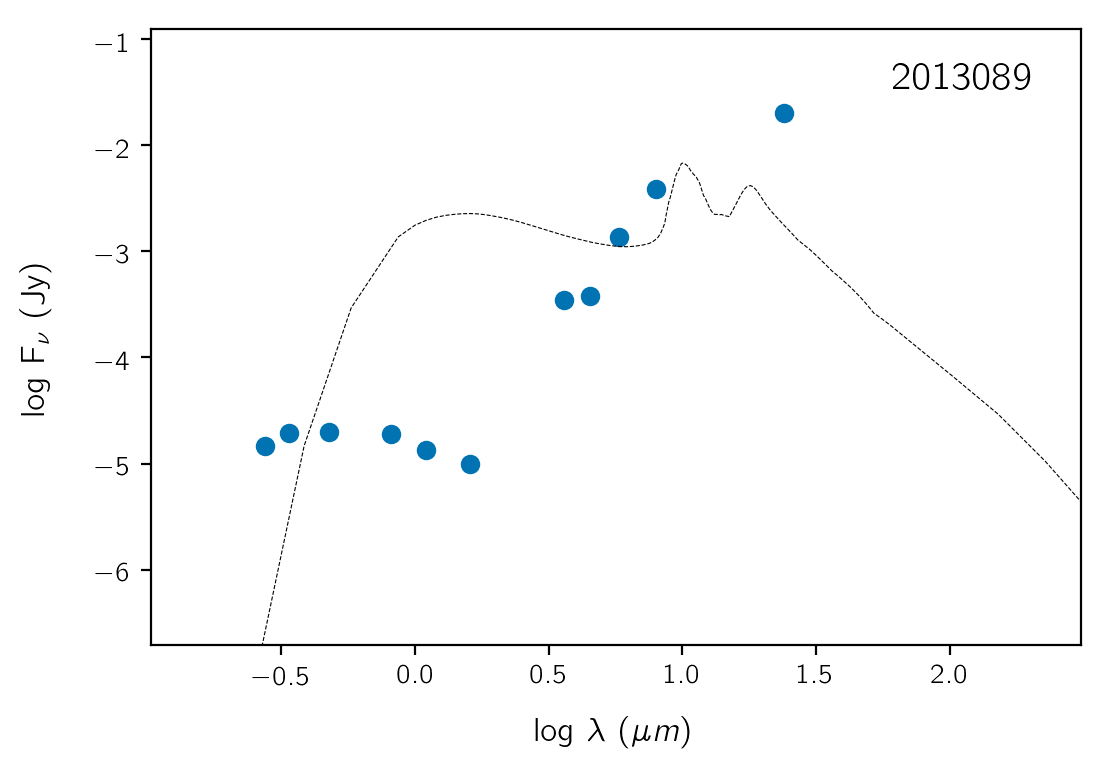}\hspace{0.3cm}
    \caption{Same as Figure \ref{fig:d_cluster}, but showing the mismatched cluster x-AGB candidate 1765755 (top) and a non-AGB source that we have incorrectly classified (bottom).\\ }
    \label{fig:dusty_cluster_mismatch}
\end{figure}

\begin{figure}
    \centering
    \includegraphics[height=\dclusterfigsize]{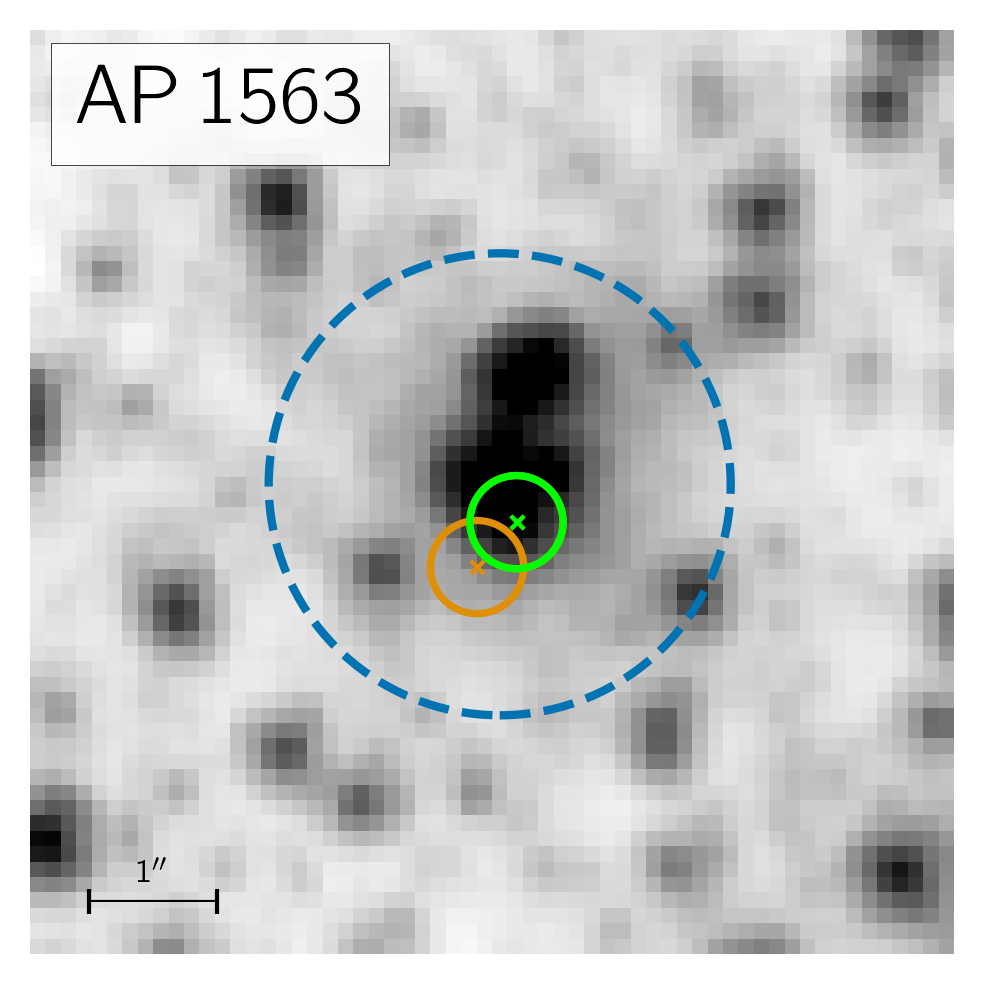}
    \includegraphics[height=\dclusterfigsize]{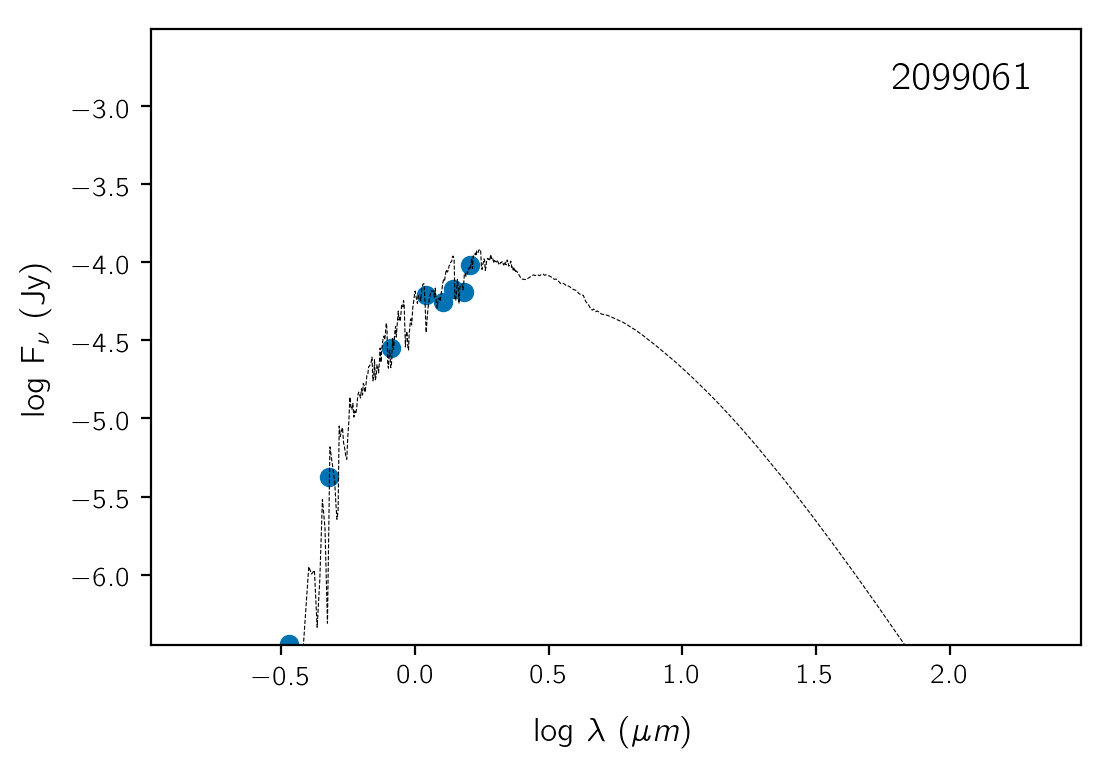}
    \caption{Same as Figure \ref{fig:d_cluster}, but showing our carbon-rich cluster AGB candidate (green circle). The SED was fit with the carbon-rich J1000-LMC dust growth radiative transfer model grid from \citet{Nanni2019} using the \desk\ and are shown with assumed distances of 776\,kpc. The age of the cluster (Log Age\,[yr] = 6.9) is likely too young to host an AGB star, and is likely a spatially co-incident carbon-rich field AGB star.\\ }
    \label{fig:cluster_carbon}
\end{figure}

\paragraph{\bf 1740537} This is the only cluster x-AGB candidate for which we have a measured metallicity for the cluster ([Fe/H]$ = -1.5$).\\

\paragraph{\bf 1765755} The automatic matching resulted in what appears to be a mismatch, but if we instead match the \spitz\ photometry to the brightest near-IR \hst\ cluster source (only $0\overset{\prime\prime}{.}5$ away), the resulting shape of the SED is more consistent with an AGB star (center and right of the top row of Figure \ref{fig:cluster_carbon}).\\

\paragraph{\bf 2013089} This source does not have a SED consistent with an AGB star. It may instead be a dusty YSO, background galaxy, or spatially co-incident sources super-imposed.\\

\paragraph{\bf 2099061} This is our only cluster source that was previously classified by \citet{Boyer2019} as a carbon star. This source, however, lies in a young cluster (log Age = 6.9) that is likely too young to host an AGB population. If the source is associated with the cluster, it would be expected to be a high-mass evolved star with an oxygen-rich chemistry. It is more-likely that the source is a lower-mass field carbon star, spatially coincident with the cluster.

\begin{table}[]
    \centering
    \begin{tabular}{|c|c|c|}
    \hline
    \multirow{2}{*}{ID} & Log Age & $[3.6]-[4.5]$ \\
    & (yr) & (mag)\\
    \hline
    \textcolor{gray}{1474474} &  \textcolor{gray}{6.69} & \textcolor{gray}{0.27} \\
    \textcolor{gray}{1671652} &  \textcolor{gray}{7.30} & \textcolor{gray}{1.11} \\
    \textcolor{gray}{1691755} &  \textcolor{gray}{7.80} & \textcolor{gray}{0.39} \\
    1740537 & \llap{1}0.14 & 0.41 \\
    1765755 &  8.50 & 0.77 \\
    1776677 &  9.19 & 0.70 \\
    1958200 & \llap{1}0.12 & 0.60 \\
    \textcolor{gray}{1998948} &  \textcolor{gray}{6.90} & \textcolor{gray}{0.76} \\
    \textcolor{gray}{2013089} &  \textcolor{gray}{6.80} & \textcolor{gray}{0.58} \\
    2070778 &  9.11 & 0.28 \\
    2088165 &  8.50 & 0.70 \\
    \textcolor{gray}{2165440} &  \textcolor{gray}{7.77} & \textcolor{gray}{0.75} \\
    2200851 &  8.39 & 1.35 \\
    2227653 &  8.80 & 0.57 \\
    2259397 &  8.19 & 0.35 \\
    2325716 &  8.43 & 0.68 \\
    \textcolor{gray}{2325752} &  \textcolor{gray}{6.69} & \textcolor{gray}{0.82} \\
    \hline
    \end{tabular}
    \vspace{0.3cm}
    \caption{The cluster x-AGB candidates, the cluster ages, and the \spitz\ IRAC [3.6]$-$[4.5] color, indicative of strong dust production. Several of our x-AGB candidates lie in clusters that are expected to be too young (Log age $<$ 8.0) to host AGB stars (shown in gray). These are likely M31 field AGB stars, supergiants, or foreground stars. }
    \label{table:cluster_ages}
\end{table}
\end{document}